\documentclass[sigconf, nonacm]{acmart}
\usepackage[utf8]{inputenc}
\pdfoutput=1
\newcommand\vldbdoi{10.14778/3626292.3626296}
\newcommand\vldbpages{119 - 133}
\newcommand\vldbvolume{17}
\newcommand\vldbissue{2}
\newcommand\vldbyear{2023}
\newcommand\vldbauthors{\authors}
\newcommand\vldbtitle{\shorttitle} 
\newcommand\vldbavailabilityurl{https://github.com/illinoisdata/ElasticNotebook}
\newcommand\vldbpagestyle{empty} 

\usepackage{bm}
\usepackage[linesnumbered,ruled]{algorithm2e}
\usepackage{algorithmic}
\usepackage[capitalise]{cleveref}
\crefname{section}{§}{§§}
\Crefname{section}{§}{§§}
\crefformat{section}{\S#2#1#3}
\crefname{figure}{Fig}{Figs}
\Crefname{figure}{Fig}{Figs}
\crefname{problem}{Prob}{Probs}
\Crefname{problem}{Prob}{Probs}
\usepackage{subfiles} 
\usepackage{multirow}
\usepackage{tabularx}
\usepackage{xspace}
\usepackage{subcaption}
\usepackage{pgfplots}
\usepackage{pgfplotstable}
\usepackage{ifthen}
\usepackage[shortlabels]{enumitem}
\setlist[enumerate]{leftmargin=1cm,topsep=0.5mm}
\setlist[itemize]{leftmargin=0.5cm,topsep=0.5mm}

\usepackage{tikz}
\usetikzlibrary{calc}
\usetikzlibrary{fit}
\usetikzlibrary{patterns}
\usetikzlibrary{decorations.pathreplacing}
\usetikzlibrary{arrows,calc,arrows.meta}
\usepackage{xcolor}
\usetikzlibrary{matrix,positioning}
\usetikzlibrary{shapes.geometric}

\definecolor{Ours0Color}{HTML}{ABDDA4}
\definecolor{Ours16Color}{HTML}{72C166}
\definecolor{Ours32Color}{HTML}{38A528}
\definecolor{PrestoColor}{HTML}{999999}
\definecolor{PostgresColor}{HTML}{6D6D6D}
\definecolor{GreenColor}{HTML}{38A528}
\definecolor{YellowColor}{HTML}{ffb570}
\definecolor{BlueColor}{HTML}{7081ff}
\definecolor{PinkColor}{HTML}{ffb0c2}

\definecolor{ComputeColor}{HTML}{ffb0c2}
\definecolor{ReadColor}{HTML}{cf3457}
\definecolor{WriteColor}{HTML}{ffb570}

\definecolor{vintagegreen}{HTML}{ABDDA4}
\definecolor{OursColor}{HTML}{38A528}
\definecolor{GreedyColor}{HTML}{7081ff}
\definecolor{RandomColor}{HTML}{ffb570}
\definecolor{NoneColor}{HTML}{6D6D6D}

\definecolor{Redborder}{HTML}{805861}
\definecolor{Greenborder}{HTML}{384180}
\definecolor{Blueborder}{HTML}{566F52}
\definecolor{Greyborder}{HTML}{4D4D4D}
\definecolor{Lightgrey}{HTML}{dadada}
\definecolor{ExampleColor1}{HTML}{7081ff}
\definecolor{ExampleColor2}{HTML}{ffb0c2}
\definecolor{Lightred}{HTML}{ffb09c}
\definecolor{Lightblue}{HTML}{b8e2f2}
\definecolor{FlagColor}{HTML}{CCCCCC}

\definecolor{vintageblue}{HTML}{7081ff}
\definecolor{vintagered}{HTML}{ffb0c2}

\definecolor{NoOptColor}{HTML}{264653}
\definecolor{LRUColor}{HTML}{777777}
\definecolor{RandomColor}{HTML}{2a9d8f}
\definecolor{GreedyColor}{HTML}{e9c46a}
\definecolor{HeuristicColor}{HTML}{f4a261}
\definecolor{SCColor}{HTML}{e76f51}
\definecolor{AllColor}{HTML}{CCCCCC}

\definecolor{SAColor}{HTML}{ffb0c2}
\definecolor{SeparatorColor}{HTML}{9b5de5}
\definecolor{BlueColor}{HTML}{0081a7}

\newcommand{\midsepremove}{\aboverulesep = 0.3mm \belowrulesep = 0.3mm}
    \midsepremove
    \newcommand{\midsepdefault}{\aboverulesep = 0.605mm \belowrulesep = 0.984mm}
    \midsepdefault

\usepackage{amsthm}
\theoremstyle{definition}

\newtheorem{definition}{Definition}
\newtheorem{problem}{Problem}

\newcommand{\system}{{\sf ElasticNotebook}\xspace}
\newcommand{\systembf}{{\sffamily\bfseries ElasticNotebook}\xspace}
\newcommand{\systemnosf}{ElasticNotebook\xspace}
\newcommand{\noidgraph}{{\sf EN (No ID graph)}\xspace}
\newcommand{\noidgraphbf}{{\sffamily\bfseries EN (No ID graph)}\xspace}

\newcommand{\kaggle}{{\sf Kaggle}\xspace}
\newcommand{\jwst}{{\sf JWST}\xspace}
\newcommand{\tutorial}{{\sf Tutorial}\xspace}
\newcommand{\homework}{{\sf Homework}\xspace}
\newcommand{\checkpoint}{{\sf RerunAll}\xspace}

\newcommand{\criu}{{\sf CRIU}\xspace}
\newcommand{\criubf}{{\sffamily\bfseries CRIU}\xspace}
\newcommand{\storedill}{{\sf \%Store}\xspace}
\newcommand{\storedillbf}{{\sffamily\bfseries \%Store}\xspace}
\newcommand{\dumpsession}{{\sf DumpSession}\xspace}
\newcommand{\dumpsessionbf}{{\sffamily\bfseries DumpSession}\xspace}

\newcommand{\helix}{{\sf ElasticNotebook + Helix}\xspace}
\newcommand{\helixbf}{{\sffamily\bfseries ElasticNotebook + Helix}\xspace}

\definecolor{vintagegray}{HTML}{e5e5e5}
\newtheoremstyle{abcd}
  {}
  {}
  {\itshape}
  {}
  {\bfseries}
  {.}
  {.5em}
  {}
\theoremstyle{abcd}
\usepackage{pgfplots}
\setenumerate{label={\arabic*.},noitemsep,topsep=2pt,leftmargin=14pt}

\makeatletter
\newcommand\resetstackedplots{
\makeatletter
\pgfplots@stacked@isfirstplottrue
\makeatother

\addplot [forget plot,draw=none] coordinates{(1,0) (2,0) (3,0)};
}

\makeatletter
\newcommand{\problemtitle}[1]{\gdef\@problemtitle{#1}}
\newcommand{\probleminput}[1]{\gdef\@probleminput{#1}}
\newcommand{\problemoutput}[1]{\gdef\@problemoutput{#1}}
\newcommand{\problemobjective}[1]{\gdef\@problemobjective{#1}}
\newcommand{\problemconstraint}[1]{\gdef\@problemconstraint{#1}}
\NewEnviron{myproblem}{
  \problemtitle{}\probleminput{}\problemquestion{}
  \BODY
  \par\addvspace{.5\baselineskip}
  \noindent
  \begin{tabularx}{\textwidth}{@{\hspace{\parindent}} l X c}
    \multicolumn{2}{@{\hspace{\parindent}}l}{\@problemtitle} \\
    \textbf{Input:} & \@probleminput \\
    \textbf{Output:} & \@problemoutput
    \textbf{Objective Function:} & \@problemobjective
    \textbf{Constraint:} & \@problemconstraint
  \end{tabularx}
  \par\addvspace{.5\baselineskip}
}
\makeatother

\begin{document}
\setlength{\abovedisplayskip}{3pt}
\setlength{\belowdisplayskip}{3pt}

\title[ElasticNotebook: Enabling Live Migration for Computational Notebooks]{ElasticNotebook: Enabling Live Migration for \\Computational Notebooks}

\author[Zhaoheng Li, Pranav Gor, Rahul Prabhu, 
    Hui Yu, Yuzhou Mao, Yongjoo Park]{Zhaoheng Li$^*$, Pranav Gor$^*$, Rahul Prabhu$^*$, 
    Hui Yu$^*$, Yuzhou Mao$^+$, Yongjoo Park$^*$}
\affiliation{%
  \institution{University of Illinois at Urbana-Champaign$^*$ 
    \quad University of Michigan$^+$}
}
\email{{zl20,gor2,rprabhu5,huiy3,yongjoo}@illinois.edu, yuzhom@umich.edu}








\begin{abstract}
Computational notebooks (e.g., Jupyter, Google Colab) are widely used for interactive data science and machine learning. 
In those frameworks, users can start a \emph{session}, then execute \emph{cells} (i.e., a set of statements) to create variables, train models, visualize results, etc.
Unfortunately, existing notebook systems do not offer live migration: when a notebook launches on a new machine, it loses its \emph{state}, preventing users from continuing their tasks from where they had left off.
This is because, unlike DBMS, the sessions directly rely on underlying kernels (e.g., Python/R interpreters)
without an additional data management layer.
Existing techniques for preserving states, such as copying all variables or OS-level checkpointing, are unreliable (often fail), inefficient, and platform-dependent. 
Also, re-running code from scratch can be highly time-consuming.

In this paper, we introduce a new notebook system, \system, that offers 
    live migration via checkpointing/restoration
using a novel mechanism that is reliable, efficient, and platform-independent.
Specifically, by observing all cell executions via transparent, lightweight monitoring, \system can find a reliable and efficient way (i.e., \emph{replication plan}) for reconstructing the original session state,
considering variable-cell dependencies, observed runtime, variable sizes, etc.
To this end, our new graph-based optimization problem finds how to reconstruct all variables (efficiently) from a subset of variables that can be transferred across machines.
We show that \system reduces 
    end-to-end migration and restoration times by 85\%-98\%
    and 94\%-99\%, respectively, 
on a variety (i.e., Kaggle, JWST, and Tutorial) of notebooks 
with negligible runtime and memory overheads of <2.5\% and <10\%.


\end{abstract}

\maketitle

\pagestyle{\vldbpagestyle}
\begingroup\small\noindent\raggedright\textbf{PVLDB Reference Format:}\\
\vldbauthors. \vldbtitle. PVLDB, \vldbvolume(\vldbissue): \vldbpages, \vldbyear.\\
\href{https://doi.org/\vldbdoi}{doi:\vldbdoi}
\endgroup
\begingroup
\renewcommand\thefootnote{}\footnote{\noindent
This work is licensed under the Creative Commons BY-NC-ND 4.0 International License. Visit \url{https://creativecommons.org/licenses/by-nc-nd/4.0/} to view a copy of this license. For any use beyond those covered by this license, obtain permission by emailing \href{mailto:info@vldb.org}{info@vldb.org}. Copyright is held by the owner/author(s). Publication rights licensed to the VLDB Endowment. \\
\raggedright Proceedings of the VLDB Endowment, Vol. \vldbvolume, No. \vldbissue\ %
ISSN 2150-8097. \\
\href{https://doi.org/\vldbdoi}{doi:\vldbdoi} \\
}\addtocounter{footnote}{-1}\endgroup


\section{Introduction}
\label{sec:intro}

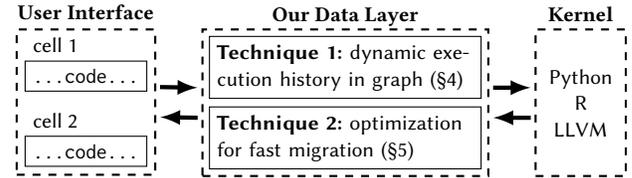
\begin{figure}[t]

\tikzset{
codenode/.style={
    draw=black,minimum width=16mm,
    font=\small\ttfamily,
},
cellnode/.style={
    font=\small\sffamily,
    anchor=south west,inner ysep=0,
},
technode/.style={
    draw=black,text width=34mm,
    align=left,
    font=\small\sffamily,
}
}

\centering
\begin{tikzpicture}

\def\g{0.6}

\node[codenode] (C1) at (-3,0) {...code...};
\node[cellnode] (T1) at ($(C1.north west)+(0,0.1)$) {cell 1};
\node[codenode] (C2) at ($(C1)+(0,-1.0)$) {...code...};
\node[cellnode] (T2) at ($(C2.north west)+(0,0.1)$) {cell 2};

\node[draw=black,dashed,thick,fit={(T1) (C2)}] (B) {};
\node[anchor=south,font=\small\bfseries] at (B.north) 
    {User Interface};

\node[draw=black,dashed,thick,anchor=north west,
    minimum width=38mm,minimum height=19.5mm] 
    (B2) at ($(B.north east)+(\g,0)$) {};
\node[anchor=south,font=\small\bfseries] 
    at ($(B2.north)+(0,-0.08)$)
    {Our Data Layer};
\node[anchor=north west,technode] 
    (T1) at ($(B2.north west)+(0.1,-0.1)$)
    {\textbf{Technique 1:}
        dynamic execution history in 
        graph (\cref{sec:data_modeling})};
\node[anchor=north west,technode] 
    (T2) at ($(T1.south west)+(0,-0.1)$)
    {\baselineskip=0pt \textbf{Technique 2:}
        optimization for fast migration (\cref{sec:algorithm})};

\node[draw=black,dashed,thick,anchor=north west,
    minimum width=12mm,minimum height=19.5mm,align=center,
    font=\small\sffamily] 
    (B3) at ($(B2.north east)+(\g,0)$) {
        Python \\
        R \\
        LLVM
    };
\node[anchor=south,font=\small\bfseries] 
    at ($(B3.north)+(0,-0.02)$)
    {Kernel};

\draw[-latex, ultra thick] 
    ($(B.east)+(0.05,0.2)$) -- ($(B2.west)+(-0.05,0.2)$);
\draw[latex-, ultra thick] 
    ($(B.east)+(0.05,-0.2)$) -- ($(B2.west)+(-0.05,-0.2)$);
\draw[-latex, ultra thick] 
    ($(B2.east)+(0.05,0.2)$) -- ($(B3.west)+(-0.05,0.2)$);
\draw[latex-, ultra thick] 
    ($(B2.east)+(0.05,-0.2)$) -- ($(B3.west)+(-0.05,-0.2)$);

\end{tikzpicture}

\vspace{-2mm}
\caption{Our transparent data layer (in the middle)
    enables robust, efficient, and platform-independent
    live migration.}
\vspace{-2mm}
\end{figure}


\begin{table*}[t]
\centering
\caption{Comparison between our \systembf and 
    other possible approaches to saving/restoring session states}
\label{tbl:existing_work}

\vspace{-3mm}
\small
\begin{tabular}{lll}
\toprule
\textbf{Approach}              &  \textbf{Mechanism}           \\ \midrule
Serialization-based tools~\cite{dumpsession, jupyterstore, marshal, dill, pickle}  & Serializes and stores variables during computing session (fails with unserializable variables)\\

System-level checkpointing~\cite{criu, juric2021checkpoint, ansel2009dmtcp, garg2018crum, jain2020crac} & Saves memory dump of computing session (high network cost and low portability) \\

Notebook Versioning and Replay~\cite{nodebooks, brachmann2020your, manne2022chex} & Enable re-execution of versioned notebook snapshots for result verification \\

   
 
Execution environment migration~\cite{wannipurage2022framework, ahmad2022reproducible} 
  
  & Migrates installed modules; useful in conjunction with (but orthogonal to) session state replication\\    

\textbf{Ours (\system)} 
& \textbf{Optimally combines copy/recompute
    for reliability, efficiency, and platform independence} \\
\bottomrule
\end{tabular}
\vspace{-2mm}
\end{table*}


\noindent
Computational notebooks%
\footnote{%
In this work, we use the term a ``notebook'' to mean either a system serving the notebook or the contents of the notebook,
    depending on the context.
}
(e.g., Jupyter~\cite{jupyter, ipython}, Rstudio~\cite{rstudio}) are widely used in data science and machine learning for
interactive tutorials~\cite{johnson2020benefits},
data exploration~\cite{crotty2015vizdom, zgraggen2014panoramicdata, dunne2012graphtrail}, 
visualization~\cite{eichmann2020idebench}, 
model tuning and selection~\cite{wagenmakers2004aic, bergstra2012random}, etc.
Cloud providers offer Software-as-a-Services 
(e.g., AWS hub~\cite{awsjupyterhub}, Azure ML studio~\cite{azuremlstudio}, Google Colab~\cite{colab}, 
IBM Watson studio~\cite{ibmwatson}) 
with commonly used libraries (e.g., Pandas, PyTorch).
A notebook workflow begins with a user starting 
    a \textit{computing session}.
Then, the user can execute a \emph{cell} 
    (i.e., a set of statements),
    one by one,
    to load datasets, create variables, train models, visualize results, etc.
The session can be terminated manually or automatically
    to save resources and costs.


\paragraph{Limitation: No Live Replication} 

Unfortunately, existing notebooks
    do not offer transparent infrastructure scaling
        (independent of applications),
which are becoming increasingly popular in the cloud
    for instant scalability and cost reduction
    (e.g., auto-scaling DBMS~\cite{verbitski2017amazon,poppe2022moneyball}, 
        micro-service orchestration~\cite{kubernetes,docker-swarm}).
That is,
    if we copy a notebook file
        to a new VM (e.g., for larger memory)
        or suspend a session to save costs,
    the resumed notebook loses its \emph{state} (i.e., a set of variables),
        having only code and outputs.
In other words,
    the user cannot resume their task from where they had previously left off.
This is
    because the notebooks directly rely on underlying kernels
        (e.g., Python/R interpreters, C++ REPL)
            without an additional data management layer.
Accordingly, the variables residing in processes
    are erased as they terminate with sessions.
To address this, we can potentially 
    save those variables and restore them
        on a new environment.
However, existing techniques 
    such as serializing all variables~\cite{dill,pickle,pythonreduce}
    and checkpointing OS processes~\cite{criu, ansel2009dmtcp, garg2018crum, jain2020crac}
    may fail, are inefficient, and platform-dependent
    (discussed shortly).
Finally, re-running code from scratch 
    can be time-consuming.

\paragraph{Our Goal.} 

We propose \system,
    a notebook system that offers
        live state migration via checkpointing/restoration
using a reliable, efficient, and platform-independent state replication mechanism.
\textbf{\emph{Reliability:}} It enables correct/successful replication for (almost) all notebooks.
\textbf{\emph{Efficiency:}} It is significantly more efficient than others.
\textbf{\emph{Platform-independence:}} It does not rely on platform-/architecture-specific features.
That is, \system enables \emph{live notebook replication}
    for potentially all notebook workloads
        by introducing a novel data management layer.
For example,
    if a user specifies a new machine to run a currently active notebook,
        the system transparently replicates the notebook,
    including all of its variables,
    as if the notebook has been running on the new machine.
If we can provide this capability with little to no modifications
    to existing systems (e.g., Jupyter),
we can offer benefits to a large number of data scientists
    and educators who use notebooks.
To achieve this,
    we must overcome the following technical challenges.

\paragraph{Challenge}


Creating a reliable, efficient, and platform-independent 
    replication mechanism is challenging.
First, the mechanism must offer high coverage.
    That is, for almost all notebooks people create,
        we should be able to successfully replicate them across machines.
Second, the mechanism should be 
    significantly faster than 
    straightforward approaches---rerunning all the cells exactly as they were run 
        in the past, or copying, if possible, all the variables 
            with serialization/deserialization.
Third, the mechanism should integrate with existing notebook systems with clean separation
    for sustainable development and easier adoption.

\paragraph{Our Approach}

Our core idea is that
by observing the evolution of session states
    via lightweight monitoring,
we can address the three important challenges---reliability, efficiency, and 
    platform-independence---by combining
program language techniques (i.e., on-the-fly code analyses) and 
    novel algorithmic solutions (i.e., graph-based mathematical optimization).
Specifically, to represent session state changes,
    we introduce the \emph{application history}, a special form of bipartite graph
        expressing the dependencies among variables and cell executions.
Using this graph, we take the following approach.

First, we achieve \emph{reliability} and \emph{platform independence} by choosing 
    a computational plan (or \emph{replication plan}) that can safely reconstruct 
        platform-dependent variables (e.g., Python \texttt{generators}, incompletely defined custom classes)
        based on the other platform-independent variables.
That is, in the presence of variables that cannot be serialized
    for platform-independent replication,
\system uses the application history to recompute them dynamically
    on a target machine.
In this process,
    \system optimizes for the collective cost of recomputing all such variables
        while still maintaining their correctness (\cref{sec:data_modeling}).

Second, for \emph{efficiency}, \system optimizes its replication plan
    to determine (1) the variables that will be copied, and 
    (2) the variables that will be recomputed based on the copied variables,
to minimize the end-to-end migration (or restoration) time
    in consideration of serialization costs, recomputation costs, data transfer costs, etc.
For example,
    even if a variable can be reliably transferred across machines,
        the variable may still be dynamically constructed
    if doing so results in a lower total cost.
To make this decision in a principled way,
    we devise a new graph-based optimization problem, which reduces to
        a well-established min-cut problem (\cref{sec:algorithm}).

\textit{\textbf{Implementation:}}
While our contributions can apply to 
    many dynamically analyzable languages (e.g., Python/R, LLVM-based ones),
    we implement our prototype (in C and Python) 
    for the Python user interface,
        which is widely used for data science, machine learning, statistical analysis, etc.
Specifically,
    \system provides a data management layer to Jupyter
    as a hidden \emph{cell magic}~\cite{jupytermagic}
    to transparently monitor cell executions
    and offer efficient replication.

\paragraph{Difference from Existing Work} 

Compared to existing work,
we pursue a significantly different direction.
For example, there are tools that make data serialization
    more convenient~\cite{jupyterstore,dumpsession};
however, they fail if a session contains
    non-serializable variables,
        and are inefficient
because they do not consider opportunities for
    dynamic recomputation.
Alternatively, system-level checkpointing~\cite{criu,ansel2009dmtcp,garg2018crum,jain2020crac}
is platform-dependent,
    limited to checkpointing memory (e.g., not GPU),
    less efficient than ours since
        dynamic recomputation is impossible. 
Building on top of result reuse~\cite{garcia2020hindsight, xin2018helix} and 
    lineage tracing~\cite{pimentel2017noworkflow, head2019managing, shankar2022bolt},
    we introduce deeper (reference-aware) analyses (\cref{sec:data_modeling:construction})
        and novel optimization techniques to incorporate
        unique constraints such as inter-variable dependencies
            (\cref{sec:algorithm})
        and also empirically confirm their effectiveness (\cref{sec:exp_robust}).
Completely orthogonal work includes
    library migration~\cite{wannipurage2022framework, ahmad2022reproducible} and
    scalable data science~\cite{moritz2018ray,xin2021enhancing,petersohn2020towards}.
\cref{tbl:existing_work} summarizes differences.

\paragraph{Contributions}

Our contributions are as follows:
\begin{itemize}
    \item \textbf{Motivation.} We discuss alternative approaches and 
        explain the advantage of our approach. (\cref{sec:motivation})

    \item \textbf{Architecture.} We describe our system architecture 
        for achieving efficient and robust session replication. (\cref{sec:system_overview})

    \item \textbf{Data Model.} We introduce a novel data model (\textit{Application History Graph}) 
        for expression session history, 
        which enables efficient and accurate state replication. (\cref{sec:data_modeling})
    
    \item \textbf{Optimization Problem and Solution.} We formally define the optimization problem of minimizing state replication cost through balancing variable copying and recomputation. We propose an efficient and effective solution. (\cref{sec:algorithm})

    
    \item \textbf{Evaluation.} 
    We show \system reduces upscaling, downscaling, and restore times by 85\%-98\%, 84\%-99\%, and 94\%-99\%, respectively. 
    Overheads are negligible (<2.5\% runtime). (\cref{sec:experiments})
\end{itemize}

\section{Motivation}
\label{sec:motivation}

This section describes use cases (\Cref{sec:checkpoint_motivation}) and requirements (\cref{sec:background_abstraction}) for session replication, and our intuition for higher efficiency (\cref{sec:background_graph,sec:background_example}). 

\subsection{Why is Live Migration Useful?}
\label{sec:checkpoint_motivation}
A seamless state replication for computational notebooks
    can allow easier infrastructure scaling and 
        frequent session suspension, without interrupting user workflow,
as described below.


\paragraph{Fast Replication for Elastic Computing}

The ability to move a state across machines is useful for
    scaling resources~\cite{cunha2021context,juric2021checkpoint},
    allowing us to migrate a live session to the machines with the right equipment/resources 
        (e.g., GPU~\cite{cuda}, specific architectures~\cite{catboost}).
For interruption-free scaling,
    we can copy data $\mathcal{D}$ from a source machine to a target machine
        in a way that the original session state can be restored from $\mathcal{D}$.
In this process,
    we want to minimize the end-to-end time
        for creating $\mathcal{D}$, transferring $\mathcal{D}$ to a target machine,
            reconstructing the state from $\mathcal{D}$ on the target machine.
This is the first use case we empirically study (\cref{sec:exp_migrate}).


\paragraph{Fast Restart for On-demand Computing}

Leveraging pay-as-you-go pricing model offered by many cloud vendors~\cite{azurepayasyougo, colabpayasyougo},
    suspending sessions (and VMs) when not in use is
    an effective way for reducing charges (e.g., up to 6$\times$~\cite{wannipurage2022framework}).
With the ability to create data $\mathcal{D}$ sufficient for reconstructing the current session state,
    we can persist $\mathcal{D}$ prior to either manual or automated suspension~\cite{idleculler, kaggletimeout, colab},
        to quickly resume, when needed, the session in the same state.
This achieves on-demand, granular computing with fast session restart times
    without impacting user experience due to frequent session suspensions~\cite{kaggleforums, colabstackoverflow}.
In this process,
    we want to restore the session as quickly as possible
        by minimizing the time it takes for downloading $\mathcal{D}$ and reconstructing a state from it.
This is the second use case we empirically study (\cref{sec:exp_restore}).

\begin{figure}[t]

\centering
\begin{tikzpicture}[>={LaTeX[width=1mm,length=1mm]},->]

\node(notebook) [draw=black, anchor=north west, minimum width=24mm, minimum height=12.5mm,densely dashdotdotted]
at (0,0) {};
\node(celltext) [anchor=south, minimum width = 22mm, minimum height=3mm, inner sep = 0.5mm, font=\small]
at ($(notebook.north) + (0, 0.05)$) {\textbf{User Interface}};
\node(magictxt) [
    anchor=west, minimum height=3mm, inner ysep=1mm, font=\ttfamily\small, align=left, 
    draw=black, inner xsep=0mm, minimum width=21mm
]
at ($(notebook.west) + (0.15, 0.0)$) {%
\hspace*{-4mm} \textcolor{purple}{\%\%intercept} \\[-0.2em]
\hspace*{-4mm} code%
};
\node [anchor=south west, minimum height=2mm, inner sep = 0.5mm, font=\ttfamily]
at ($(magictxt.north west) + (0, 0)$) {...};
\node [anchor=north west, minimum height=2mm, inner sep = 0.5mm, font=\ttfamily]
at ($(magictxt.south west) + (0, -0.05)$) {...};

\node(magictxt) [
    anchor=west, minimum height=3mm, inner sep = 1.2mm, align=left, font=\ttfamily\small,
    draw=black, rounded corners=1mm,
]
at ($(notebook.east) + (1.2, 0.15)$) {
    def intercept(code): \\[-0.2em]
    \qquad \textbf{preprocess}(code)  \\[-0.2em]
    \qquad \textcolor{RandomColor}{\# regular kernel execution} \\[-0.2em]
    \qquad out = execute(code)  \\[-0.2em]
    \qquad \textbf{postprocess}(out, code) 
};

\draw[line width=1.0mm,-latex]  
    ($(notebook.east) + (0.2, 0.2)$) -- ($(notebook.east) + (1.0, 0.2)$);

\end{tikzpicture}
\vspace{-2mm}

\caption{For every cell run, 
we can inject custom pre-/post-processing logic.
``\%\%intercept'' is hidden to users.
}
\label{fig:transform}
\vspace{-2mm}
\end{figure}
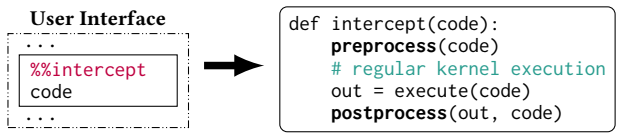

\subsection{How to Enable Data Management Layer?}
\label{sec:background_abstraction}


We discuss the pros and cons of several different approaches to enabling a data management layer.




\paragraph{OS-level Checkpointing}

To save the current session state, 
    we can checkpoint the entire memory space associated with the underlying Python/R kernels.
To make the process more efficient,
    existing tools like CRIU patch the Linux kernel
        to trace dirty pages.
However, as described in \cref{sec:intro},
    this approach is platform-independent, incurs higher space cost, and is limited to storing the state of primary memory
    (not GPU or other devices).
We empirically compare our approach to CRIU to understand reliability and efficiency (\cref{sec:experiments}).

\paragraph{Object wrappers}

Watchpoint object wrappers~\cite{gdbcheckpoints, pythonwatchpoints} are commonly used for debugging purposes~\cite{pimentel2017noworkflow} and program slicing~\cite{head2019managing,shankar2022bolt}: they maintain deep copies for objects in the session state, which are compared to check for changes after each frame execution; however, they are unsuitable for use during data science workflows due to the unacceptable \textasciitilde20$\times$ runtime overhead in our preliminary tests.



\paragraph{Monitoring Cell Executions (Ours)} 

In order to trace cell executions and their effects on variables, 
we can add a lightweight wrapper (i.e., our data management layer) that
    functions before and after each cell execution to monitor the cell code, runtime, and variable changes.
This idea is depicted conceptually in \Cref{fig:transform}.
Specifically,
    our implementation uses \emph{cell magics},
a Jupyter-native mechanism that allows
    arbitrary modification to cell statements when the cell is executed.
With this, 
    we add pre-/post-processing steps to capture cell code and resulting session state modifications.

\subsection{Fast Replication with Application History}
\label{sec:background_graph}
\label{sec:background_example}

This section describes our core idea for
    devising an efficient replication strategy
        by leveraging the ability to monitor cell executions.

\paragraph{Application History}

An \emph{application history graph} (AHG) is a bipartite graph for expressing
    session states changes with respect to cell runs.
There are two types of nodes: variables and transformations.
A transformation node connects input variables to output variables
    (see an example in \cref{fig:motivating_example}).
AHG aims to achieve two properties:

\begin{figure}[t]

\begin{subfigure}[b]{\linewidth}
\centering
\begin{tikzpicture}[>={LaTeX[width=1mm,length=1mm]},->]

\node(ahgtext) [anchor=west, minimum height=3mm, inner sep = 0.5mm]
at (0, 0) {\small \textbf{Application History}};

  \node(ce1) [draw=black, anchor=west, minimum width = 6mm, minimum height=4mm, inner sep = 0.3mm, circle,align=center]
 at ($(ahgtext.west) + (0, -0.65)$) {\footnotesize Cell 1\\[-0.35em]\footnotesize \textbf{3mins}};

\node(x1) [draw=black, anchor=center, minimum height=3mm, inner sep = 0.6mm]
at ($(ce1) + (0.9, 0)$) {\footnotesize \texttt{df}};
 \draw[->, thick] 
 ($(ce1.east)$) --
($(x1.west)$);

  \node(ce2) [draw=black, anchor=center, minimum width = 4mm, minimum height=4mm, inner sep = 0.3mm, circle,align=center]
 at ($(ce1) + (1.7,0)$) {\footnotesize Cell 2\\[-0.35em]\footnotesize \textbf{1min}};
  \draw[->, thick] 
 ($(x1.east)$) --
($(ce2.west)$);

\node(dftraing) [draw=black, anchor=center, minimum height=3mm, inner sep = 0.6mm]
at ($(ce2) + (1.3, 0.6)$) {\footnotesize \texttt{df\_train}};
\node(dftestg) [draw=black, anchor=center, minimum height=3mm, inner sep = 0.6mm]
at ($(ce2) + (1.3, -0.6)$) {\footnotesize \texttt{df\_test}};
  \draw[->, thick]
 ($(ce2.east)$) --
($(dftraing.west)$);
  \draw[->, thick]
 ($(ce2.east)$) --
($(dftestg.west)$);

  \node(ce3) [draw=black, anchor=center, minimum width = 4mm, minimum height=4mm, align=center, inner sep = 0.3mm, circle]
 at ($(ce2) + (2.6,0)$) {\footnotesize Cell 3\\[-0.35em]\footnotesize \textbf{20mins}};
  \draw[->, thick] 
 ($(dftraing.east)$) --
($(ce3.west)$);

\node(modelg) [draw=black, anchor=center, minimum height=3mm, inner sep = 0.6mm]
at ($(ce3) + (1.1, 0)$) {\footnotesize \texttt{model}};
 \draw[->, thick] 
 ($(ce3.east)$) --
($(modelg.west)$);

  \node(ce4) [draw=black, anchor=center, minimum width = 4mm, minimum height=4mm, inner sep = 0.3mm, circle,align=center]
 at ($(ce3) + (2.2,0)$) {\footnotesize Cell 4\\[-0.35em]\footnotesize \textbf{10mins}};
  \draw[->, thick] 
 ($(modelg.east)$) --
($(ce4.west)$);

  \draw[-, thick] 
 ($(dftestg.east)$) --
($(ce4) + (0, -0.6)$);

  \draw[->, thick] 
 ($(ce4) + (0, -0.6)$) --
($(ce4.south)$);

\node(plotg) [draw=black, anchor=center, minimum height=3mm, inner sep = 0.6mm]
at ($(ce4) + (1.1, 0)$) {\footnotesize \texttt{plot}};
 \draw[->, thick] 
 ($(ce4.east)$) --
($(plotg.west)$);

\node(variable) [anchor=west, minimum height=3mm, inner sep = 0.5mm]
at ($(ce1.west) + (0, -1.2)$) {\small Variable};
\node(df) [anchor=west,inner sep = 0.5mm]
at ($(variable.west) + (2.4,  0)$) {\texttt{df}};
\node(dftrain) [anchor=west, minimum height=8mm, minimum width=12mm,inner sep = 0.5mm]
at ($(df.west) + (1.3, 0)$) {\small{\texttt{df\_train}}};
\node(dftest) [anchor=west, minimum height=4mm,inner sep = 0.5mm]
at ($(dftrain.west) + (1.3, 0)$) {\small{\texttt{df\_test}}};
\node(model) [anchor=west, minimum height=3mm,inner sep = 0.5mm]
at ($(dftest.west) + (1.3, 0)$) {\small\texttt{model}};
\node(plot) [anchor=west, minimum height=3mm,inner sep = 0.5mm]
at ($(model.west) + (1.3, 0)$) {\small\texttt{plot}};

\node(storecost) [anchor=north west, minimum height=3mm, inner sep = 0.5mm]
at ($(variable.south west) + (0, -0.05)$) {\small Store cost (mins)};
\node(s1) [anchor=west, minimum height=3mm, inner sep = 0.5mm,align=center]
at ($(storecost.west) + (2.4, 0)$) {\small 8};
\node(s2) [anchor=west, minimum height=3mm, inner sep = 0.5mm,align=center]
at ($(s1.west) + (1.3, 0)$) {\small 6.4};
\node(s3) [anchor=west, minimum height=3mm, inner sep = 0.5mm,align=center]
at ($(s2.west) + (1.3, 0)$) {\small 1.6};
\node(s4) [anchor=west, minimum height=3mm, inner sep = 0.5mm,align=center]
at ($(s3.west) + (1.3, 0)$) {\small 0.2};
\node(s5) [anchor=west, minimum height=3mm, inner sep = 0.5mm,align=center]
at ($(s4.west) + (1.3, 0)$) {\small 0.1};

\node(reloadcost) [anchor=north west, minimum height=3mm, inner sep = 0.5mm]
at ($(storecost.south west) + (0, 0.05)$) {\small Reload cost (mins)};
\node(r1) [anchor=west, minimum height=3mm, inner sep = 0.5mm,align=center]
at ($(reloadcost.west) + (2.4, 0)$) {\small 2};
\node(r2) [anchor=west, minimum height=3mm, inner sep = 0.5mm,align=center]
at ($(r1.west) + (1.3, 0)$) {\small 1.6};
\node(r3) [anchor=west, minimum height=3mm, inner sep = 0.5mm,align=center]
at ($(r2.west) + (1.3, 0)$) {\small 0.4};
\node(r4) [anchor=west, minimum height=3mm, inner sep = 0.5mm,align=center]
at ($(r3.west) + (1.3, 0)$) {\small 0.2};
\node(r5) [anchor=west, minimum height=3mm, inner sep = 0.5mm,align=center]
at ($(r4.west) + (1.3, 0)$) {\small 0.1};

\node(migratecost) [anchor=north west, minimum height=3mm, inner sep = 0.5mm]
at ($(reloadcost.south west) + (0, 0.05)$) {\small Total cost (mins)};
\node(m1) [anchor=west, minimum height=3mm, inner sep = 0.5mm,align=center]
at ($(migratecost.west) + (2.4, 0)$) {\small 10};
\node(m2) [anchor=west, minimum height=3mm, inner sep = 0.5mm,align=center]
at ($(m1.west) + (1.3, 0)$) {\small 8};
\node(m3) [anchor=west, minimum height=3mm, inner sep = 0.5mm,align=center]
at ($(m2.west) + (1.3, 0)$) {\small 2};
\node(m4) [anchor=west, minimum height=3mm, inner sep = 0.5mm,align=center]
at ($(m3.west) + (1.3, 0)$) {\small 0.4};
\node(m5) [anchor=west, minimum height=3mm, inner sep = 0.5mm,align=center]
at ($(m4.west) + (1.3, 0)$) {\small 0.2};

  \draw[-, gray, opacity=0.6, thick] 
 ($(variable.north west)$) --
($(migratecost.south west)$);
  \draw[-, gray, opacity=0.6, thick] 
 ($(migratecost.south west) + (8.4, 0)$) --
($(migratecost.south west)$);
  \draw[-, gray, opacity=0.6, thick] 
 ($(variable.north west) + (8.4, 0)$) --
($(variable.north west)$);
  \draw[-, gray, opacity=0.6, thick] 
 ($(variable.north west) + (8.4, 0)$) --
($(migratecost.south west) + (8.4, 0)$);
  \draw[-, gray, opacity=0.6, thick] 
 ($(variable.south west) + (8.4, -0.05)$) --
($(variable.south west) + (0, -0.05)$);

\end{tikzpicture}
\end{subfigure}

\vspace{2mm}
\midsepremove
\footnotesize
\addtolength{\tabcolsep}{-3pt} 
\begin{tabular}{lllll}

\textbf{Method}              & \textbf{Store Vars} & \textbf{Rerun cells} & \textbf{Migration Cost} & \textbf{Restore Cost}      \\ \midrule
 Rerun all       & N/A & All &  3+1+20+10=33 &3+1+20+10=33  \\                
Store all        & All &  N/A & 10+8+2+.4+.2=20.6 &2+1.6+.4+.2+.1=4.3                 \\
\textbf{Fast-migrate}       & \texttt{model,plot} & 1, 2 & \textbf{3+1+.4+.2=4.6}   & 3+1+.2+.1=4.3                 \\
\textbf{Fast-restore}         & \texttt{df,model,plot} &  2 &  10+1+.4+.2=11.6  & \textbf{2+1+.2+.1=3.3}                   \\
\bottomrule
\end{tabular}
\addtolength{\tabcolsep}{3pt} 
\midsepdefault

\vspace{-1mm}
\caption{Example app history (top) and different replication plan costs (bottom).
Combining recompute/copy allows faster migration (Fast-migrate).
Alternatively, the optimal plan changes if the restoration is prioritized (Fast-restore).}
\label{fig:motivating_example}
\vspace{-2mm}
\end{figure}

\begin{itemize}
    \item \textbf{Completeness:} No false negatives. All input/output variable for each transformation must be captured.
    \item \textbf{Minimal:} Minimal false positives. 
        The number of variables that are incorrectly identified as accessed/modified,
            while variables are not actually accessed/modified,
            must be minimized.
\end{itemize}
These properties are required for correct state reconstruction (\cref{sec:data_modeling}).

\paragraph{Core Optimization Idea}
AHG allows for efficient state replication 
    with a combination of (1) recompute and (2) copy.
\emph{Motivating Example.}
Suppose a data analyst fitting a regression model (\cref{fig:motivating_example}).
The notebook contains 4 cell runs: 
    data load (Cell 1), train-test split (Cell 2), fitting (Cell 3), and evaluation (Cell 4).
After fitting, 
    the analyst decides to move the session to a new machine for GPU.
Simply rerunning the entire notebook incurs
    \textbf{\textit{33 minutes}}.
Alternatively, serializing/copying variables takes 
    \textbf{\textit{20.6 minutes}}.

However, there is a more efficient approach.
By copying only \texttt{model} and \texttt{plot}
    and recomputing others on a new machine (Fast-migrate),
\textit{\textbf{we can complete end-to-end migration in 4.6 minutes.}}
Or, if we prioritize restoration time (to reduce user-perceived restart time for on-demand computing),
    our optimized plan (Fast-restore) takes 3.3 minutes.
%
This example illustrates significant optimization opportunities in session replication. 
Our goal is to have the ability to find the best replication plan for arbitrarily complex AHGs.

\section{System Overview}
\label{sec:system_overview}

This section presents \system at a high level
    by describing its components (\cref{sec:session_lifecycle}) and operations (\cref{sec:system_workflow}).


\subsection{\systemnosf Components}
\label{sec:session_lifecycle}

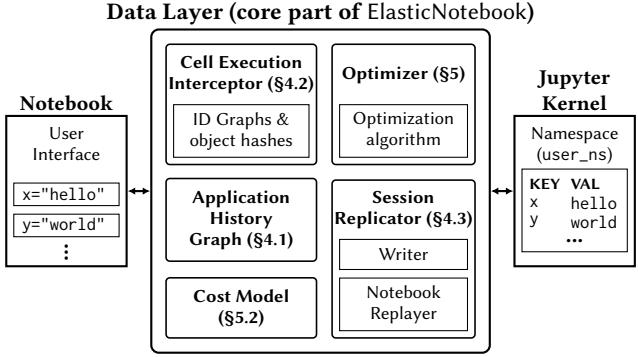
\begin{figure}[t]
\usetikzlibrary{calc}

\begin{subfigure}{\columnwidth}
\centering

\tikzset{
mylabel/.style={
    font=\footnotesize\sffamily\bfseries,
    align=center,
},
mylabel2/.style={
    font=\footnotesize\sffamily,
    align=center,
},
mycomponent/.style={
    semithick, rounded corners=0.5mm,
}
}

\begin{tikzpicture}[>={LaTeX[width=1mm,length=1mm]},->]
\node(notebook) [draw=black, align=center, anchor=north west, thick, minimum width = 16mm, minimum height=20mm,inner sep=1mm]
at (0,0) {};
\node(notebooktxt) [anchor=south, minimum height=3mm, inner sep = 0.5mm, font=\bfseries\small]
at ($(notebook.north)$) {Notebook};
\node(notebooktxt) [anchor=north, align=center, minimum height=3mm, inner sep = 0.5mm, font=\footnotesize\sffamily]
at ($(notebook.north) + (0, -0.1)$) {User \\ Interface};

\node[rotate=90] (jupyterdots) at ($(notebook.south) + (0, 0.2)$) {\bf ...};
\node(cell2) [draw=black, align=left, anchor=south, minimum width = 14mm, minimum height=3mm,inner sep=0.3mm]
at ($(notebook.south) + (0, 0.4)$) {};
\node(cell1) [draw=black, align=left, anchor=south, minimum width = 14mm, minimum height=3mm,inner sep=0.3mm]
at ($(cell2.north) + (0, 0.1)$) {};

\node(codetxt2) [anchor=west, minimum height=3mm, inner sep = 0.6mm]
at ($(cell2.west) + (0.05,0)$) {\footnotesize \texttt{y="world"}};
\node(codetxt1) [anchor=west, minimum height=3mm, inner sep = 0.6mm]
at ($(cell1.west) + (0.05,0)$) {\footnotesize \texttt{x="hello"}};

\node(elastic) [minimum height=43mm, minimum width=45mm,
  draw=black, anchor=west, thick, rounded corners=1mm]
at ($(notebook.east) + (0.3, 0)$)  {};
\node(elastictxt) [anchor=south, minimum height=3mm, inner sep = 0.5mm]
at ($(elastic.north)$) {\textbf{Data Layer (core part of \system)}};

\node(shell) [draw=black, anchor=west, minimum width = 16mm, minimum height=20mm, thick]
at ($(elastic.east) + (0.3, 0)$)  {};
\node(shelltxt) [anchor=south, align = center, minimum height=3mm, inner sep = 0.5mm, font=\bfseries\small]
at ($(shell.north) + (0, 0)$) {Jupyter \\ Kernel};
\node(ns) [draw=black, anchor=south, minimum width = 14mm, minimum height=11mm]
at ($(shell.south) + (0, 0.2)$)  {};
\node(nstxt2) [anchor=south, align = center, minimum height=3mm, inner sep = 0.5mm, font=\footnotesize\sffamily]
at ($(ns.north) + (0, 0)$)  {Namespace \\ (\texttt{user\_ns}) };

\node(key) [anchor=north west, align = center, minimum height=3mm, inner sep = 0.5mm, font=\scriptsize\sffamily\bfseries]
at ($(ns.north west) + (0.05, -0.05)$)  {KEY};
\node(val) [anchor=north west, align = center, minimum height=3mm, inner sep = 0.5mm, font=\scriptsize\sffamily\bfseries]
at ($(ns.north west) + (0.6, -0.05)$)  {VAL};
\node(key1) [anchor=north west, align = center, minimum height=3mm, inner sep = 0.5mm]
at ($(ns.north west) + (0.05, -0.3)$)  {\footnotesize \texttt{x}};
\node(val1) [anchor=north west, align = center, minimum height=3mm, inner sep = 0.5mm]
at ($(ns.north west) + (0.6, -0.3)$)  {\footnotesize \texttt{hello}};
\node(key2) [anchor=north west, align = center, minimum height=3mm, inner sep = 0.5mm]
at ($(ns.north west) + (0.05, -0.55)$)  {\footnotesize \texttt{y}};
\node(val2) [anchor=north west, align = center, minimum height=3mm, inner sep = 0.5mm]
at ($(ns.north west) + (0.6, -0.55)$)  {\footnotesize \texttt{world}};
\node[] (nsdots) at ($(ns.north) + (0, -0.95)$) {\bf ...};

\draw[<->, thick] 
($(notebook.east)$) --
($(elastic.west)$);
\draw[<->, thick] 
($(elastic.east)$) --
($(shell.west)$);

\node(intercepter) [draw=black, fill = white, anchor=north west, minimum width = 20mm, minimum height=16mm, mycomponent]
at ($(elastic.north west) + (0.2, -0.2)$)  {};

\node(interceptertxt) [anchor=north, align = center, minimum height=3mm, inner sep = 0.5mm, mylabel]
at ($(intercepter.north) + (0, -0.1)$) {Cell Execution \\ Interceptor (\cref{sec:data_modeling:construction})};

\node(idgraph) [draw=black, fill = white, anchor=south, minimum width = 18mm, minimum height=7mm, align=center, inner sep = 0.5mm, mylabel2]
at ($(intercepter.south) + (0, 0.1)$) {ID Graphs \& \\ object hashes};

\node(cellcode) [draw=black, fill = white, anchor=south west, minimum width = 20mm, minimum height=8mm, mycomponent]
at ($(elastic.south west) + (0.2, 0.2)$)  {};

\node(cellcodetxt) [anchor=center, align = center, minimum height=3mm, inner sep = 0.5mm, mylabel]
at ($(cellcode)$) {Cost Model\\ (\Cref{sec:cost_model})};

\node(metadata) [draw=black, fill = white, anchor=south, minimum width = 20mm, minimum height=11mm, mycomponent]
at ($(cellcode.north) + (0, 0.2)$)  {};

\node(metadatatxt) [anchor=center, align = center, minimum height=3mm, inner sep = 0.5mm, mylabel]
at ($(metadata)$) {Application\\ History\\ Graph (\cref{sec:ahg_graph})};


\node(optimizer) [draw=black, fill = white, anchor=north east, minimum width = 19mm, minimum height=16mm, mycomponent]
at ($(elastic.north east) + (-0.2, -0.2)$)  {};

\node(optimizertxt) [anchor=north, align = center, minimum height=3mm, inner sep = 0.5mm, mylabel]
at ($(optimizer.north) + (0, -0.25)$) {Optimizer (\Cref{sec:algorithm})};

\node(optalgm) [draw=black, fill = white, anchor=south, minimum width = 17mm, minimum height=7mm, align=center, inner sep = 0.5mm, mylabel2]
at ($(optimizer.south) + (0, 0.1)$) {Optimization \\ algorithm};

\node(replicator) [draw=black, fill = white, anchor=south east, minimum width = 19mm, minimum height=21mm, mycomponent]
at ($(elastic.south east) + (-0.2, 0.2)$)  {};

\node(replicatortxt) [anchor=north, align = center, minimum height=3mm, inner sep = 0.5mm, mylabel]
at ($(replicator.north) + (0, -0.1)$) {Session \\ Replicator (\cref{sec:ahg_problem})};

\node(replayer) [draw=black, fill = white, anchor=south, minimum width = 17mm, minimum height=7mm, align=center, inner sep = 0.5mm, mylabel2]
at ($(replicator.south) + (0, 0.1)$) {Notebook \\ Replayer};
\node(writer) [draw=black, fill = white, anchor=south, minimum width = 17mm, minimum height=4mm, align=center, inner sep = 0.5mm, mylabel2]
at ($(replayer.north) + (0, 0.1)$) {Writer};

\end{tikzpicture}
\vspace{-2mm}
\end{subfigure}

\caption{\systemnosf architecture. Its data layer acts as a gateway between the user interface and the kernel: 
    cell executions are intercepted to observe session state changes.}
\label{fig:system_overview}
\vspace{-2mm}
\end{figure}

\system introduces a unique data layer that acts as a gateway 
    between the user and the kernel (See \cref{fig:system_overview}): 
it monitors every cell execution, observing code and resulting session state changes.


\paragraph{Cell Execution Interceptor} 

The Cell Execution Interceptor intercepts cell execution requests and adds pre-/post-processing scripts 
    before rerouting it into the underlying kernel for regular execution.
The added scripts perform (1) cell code analyses and the AHG updates, and (2) cell runtime recordings.

\paragraph{Application History Graph (AHG)}
The AHG is incrementally built by the Cell Execution Interceptor 
    to record how variables have been accessed/modified by each cell execution (\cref{sec:data_modeling}).
The AHG is used by the Optimizer 
    to compute replication plans (\Cref{sec:algorithm}).

\paragraph{Cost Model}
The cost model stores profiled metrics (i.e., cell runtimes, variable sizes, network bandwidth),
    serving as the hyperparameters 
    for the Optimizer (\Cref{sec:cost_model}).

\paragraph{Optimizer}
The Optimizer uses the AHG and the Cost Model to determine the most efficient replication plan consisting of (1) variables to store and (2) cells to re-run. We discuss \system's cost model and optimization in detail in \Cref{sec:algorithm}.

\paragraph{Session Replicator}
The Session Replicator replicates a notebook session according to the Optimizer's plan.
Specifically, the Writer creates and writes a checkpoint file to storage (e.g., SSD, cloud storage), 
    while the Notebook Replayer reads the file and restores the session, both following the replication plan.
We discuss \system's session replication in detail in \Cref{sec:system_workflow}.

\subsection{\systemnosf Workflow}
\label{sec:system_workflow}

This section describes \system's operations. 
\system monitors every cell execution during a session lifecycle, 
    then performs on-request replication of the session in two steps: 
    \textit{checkpointing} (writing to the checkpoint file) and \textit{restoration}.

\paragraph{Monitoring Cell Executions} 
Upon each cell execution by the user, 
    \system performs the following steps:
\begin{enumerate}
    \item Accessed variables of the cell execution are identified via AST analysis (described in \Cref{sec:identifying_access}).
    \item The cell code is executed by the Jupyter kernel.
    \item Variable changes (i.e., creation/deletion/modification) are identified within the global namespace (\Cref{sec:identifying_modify}).
    \item The AHG is updated using (1) the cell code and (2) modified variables by the cell execution.
    \item The Cost Model is updated to record cell runtime.
\end{enumerate}

\paragraph{Initiating Replication}

When replication is requested, \system creates and writes a \textit{checkpoint file} to storage, which can be restored later 
    to exactly and efficiently reconstruct the current session.
\system first completes the Cost Model by profiling variable sizes and network bandwidth to storage; then, the Optimizer utilizes the AHG and Cost model to compute a replication plan, according to which the Writer creates the checkpoint file:
it consists of (1) a subset of stored variables from the session state, (2) cells to rerun, (3) the AHG, and (4) the Cost Model.

\paragraph{Restoring a Session}

When requested, \system restores the notebook session from the checkpoint file according to the replication plan.
The Notebook Replayer 
    reconstructs variables in the order they appeared in the original session
    by combining (1) cell reruns and (2) data deserialization followed by variable re-declaration (into the kernel).
Finally, \system loads the AHG and Cost Model for future replications.



\emph{Accuracy Guarantee:}
\system's state reconstructing is effectively the same
    as re-running all the cells from scratch
        exactly in the order they were run in the past.
That is,
    \system shortens the end-to-end reconstruction time
        by loading saved variables (into the kernel namespace) if doing so achieves time savings.
\cref{sec:ahg_problem} presents formal correctness analysis.
\cref{sec:fault_tolerance} discusses how we address external resources, side effects,
    and deserialization failures.

\section{Application History Graph}
\label{sec:data_modeling}

\begin{table}[t]
\caption{Notations and their meaning}
\label{tbl:symbols}

\vspace{-3mm}
\small
\midsepremove
\begin{tabular}{ll}
\toprule
\textbf{Symbols}              & \textbf{Definition}           \\ \midrule
$\mathcal{X}$            & Set of Variables       \\
$\mathcal{V}$            & Set of Variable Snapshots (VSs)       \\
$\mathcal{V}_a$ & Set of Active Variable Snapshots\\
$\mathcal{C}$ (= $c_{t_1}, c_{t_2}, \ldots$)            & Set of Cell Executions (CEs)            \\
$\mathcal{E}_w$            & Set of write dependencies          \\
$\mathcal{E}_r$            & Set of read dependencies          \\
$\mathcal{G} := \{\mathcal{V}\cup\mathcal{C}, \mathcal{E}_w\cup\mathcal{E}_r\}$         & Application History Graph (AHG)                    \\
\hline
$req: \mathcal{X} \rightarrow 2^{\mathcal{C}}$            & Reconstruction mapping function            \\
$w_{store}: \mathcal{X} \rightarrow \mathbb{R}^+$        & Variable storage cost \\
$w_{rerun}: \mathcal{C} \rightarrow \mathbb{R}^+$        & Cell Rerun cost    \\
\hline
$w_{\mathcal{M}}: 2^{\mathcal{X}} \rightarrow \mathbb{R}^+$        & Migration cost function \\
$w_{\mathcal{R}}: 2^{\mathcal{X}} \rightarrow \mathbb{R}^+$        & Recomputation cost function    \\
$\mathcal{L}\subseteq \mathcal{X}\times\mathcal{X}$        & Pairs of linked variables    \\
\hline
$\mathcal{H} := \{\mathcal{V}_H, \mathcal{E}_H\}$         & Flow graph                   \\
$c: \mathcal{E}_H \rightarrow \mathbb{R}^+$ & Flow graph edge capacity function \\
\bottomrule
\end{tabular}
\midsepdefault
\vspace{-2mm}
\end{table}

This section formally defines the Application History Graph (\cref{sec:ahg_graph}), 
    and describes how we achieve exact state replication (\cref{sec:ahg_problem}).



\subsection{AHG Formal Definition}
\label{sec:ahg_graph}

\begin{figure}[t]

\centering
\begin{tikzpicture}[>={LaTeX[width=1mm,length=1mm]},->]

\node(notebook) [draw=black, anchor=north west, minimum width = 33mm, minimum height=49.5mm,densely dashdotdotted]
at (0,0) {};
\node(celltext) [anchor=south west, minimum height=3mm, inner sep = 0.5mm]
at ($(notebook.north west) + (0, 0.05)$) {\large \textbf{Notebook}};
\node(cell) [draw=black, anchor=north, minimum width = 31mm, minimum height=3.5mm]
at ($(notebook.north) + (0, -0.4)$) {};
\node(celltext) [anchor=south west, minimum height=3mm, inner sep = 0.2mm]
at ($(cell.north west) + (0, 0)$) {\small Cell 1 ($c_{t_1}$)};
\node(codetxt) [anchor=west, minimum height=3mm, inner sep = 0.6mm]
at ($(cell.west) + (0.05,0)$) {\footnotesize \texttt{x, y = 1}};
\node(cell2) [draw=black, anchor=north, minimum width = 31mm, minimum height=9mm]
at ($(cell.south) + (0, -0.4)$) {};
\node(celltext2) [anchor=south west, minimum height=3mm, inner sep = 0.2mm]
at ($(cell2.north west) + (0, 0)$) {\small Cell 2 ($c_{t_2}$)};
\node(codetxt2) [anchor=west, minimum height=3mm, inner sep = 0.6mm, align=left]
at ($(cell2.west) + (0.05, 0)$) {\footnotesize \texttt{z = y}\\[-0.35em]\footnotesize \texttt{if False:}\\[-0.35em]\footnotesize \texttt{\,\,\,\,print(x)}};

\node(cell3) [draw=black, anchor=north, minimum width = 31mm, minimum height=6mm]
at ($(cell2.south) + (0, -0.4)$) {};
\node(celltext3) [anchor=south west, minimum height=3mm, inner sep = 0.2mm]
at ($(cell3.north west) + (0, 0)$) {\small Cell 3 ($c_{t_3}$)};
\node(codetxt3) [anchor=west, minimum height=3mm, inner sep = 0.6mm, align=left]
at ($(cell3.west) + (0.05, 0)$) {\footnotesize \texttt{x += 1}\\[-0.35em]\footnotesize \texttt{l1 = [z, 2, 3]}};

\node(cell4) [draw=black, anchor=north, minimum width = 31mm, minimum height=6mm]
at ($(cell3.south) + (0, -0.4)$) {};
\node(celltext4) [anchor=south west, minimum height=3mm, inner sep = 0.2mm]
at ($(cell4.north west) + (0, 0)$) {\small Cell 4 ($c_{t_4}$)};
\node(codetxt4) [anchor=west, minimum height=3mm, inner sep = 0.6mm,align=left]
at ($(cell4.west) + (0.05, 0)$) {\footnotesize \texttt{gen=(i for i in l1)}\\[-0.35em]\footnotesize \texttt{2dlist = [l1]}};

\node(cell5) [draw=black, anchor=north, minimum width = 31mm, minimum height=3.5mm]
at ($(cell4.south) + (0, -0.4)$) {};
\node(celltext5) [anchor=south west, minimum height=3mm, inner sep = 0.2mm]
at ($(cell5.north west) + (0, 0)$) {\small Cell 5 ($c_{t_5}$)};
\node(codetxt5) [anchor=west, minimum height=3mm, inner sep = 0.6mm]
at ($(cell5.west) + (0.05, 0)$) {\footnotesize \texttt{print(gen)}};

  \node(ce1) [draw=purple, thick, anchor=north west, minimum width = 4mm, minimum height=4mm, circle, inner sep = 0.4mm]
 at ($(notebook.north east) + (2.2, 0.3)$) {\footnotesize \textcolor{purple}{\bm{$c_{t_1}$}}};

\node(x1) [draw=purple,densely dashed,thick, fill=Lightgrey, opacity = 0.7, anchor=center, minimum height=3mm, inner sep = 0.6mm]
at ($(ce1) + (-0.8, -0.55)$) {\footnotesize \textcolor{purple}{\textbf{(}\texttt{\textbf{x}}\textbf{,} \bm{$t_1$}\textbf{)}}};
\node(y1) [draw=black, anchor=center, minimum height=3mm, inner sep = 0.6mm]
at ($(ce1) + (0.8, -0.55)$) {\footnotesize (\texttt{y}, $t_1$)};

  \node(ce2) [draw=black, anchor=center, minimum width = 4mm, minimum height=4mm, circle, inner sep = 0.4mm]
 at ($(ce1) + (0, -1.1)$) {\footnotesize $c_{t_2}$};

\node(z1) [draw=black, anchor=center,  minimum height=3mm, inner sep = 0.6mm]
at ($(ce2) + (0, -0.55)$) {\footnotesize (\texttt{z}, $t_2$)};

  \node(ce3) [draw=purple, thick, anchor=center, minimum width = 4mm, minimum height=4mm, circle, inner sep = 0.4mm]
 at ($(z1) + (0, -0.55)$) {\footnotesize \small \textcolor{purple}{\bm{$c_{t_3}$}}};

\node(x2) [draw=purple,anchor=center, thick, minimum height=3mm, inner sep = 0.6mm]
at ($(ce3) + (-0.8, -0.55)$) {\footnotesize \textcolor{purple}{\textbf{(}\texttt{\textbf{x}}\textbf{,} \bm{$t_3$}\textbf{)}}};

\node(l1) [draw=black,anchor=center,  minimum height=3mm, inner sep = 0.6mm]
at ($(ce3) + (0.8, -0.55)$) {\footnotesize (\texttt{l1},$t_3$)};

  \node(ce4) [draw=black, anchor=center, minimum width = 4mm, minimum height=4mm, circle, inner sep = 0.4mm]
 at ($(ce3) + (0, -1.1)$) {\footnotesize  $c_{t_4}$};

 \node(gen1) [draw=black,densely dashed, fill=Lightgrey, opacity = 0.7, anchor=center, minimum height=3mm, inner sep = 0.6mm]
at ($(ce4) + (-0.8, -0.55)$) {\footnotesize (\texttt{gen}, $t_4$)};
\node(2dlist) [draw=black, anchor=center,  minimum height=3mm, inner sep = 0.6mm]
at ($(ce4) + (0.8, -0.55)$) {\footnotesize (\texttt{2dlist}, $t_4$)};

   \node(ce5) [draw=black, anchor=center, minimum width = 4mm, minimum height=4mm, circle, inner sep = 0.4mm]
 at ($(ce4) + (0, -1.1)$) {\footnotesize $c_{t_5}$};

 \node(gen2) [draw=black, anchor=center, minimum height=3mm, inner sep = 0.6mm]
at ($(ce5) + (0, -0.55)$) {\footnotesize (\texttt{gen}, $t_5$)};

 \draw[->,purple,thick] 
 ($(ce1.south)$) --
($(x1.north)$);
 \draw[->] 
 ($(ce1.south)$) --
($(y1.north)$);

 \draw[->] 
 ($(x1.south)$) --
($(ce2.north)$);
 \draw[->] 
 ($(y1.south)$) --
($(ce2.north)$);

 \draw[->] 
 ($(ce2.south)$) --
($(z1.north)$);

 \draw[-,purple,thick]
 ($(x1.south)$) --
($(x1) + (0, -1.65)$);
 \draw[->,purple,thick] 
 ($(x1) + (0, -1.65)$) --
($(ce3.west)$);

 \draw[->] 
 ($(z1.south)$) --
($(ce3.north)$);

 \draw[->,purple,thick] 
 ($(ce3.south)$) --
($(x2.north)$);
 \draw[->] 
 ($(ce3.south)$) --
($(l1.north)$);

 \draw[->] 
 ($(x2.south)$) --
($(ce4.north)$);

 \draw[->] 
 ($(ce4.south)$) --
($(gen1.north)$);
 \draw[->] 
 ($(ce4.south)$) --
($(2dlist.north)$);

 \draw[->] 
 ($(gen1.south)$) --
($(ce5.north)$);
 \draw[->] 
 ($(ce5.south)$) --
($(gen2.north)$);



\node(examplex1) [draw=black,densely dashed, fill=Lightgrey, opacity = 0.7, anchor=north west, minimum height=3mm, inner sep = 0.6mm]
at ($(notebook.south west) + (0, -0.3)$) {\footnotesize (\texttt{x}, $t_1$)};
\node[anchor=west, align=left] at ($(examplex1.east) + (0,0)$) {\small (Overwritten/deleted)\\[-0.3em]\small Variable Snapshot};

  \node(examplex2) [draw=black, anchor=west, minimum width = 4mm, minimum height=4mm, circle, inner sep = 0.4mm]
 at ($(examplex1.east) + (2.8, 0)$) {\footnotesize $c_{t_1}$};
\node[anchor=west, align=left] at ($(examplex2.east) + (0,0)$) {\small Cell \\[-0.2em]\small Execution};

\node(examplex3) [draw=black, anchor=west, minimum height=3mm, inner sep = 0.6mm]
at ($(examplex2.east) + (1.5, 0)$) {\footnotesize(\texttt{x}, $t_1$)};
\node[anchor=west, align=left] at ($(examplex3.east) + (0,0)$) {\small Active\\[-0.3em]\small Variable Snapshot};

\end{tikzpicture}

\vspace{-2mm}
\caption{An example notebook and its corresponding Application History Graph. 
The AHG tells \systemnosf how to recompute variables; for example, 
    rerunning $c_{t_1}$ and $c_{t_3}$ is necessary for recomputing \texttt{x} (red).}
\label{fig:problem_setup_graph}
\vspace{-2mm}
\end{figure}
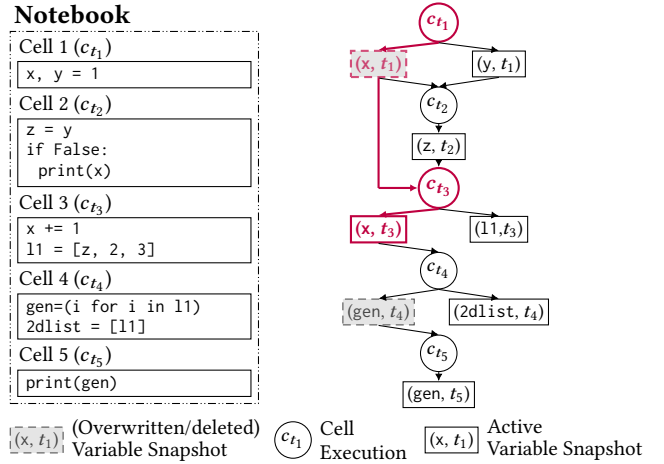

The AHG is a directed acyclic graph expressing 
    how a session state has changed with respect to cell executions.
\cref{fig:problem_setup_graph} is an example.

\begin{definition}
A \textbf{variable} is a named entity (e.g., \texttt{df})
    referencing an \textbf{object} (which can be uniquely identified by its object ID).
\end{definition}

\noindent
A variable can be primitive (e.g., int, string) or 
    complex (e.g., list, dataframe).
Multiple variables may point to the same object.
The set of all variables (i.e., $\mathcal{X}$) defined in the global namespace
    forms a session state.
Cell executions may modify the values of variables (or referenced objects) 
    without changes to their names,
        which we recognize in AHG using \emph{variable snapshot}, as follows.

\begin{definition}
A \textbf{variable snapshot} (VS) is a name-timestamp pair, ($x$, $t$), representing 
    the variable $x$ created/modified at $t$.
    We denote the set of VSes as $\mathcal{V}$.
\end{definition}

\begin{definition}
A \textbf{cell execution} (CE) $c_t$ represents a cell execution that finishes at timestamp $t$.
\end{definition}

\noindent
All cell executions are \emph{linear};
    that is, for each session, there is at most one cell running at a time,
        and their executions are totally ordered.
We denote the list of CEs by $\mathcal{C}$. 
Each CE also stores executed cell code, which can be used for re-runs (\Cref{sec:system_workflow}).

\begin{definition}
    A \textbf{write dependency} ($c_{t}$ $\rightarrow$ (\texttt{x}, $t$))
        indicates 
        CE $c_{t}$ may have modified/created at time $t$
            the object(s) reachable 
        from the variable $x$.
    We denote the set of write dependencies as $\mathcal{E}_w$.
\end{definition}

\noindent
In \cref{fig:problem_setup_graph}, 
$c_{t_3}$ modifies \texttt{x} with ``\texttt{x += 1}''; hence, 
     ($c_{t_3}$ $\rightarrow$ (\texttt{x}, $c_{t_3}$)).

\begin{definition}
    A \textbf{read dependency} ((\texttt{x}, $s$) $\rightarrow$ $c_{t}$)
        indicates CE $c_{t}$ may have accessed object(s) reachable from \texttt{x}
        last created/modified at time $s$.
    We denote 
    the set of read dependencies by $\mathcal{E}_r$.
\end{definition}

\noindent
In \cref{fig:problem_setup_graph}, 
``\texttt{gen=(i for i in l1)}'' in $C_{t_4}$ accesses elements in the list \texttt{l1} after its creation in $c_{t_3}$; hence there is ((\texttt{x} $\rightarrow$ $c_{t_3}$), $c_{t_4}$).
Note that write/read dependencies are allowed to contain false positives;
    nevertheless, our replication ensures correctness (\cref{sec:ahg_problem}).

\begin{definition}
    The \textbf{AHG} $G := \{\mathcal{V}\cup\mathcal{C}, \, \mathcal{E}_w\cup \mathcal{E}_r\}$ is a bipartite graph,
        where $\mathcal{V}$ is VSes, $\mathcal{C}$ is CEs;
            $\mathcal{E}_w$ and $\mathcal{E}_r$ are write/read dependencies, respectively.
    It models the lineage of the notebook session.
\end{definition}

\noindent
In sum, 
AHG formalizes variable accesses/modifications with respect to cell executions.
at the variable level (not object level),
    theoretically bounding the size of AHG
        to scale linearly with the number of defined variables,
    not the number of underlying objects (which can be very large for lists, dataframes, and so on).
We empirically verify AHG's low memory overhead 
in \Cref{sec:exp_overhead}.
    



\subsection{Dynamic AHG Construction}
\label{sec:data_modeling:construction}

We describe how \system constructs the AHG accurately.

\paragraph{Constructing the AHG} 

The AHG is incrementally built with 
accessed/created/modified variables by each cell execution: 
\begin{itemize}
    \item A new CE $c_t$ is created; $t$ is an execution completion time.
    \item Read dependencies are created from VSes ($x_1$, $t_{x_1}$), ..., ($x_k$, $t_{x_k}$) to $c_t$, where $x_1,...,x_k$ are variables 
        \emph{possibly} accessed by $c_t$.
    \item VSes ($y_1$, $t$), ..., ($y_k$, $t$) are created, where $y_1,...,y_k$ are variables \emph{possibly} modified and created by $c_t$. 
    Write dependencies are added from $c_t$ to each of the newly created VSes.
\end{itemize}
\cref{fig:problem_setup_graph} (right) shows an example AHG.
Identifying access/modified variables is crucial for its construction,
which we describe below.

\paragraph{ID Graph}
The ID Graph aims to
    to detect changes at the reference level (in addition to values).
For instance, conventional equality checks (e.g., based on serialization)
    will return True for ``\texttt{[a] == [b]}'' if
    \texttt{a} and \texttt{b} have the same value
        (e.g.,  \texttt{a = [1]} and \texttt{b = [1]}),
whereas we ensure it returns True only if \texttt{a} and \texttt{b} refer to the same \textit{object}, i.e., \texttt{id(a)==id(b)}, where \texttt{id} is the object's unique ID. 
This is
because for correct state replication,
    shared references (e.g. aliases) and inter-variable relationships must be captured precisely.


\paragraph{Identifying Accessed Variables}

\label{sec:identifying_access}

\system identifies both \emph{directly accessed} variables (via AST~\cite{ast} parsing) 
and \emph{indirectly accessed} variables (with ID Graphs), as follows.

\emph{Direct Accesses}: Cell code is analyzed with AST, 
stepping also into user-defined functions (potentially nested) 
to check for accesses to variables not explicitly passed in as parameters (e.g., \texttt{global x}).
        
\begin{figure}[t]

\begin{subfigure}[b]{\linewidth}
\centering
\begin{tikzpicture}[>={LaTeX[width=1mm,length=1mm]},->]

\node(notebook) [draw=black, anchor=north west, minimum width = 28mm, minimum height=21mm,densely dashdotdotted]
at (0,0) {};
\node(cell) [draw=black, anchor=north, minimum width = 26mm, minimum height=9mm]
at ($(notebook.north) + (0, -0.4)$) {};
\node(celltext) [anchor=south west, minimum height=3mm, inner sep = 0.2mm]
at ($(cell.north west) + (0, 0)$) {\small Cell 1};
\node(codetxt) [anchor=west, minimum height=3mm, inner sep = 0.2mm,align=left]
at ($(cell.west) + (0.05, 0)$) {\footnotesize\texttt{func = lambda x:...}\\[-0.35em]\footnotesize \texttt{obj1.foo = func}\\[-0.35em]\footnotesize \texttt{obj2.foo = func}};
\node(cell2) [draw=black, anchor=north, minimum width = 26mm, minimum height=3mm]
at ($(cell.south) + (0, -0.4)$) {};
\node(celltext2) [anchor=south west, minimum height=3mm, inner sep = 0.2mm]
at ($(cell2.north west) + (0, 0)$) {\small Cell 2};
\node(codetxt2) [anchor=west, minimum height=3mm, inner sep = 0.1mm,align=left]
at ($(cell2.west) + (0.05, 0)$) {\footnotesize\texttt{obj2.foo(\textcolor{purple}{"str"})}};

\node(2dlist1) [draw=black, anchor=north west, minimum height=3mm, inner sep = 0.6mm]
at ($(cell.north east) + (2.0, -0.3)$) {\footnotesize\texttt{\&obj1}};
\node(list1) [draw=purple, anchor=north, minimum height=3mm, inner sep = 0.6mm]
at ($(2dlist1.south east) + (0.3, -0.5)$) {\footnotesize\texttt{\textcolor{purple}{\&func}}};
 \draw[->] 
 ($(2dlist1.south)$) --
($(list1.north)$);

\node(2dlist2) [draw=purple, anchor=west, minimum height=3mm, inner sep = 0.6mm]
at ($(2dlist1.east) + (0.6, 0)$) {\footnotesize\texttt{\textcolor{purple}{\&obj2}}};
 \draw[->] 
 ($(2dlist2.south)$) --
($(list1.north)$);

\node(valuetext) [anchor=east, minimum height=3mm, inner sep = 0.2mm, font=\sffamily]
at ($(2dlist1.south west) + (-0.2, -0.2)$) {ID Graph};

\end{tikzpicture}
\vspace{-1mm}
\caption{Detecting indirect variable accesses from aliases}
\label{fig:id_graph2}
\end{subfigure}

\vspace{3mm}

\begin{subfigure}[b]{\linewidth}
\centering
\begin{tikzpicture}[>={LaTeX[width=1mm,length=1mm]},->]

\node(notebook) [draw=black, anchor=north west, minimum width = 28mm, minimum height=24mm,densely dashdotdotted]
at (0,0) {};
\node(cell) [draw=black, anchor=north, minimum width = 26mm, minimum height=9mm]
at ($(notebook.north) + (0, -0.4)$) {};
\node(celltext) [anchor=south west, minimum height=3mm, inner sep = 0.2mm]
at ($(cell.north west) + (0, 0)$) {\small Cell 1};
\node(codetxt) [anchor=west, minimum height=3mm, inner sep = 0.2mm,align=left]
at ($(cell.west) + (0.05, 0)$) {\footnotesize\texttt{list1 = [1, 2, 3]}\\[-0.35em]\footnotesize \texttt{2dlist1 = [list1]}\\[-0.35em]\footnotesize \texttt{2dlist2 = [list1]}};
\node(cell2) [draw=black, anchor=north, minimum width = 26mm, minimum height=6mm]
at ($(cell.south) + (0, -0.4)$) {};
\node(celltext2) [anchor=south west, minimum height=3mm, inner sep = 0.2mm]
at ($(cell2.north west) + (0, 0)$) {\small Cell 2};
\node(codetxt2) [anchor=west, minimum height=3mm, inner sep = 0.2mm,align=left]
at ($(cell2.west) + (0.05, 0)$) {\footnotesize\texttt{list2 = [1, 2, 3]}\\[-0.35em]\footnotesize \texttt{2dlist1[0] = list2}};

\node(before) [anchor=west, minimum height=3mm, inner sep = 0.2mm]
at ($(notebook.north east) + (1.9, -0.2)$) {Before Cell 2};
\node(after) [anchor=west, minimum height=3mm, inner sep = 0.2mm]
at ($(before.east) + (0.4, -0)$) {After Cell 2};

\node(2dlist1) [draw=black, anchor=north, minimum height=3mm, inner sep = 0.6mm]
at ($(before.south) + (0, -0.3)$) {\footnotesize\texttt{\&2dlist1}};
\node(list1) [draw=black, anchor=north, minimum height=3mm, inner sep = 0.6mm]
at ($(2dlist1.south) + (0, -0.5)$) {\footnotesize\texttt{\&list1}};
\node(val1) [anchor=north, minimum height=3mm, inner sep = 0.3mm]
at ($(list1.south) + (0, -0.3)$) {\texttt{[[1,2,3]]}};
 \draw[->] 
 ($(2dlist1.south)$) --
($(list1.north)$);

\node(2dlist2) [draw=black, anchor=north, minimum height=3mm, inner sep = 0.6mm]
at ($(after.south) + (0, -0.3)$) {\footnotesize\texttt{\&2dlist1}};
\node(list2) [draw=purple, anchor=north, minimum height=3mm, inner sep = 0.6mm]
at ($(2dlist2.south) + (0, -0.5)$) {\footnotesize\texttt{\textcolor{purple}{\&list2}}};
\node(val2) [anchor=north, minimum height=3mm, inner sep = 0.3mm]
at ($(list2.south) + (0, -0.3)$) {\texttt{[[1,2,3]]}};
 \draw[->] 
 ($(2dlist2.south)$) --
($(list2.north)$);

\node(valuetext) [anchor=east, minimum height=3mm, inner sep = 0.2mm, font=\sffamily]
at ($(val1.west) + (-0.1, 0.05)$) {Value};
\node(idgraphtext) [anchor=south east, minimum height=3mm, inner sep = 0.2mm, font=\sffamily]
at ($(valuetext.north east) + (0, 0.8)$) {ID Graph};

\node(eq) [anchor=west, minimum height=3mm, inner sep = 0.2mm]
at ($(val1.east) + (0.12, 0.02)$) {$=$};
\node(neq) [anchor=south west, minimum height=3mm, inner sep = 0.2mm]
at ($(eq.north west) + (0, 0.8)$) {$\neq$};

\end{tikzpicture}
\vspace{-1mm}
\caption{Detecting structural variable modifications}
\label{fig:id_graph}
\end{subfigure}

\vspace{-2mm}
\caption{Two uses of the ID Graph during AHG construction.}
\vspace{-2mm}
\end{figure}
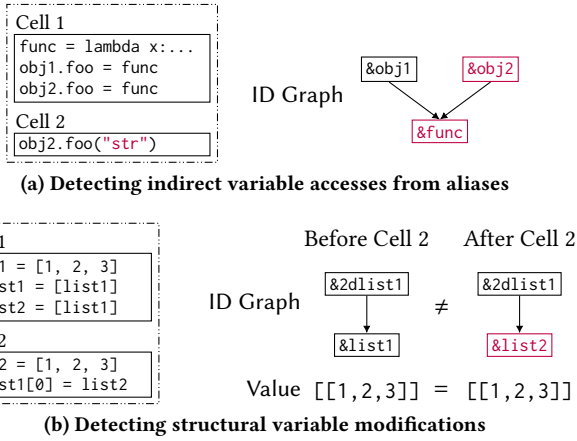

\emph{Indirect Accesses}: 
The object(s) reachable from a variable \texttt{X} may be accessed indirectly via another variable \texttt{Y}
if \texttt{X} and \texttt{Y} reference common object(s) (e.g., when aliases exist, \Cref{fig:id_graph2}),
    which cannot be identified via parsing only.
To recognize indirect accesses, we check the existence of
    overlaps between the ID Graphs of \texttt{X} and \texttt{Y}.

Our approach is conservative; that is, 
    it may over-identify variables
        by including, for example, 
        ones reachable from control flow branches 
            that were not taken during cell executions.
However, these false positives do not affect accuracy of state replication (\Cref{sec:ahg_problem}).



\paragraph{Identifying Modified Variables} 
\label{sec:identifying_modify}

Variable modifications are identified using a combination of (1) object hashes and (2) ID Graphs.

\textit{Value Changes:}
\system identifies value modifications
    by comparing hashes (by xxHash~\cite{xxhash})
        before and after each cell execution
    while using deep copy as a fallback.
If the deep copy fails 
    (e.g., unserializable or uncomparable variables),
    we consider them to be modified-on-access using results from AST and ID Graph (\cref{sec:fault_tolerance}).
This may result in false positives; however, as previously mentioned,
these false positives do not affect the accuracy.

\textit{Structural Changes:}
The ID Graph enables detecting structural changes (\Cref{fig:id_graph}).
After each cell execution, the current variables' ID Graphs are compared to the ones created before to identify reference swaps. 
In \Cref{fig:id_graph}, while the value of \texttt{2dlist1} remains unchanged after execution after executing Cell 2, the memory address of its nested list has been changed,
    no longer referencing \texttt{list1}.

\subsection{State Reconstruction with AHG}
\label{sec:ahg_problem}

This section describes how we reconstruct variable(s).
We focus on reconstructing the latest version of each variable,
    as defined in \textit{active variable snapshot} (VS) in an AHG.
    

\begin{definition}
    VS ($x$, $t_i$) is \textbf{active} if 
        $x$ is in the system (i.e., not deleted), and 
        there is no VS ($x$, $t_j$) such that $t_i < t_j$.
\end{definition}

\noindent
An active VS, ($x$, $t_i$), represents 
    the current version of $x$.
For example, even if we checkpoint after $c_{t_5}$ (in \cref{fig:problem_setup_graph}), 
``(\texttt{x}, $t_3$)'' is active since \texttt{x} was last modified by $c_{t_3}$. 
We denote the set of active VSes as $\mathcal{V}_a$.

\paragraph{Reconstruction Algorithm}

Our goal is
    to identify the most efficient computation strategy
        for reconstructing one or more active variables.
Note that we do not reconstruct non-active variables
    since they are not part of the current session state.
In achieving this goal, the AHG allows us to
    avoid unnecessary cell executions (e.g., because their outcomes have been overwritten)
        and to learn proper execution orders.
Moreover, this process can be extended to reconstruct a set of variables
    more efficiently than computing them one by one.
        while still ensuring correctness.

Specifically,
    to recompute VS ($x$, $t$), we traverse back to its ancestors in the AHG (e.g., using the breadth-first search),
        collecting all CEs into a list $req(x, t)$,
            until we find a \emph{ground variable} for every path,
    where the ground variable is a variable whose value is available in the system,
        i.e., either another active VS or copied variable.
By rerunning all the CEs in $req(x, t)$ in the order of their completion times, 
    we can obtain the target VS ($x$, $t$).\
To extend this algorithm to multiple VSes, say ($x1$, $t_{x1}$), ($x2$, $t_{x2}$), and ($x3$, $t_{x3}$),
    we obtain $req$ for each VS and union them into a merged set (that is, identical CEs collapse into one).
By rerunning all the CEs in the merged set,
    we obtain all target VSes.
\Cref{fig:problem_setup_graph} shows an example.
To recompute ($x$, $t_3$), we rerun $c_{t_3}$ which requires the previous version (\texttt{x}, $t_1$) as input, which in turn requires $c_{t_1}$ to be rerun.
Notably, it is not necessary to rerun $c_{t_2}$ as its output \texttt{z} is available in the namespace.
Finally,
    \cref{sec:fault_tolerance} discusses
        how this approach can recover even if 
    some ground variables are unexpectedly unobtainable.


\emph{Why Only Use Active VSes?} 
Theoretically, it is possible to use non-active variables as ground variables. 
That is,
by preserving deleted/overwritten variables (e.g., in a cache),
we may be able to speed up the recomputation of active variables~\cite{garcia2020hindsight, xin2018helix}.
However, we don't consider this approach as many data science workloads
    are memory-hungry with large training data and model sizes.
Still, there might be cases where we can speed up recomputation by storing small overwritten variables,
which we leave as future work.


\paragraph{Correctness of Reconstruction} 

As stated in \Cref{sec:background_graph}, the AHG is allowed to have 
    false positives, meaning it may indicate
        a cell accessed/modified variables that were not actually accessed/modified.
While the false positives have a performance impact, they do not affect the correctness of identification.

\begin{theorem}
\label{theorem:correctness}
    Given the approximate AHG $\mathcal{G}$ of \system with false positives, and the true AHG $\mathcal{G}^*$, there is $req^*(x, t^*) \subseteq req(x, t)$ for any variable $x \in \mathcal{X}$, where $(x, t)$ and $(x, t^*)$, $req$ and $req^*$ are the active VSs of $x$ and reconstruction mapping functions defined on $\mathcal{G}$ and $\mathcal{G}^*$ respectively.
\end{theorem}

\noindent
That is, for any arbitrary variable $x$, while $req(x, t)$ may contain cell executions unnecessary for recomputing $x$, it will never miss any necessary cell executions (i.e., those in $req(x, t^*)$). The proof is presented in \Cref{sec:appendix_proof}.

\section{Correct \& Efficient Replication}
\label{sec:algorithm}

\begin{figure}[t]

\begin{subfigure}[b]{0.3\linewidth}
\centering
\begin{tikzpicture}

\node(notebook) [draw=black, anchor=north west, minimum width = 23mm, minimum height=16mm,densely dashdotdotted]
at (0,0) {};
\node(celltext) [anchor=south west, minimum height=3mm, inner sep = 0.5mm]
at ($(notebook.north west) + (0, 0.05)$) {\textbf{Notebook}};

\node(cell3) [draw=black, anchor=north, minimum width = 21mm, minimum height=3.5mm]
at ($(notebook.north) + (0, -0.4)$) {};
\node(celltext3) [anchor=south west, minimum height=3mm, inner sep = 0.2mm]
at ($(cell3.north west) + (0, 0)$) {\small Cell 3 ($c_{t_3}$)};
\node(codetxt3) [anchor=west, minimum height=3mm, inner sep = 0.6mm]
at ($(cell3.west) + (0.05,0)$) {\footnotesize \texttt{l1 = [z, 2, 3]}};

\node(cell4) [draw=black, anchor=north, minimum width = 21mm, minimum height=3.5mm]
at ($(cell3.south) + (0, -0.4)$) {};
\node(celltext4) [anchor=south west, minimum height=3mm, inner sep = 0.2mm]
at ($(cell4.north west) + (0, 0)$) {\small Cell 4 ($c_{t_4}$)};
\node(codetxt4) [anchor=west, minimum height=3mm, inner sep = 0.2mm]
at ($(cell4.west) + (0.05, 0)$) {\footnotesize \texttt{2dlist = [l1]}};







\end{tikzpicture}
\vspace{-7.2mm}
\end{subfigure}
\hfill
\begin{subfigure}[b]{0.68\linewidth}
\centering
\begin{tikzpicture}
    \node(celltext) [anchor=south west, minimum height=3mm, inner sep = 0.5mm]
at (0, 0) {\textbf{Replication Plan}};
\end{tikzpicture}
\midsepremove
\small
\addtolength{\tabcolsep}{-1pt} 

\begin{tabular}{ccc}
\toprule

\texttt{l1}               & \texttt{2dlist} & \texttt{\&l1==\&2dlist1[0]}    \\ \midrule
 Migrate       & Migrate & \textcolor{black!30!RandomColor}{\texttt{\textbf{True}}} \\                
 Recompute       & Recompute & \textcolor{black!30!RandomColor}{\texttt{\textbf{True}}} \\ 
  Recompute       & Migrate & \textcolor{purple}{
    \texttt{\textbf{False}}} \\  
\bottomrule
\end{tabular}
\addtolength{\tabcolsep}{1pt} 
\end{subfigure}
\vspace{-2mm}
\caption{Two variables sharing references (in \Cref{fig:problem_setup_graph}). They must be migrated/recomputed together for the correct replication,
    serving as constraints to our opt problem (see \cref{sec:opt_problem}).
}
\label{fig:constraint}
\vspace{-2mm}
\end{figure}

This section covers how \system computes an efficient and correct plan for state replication with the AHG and profiled metrics.
We describe correctness requirements in \Cref{sec:problem_statement}, the cost model in \Cref{sec:cost_model}, the optimization problem in \Cref{sec:opt_problem}, and our solution in \cref{sec:algorithm_desc}.

\subsection{Correctness Requirements}
\label{sec:problem_statement}
\system aims to \textit{correctly} replicate session states. which we define the notion of in this section:
\begin{definition}
A replication of state $\mathcal{X}$ is \textbf{value-equivalent} if $\forall\!x\!\in\!\mathcal{X},\!x\!=\!new(x)$, where $new(x)$ is the value of $x$ post-replication.
\end{definition}

\noindent
A value-equivalent replication preserves the value of each individual variable and is guaranteed by the correct identification of $req(x, t)$ for each variable $x$ (\Cref{sec:ahg_problem}).
However, it is additionally important that shared references are preserved,
    as defined below.
\begin{definition}
A value-equivalent replication of a session state $\mathcal{X}$ is additionally \textbf{isomorphic} if $\forall a, b,\, id(a) = id(b) \rightarrow id\_new(a) = id\_new(b)$, where $a, b$ are arbitrary references (e.g., \texttt{x[0][1]}, \texttt{y.foo}), and $id(a), id\_new(a)$ are the unique IDs (i.e., memory addresses) of the objects pointed to by $a$ before and after replication.
\end{definition}

\noindent
\system defines replication as 'correct' only if it is isomorphic, requiring all shared references to be preserved: two references pointing to the same object pre-replication will still do so post-replication.
That is, inter-object relations are identical (analogous to graph isomorphism). We describe how \system ensures isomorphic replication via its \textit{linked variable constraint} in \Cref{sec:opt_problem}.
\subsection{Cost Model}
\label{sec:cost_model}

Our model captures the costs associated with
    (1) serializing variables,
    (2) writing byte data into storage (e.g., local SSD, cloud storage)
    and (3) rerunning cell executions.
These costs are computed using the AHG and profiled system metrics.
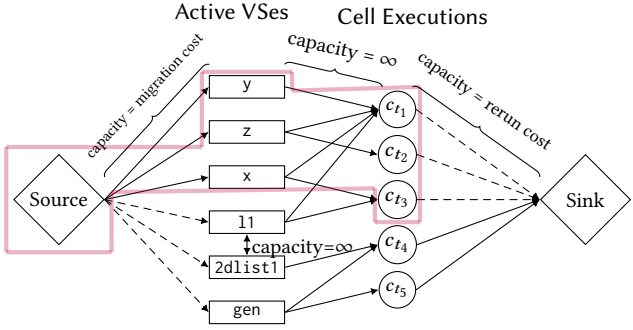
\begin{figure}[t]
\usetikzlibrary{calc}
\begin{subfigure}{\columnwidth}
\centering
\begin{tikzpicture}

\tikzset{
cutline/.style={
    draw=purple, ultra thick, opacity=0.3,
},
}

\node(ce1) [draw=black, anchor=center, minimum width = 4mm, minimum height=4mm, circle, inner sep = 0.4mm]
 at (0, 0) {\footnotesize $c_{t_1}$};
  \node(ce2) [draw=black, anchor=center, minimum width = 4mm, minimum height=4mm, circle, inner sep = 0.4mm]
 at ($(ce1) + (0, -0.6)$) {\small $c_{t_2}$};
  \node(ce3) [draw=black, anchor=center, minimum width = 4mm, minimum height=4mm, circle, inner sep = 0.4mm]
 at ($(ce2) + (0, -0.6)$) {\small $c_{t_3}$};
  \node(ce4) [draw=black, anchor=center, minimum width = 4mm, minimum height=4mm, circle, inner sep = 0.4mm]
 at ($(ce3) + (0, -0.6)$) {\small $c_{t_4}$};
  \node(ce5) [draw=black, anchor=center, minimum width = 4mm, minimum height=4mm, circle, inner sep = 0.4mm]
 at ($(ce4) + (0, -0.6)$) {\footnotesize $c_{t_5}$};

 \node(y) [draw=black, anchor=center, minimum height=3mm, minimum width = 10mm, inner sep = 0.6mm]
at ($(ce1) + (-2.0, 0.3)$) {\footnotesize\texttt{y}};
 \node(z) [draw=black, anchor=center, minimum height=3mm, minimum width = 10mm, inner sep = 0.6mm]
at ($(y) + (0, -0.6)$) {\footnotesize\texttt{z}};
 \node(x) [draw=black, anchor=center, minimum height=3mm, minimum width = 10mm, inner sep = 0.6mm]
at ($(z) + (0, -0.6)$) {\footnotesize\texttt{x}};
 \node(l1) [draw=black, anchor=center, minimum height=3mm, minimum width = 10mm, inner sep = 0.6mm]
at ($(x) + (0, -0.6)$) {\footnotesize\texttt{l1}};
 \node(2dlist1) [draw=black, anchor=center, minimum height=3mm, minimum width = 10mm, inner sep = 0.6mm]
at ($(l1) + (0, -0.6)$) {\footnotesize\texttt{2dlist1}};
 \node(gen) [draw=black, anchor=center, minimum height=3mm, minimum width = 10mm, inner sep = 0.6mm]
at ($(2dlist1) + (0, -0.6)$) {\footnotesize\texttt{gen}};

   \node(sink) [draw=black, anchor=center, diamond, minimum width = 12mm, minimum height=12mm,inner sep = 0.6mm]
 at ($(ce3) + (2.5, 0)$) {\small Sink};
    \node(source) [draw=black, anchor=center, diamond, minimum width = 12mm, minimum height=12mm,inner sep = 0.6mm]
 at ($(x) + (-2.5, -0.3)$) {\small Source};

\draw[->,>={LaTeX[width=1mm,length=1mm]}, densely dashed] 
($(ce1.east)$) --
($(sink.west)$);
\draw[->,>={LaTeX[width=1mm,length=1mm]}, densely dashed] 
($(ce2.east)$) --
($(sink.west)$);
\draw[->,>={LaTeX[width=1mm,length=1mm]}, densely dashed] 
($(ce3.east)$) --
($(sink.west)$);
\draw[->,>={LaTeX[width=1mm,length=1mm]}] 
($(ce4.east)$) --
($(sink.west)$);
\draw[->,>={LaTeX[width=1mm,length=1mm]}] 
($(ce5.east)$) --
($(sink.west)$);

\draw[->,>={LaTeX[width=1mm,length=1mm]}] 
($(source.east)$) --
($(y.west)$);
\draw[->,>={LaTeX[width=1mm,length=1mm]}] 
($(source.east)$) --
($(z.west)$);
\draw[->,>={LaTeX[width=1mm,length=1mm]}] 
($(source.east)$) --
($(x.west)$);
\draw[->,>={LaTeX[width=1mm,length=1mm]}, densely dashed] 
($(source.east)$) --
($(l1.west)$);
\draw[->,>={LaTeX[width=1mm,length=1mm]}, densely dashed] 
($(source.east)$) --
($(2dlist1.west)$);
\draw[->,>={LaTeX[width=1mm,length=1mm]}, densely dashed] 
($(source.east)$) --
($(gen.west)$);

\draw[->,>={LaTeX[width=1mm,length=1mm]}] 
($(y.east)$) --
($(ce1.west)$);

\draw[->,>={LaTeX[width=1mm,length=1mm]}] 
($(z.east)$) --
($(ce1.west)$);
\draw[->,>={LaTeX[width=1mm,length=1mm]}] 
($(z.east)$) --
($(ce2.west)$);

\draw[->,>={LaTeX[width=1mm,length=1mm]}] 
($(x.east)$) --
($(ce1.west)$);
\draw[->,>={LaTeX[width=1mm,length=1mm]}] 
($(x.east)$) --
($(ce3.west)$);
\draw[->,>={LaTeX[width=1mm,length=1mm]}] 
($(l1.east)$) --
($(ce1.west)$);
\draw[->,>={LaTeX[width=1mm,length=1mm]}] 
($(l1.east)$) --
($(ce3.west)$);

\draw[->,>={LaTeX[width=1mm,length=1mm]}] 
($(2dlist1.east)$) --
($(ce4.west)$);
\draw[->,>={LaTeX[width=1mm,length=1mm]}] 
($(gen.east)$) --
($(ce4.west)$);
\draw[->,>={LaTeX[width=1mm,length=1mm]}] 
($(gen.east)$) --
($(ce5.west)$);
\draw[<->,>={LaTeX[width=1mm,length=1mm]}] 
($(2dlist1.north)$) --
($(l1.south)$);

\draw[-, color=purple, ultra thick, cutline] 
($(y) + (-0.6, 0.2)$) --
($(y) + (0.6, 0.2)$);
\draw[-, color=purple, ultra thick, cutline] 
($(y) + (0.6, -0.05)$) --
($(y) + (0.6, 0.2)$);
\draw[-, color=purple, ultra thick, cutline] 
($(y) + (0.6, -0.05)$) --
($(ce1) + (0.3, 0.3)$);
\draw[-, color=purple, ultra thick, cutline] 
($(ce3) + (0.3, -0.3)$) --
($(ce1) + (0.3, 0.3)$);
\draw[-, color=purple, ultra thick, cutline] 
($(ce3) + (0.3, -0.3)$) --
($(ce3) + (-0.3, -0.3)$);
\draw[-, color=purple, ultra thick, cutline] 
($(ce3) + (-0.3, -0.3)$) --
($(ce3) + (-0.3, 0.15)$);
\draw[-, color=purple, ultra thick, cutline] 
($(ce3) + (-0.3, 0.15)$) --
($(x) + (-1.8, -0.2)$);
\draw[-, color=purple, ultra thick, cutline] 
($(source) + (0.7, -0.7)$) --
($(x) + (-1.8, -0.2)$);
\draw[-, color=purple, ultra thick, cutline] 
($(source) + (0.7, -0.7)$) --
($(source) + (-0.7, -0.7)$);
\draw[-, color=purple, ultra thick, cutline] 
($(source) + (-0.7, 0.7)$) --
($(source) + (-0.7, -0.7)$);
\draw[-, color=purple, ultra thick, cutline] 
($(source) + (-0.7, 0.7)$) --
($(source) + (1.9, 0.7)$);
\draw[-, color=purple, ultra thick, cutline] 
($(y) + (-0.6, 0.2)$) --
($(source) + (1.9, 0.7)$);

\node[anchor=center, align=left, font=\sffamily] at ($(y) + (-0.2, 1.0)$) {Active VSes};
\node[anchor=center, align=left, font=\sffamily] at ($(ce1) + (0.2, 1.25)$) {Cell Executions};

\draw [decorate,
    decoration = {brace}] ($(source.east) + (0, 0.3)$) --  ($(y)+ (-0.5, 0.3)$);
\node[anchor=center, align=left, rotate= 46.975] at ($(source) + (1.15, 1.375)$) {\scriptsize capacity = migration cost};
\draw [decorate,
    decoration = {brace}] ($(y)+ (0.5, 0.3)$) --  ($(ce1)+ (-0.2, 0.3)$);
\node[anchor=center, align=left, rotate=-12.995] at ($(ce1) + (-0.73, 0.775)$) {\small capacity = $\infty$};
\draw [decorate,
    decoration = {brace}] ($(ce1)+ (0.2, 0.3)$) --  ($(sink.west)+ (0, 0.3)$);
\node[anchor=center, align=left, rotate=-35.21] at ($(sink)+ (-1.35, 1.3)$) {\footnotesize capacity = rerun cost};
\node[anchor=west, align=left] at ($(l1) + (-0.05, -0.35)$) {\small capacity=$\infty$};

\end{tikzpicture}
\vspace{-2mm}
\end{subfigure}
    \caption{Running min-cut on the flow graph constructed from the AHG in \cref{fig:problem_setup_graph}. The partition (red) defined by the minimum cut (dashed edges) determines the replication plan.
}
\label{fig:algorithm}
\vspace{-2mm}
\end{figure}
\paragraph{Variable Migration Cost}

\textit{Migrating} a variable (from one session to another) 
    includes serializing it to the checkpoint file, 
    then loading it into a new session. 
Given a subset of variables to migrate $\mathcal{S}\subseteq \mathcal{X}$, the migration cost $w_M$ can be expressed as follows:
\begin{equation}
\label{eq:migrate_cost}
w_M(\mathcal{S}) = \sum_{x\in\mathcal{S}} \alpha \times w_{store}(x) + w_{load}(x)
\end{equation}
Where $w_{store}(x)$ and $w_{load}(x)$ are the time costs for serializing the value of $x$ at checkpointing time into a file and unpacking into the new session, respectively.
These times are estimated using the size of $x$ and storage latency/bandwidth from \system's Profiler (\Cref{sec:session_lifecycle}).
The time costs for unserializable variables are set to infinity.
$\alpha$ is a coefficient for adjusting the time cost of storage; for example, if \system is to be invoked upon auto-suspension, $\alpha$ can be set to a low value to discount the user-perceived time of storing variables prior to completely suspending a session (as the user is likely away).


\paragraph{Variable Recomputation Cost}

The Interceptor records cell runtimes during a session lifecycle (\Cref{sec:session_lifecycle}).
Combined with the reconstruction mapping $req()$ for the AHG 
(\cref{sec:ahg_problem}), 
    the cost $w_R$ for recomputing a subset of variables $\mathcal{S}\subseteq \mathcal{X}$ can be defined as follows:
\begin{equation}
\label{eq:recompute_cost}
w_R(\mathcal{S})\!=\!\sum_{c\in req(\mathcal{S})}w_{rerun}(c), \,\,\text{where}\,\,req(\mathcal{S})\!=\!\bigcup_{x \in \mathcal{S}}req(x, t)
\end{equation}
where $(x, t)$ is the active VS of $x$ and $w_{rerun}(c):\mathcal{C}\rightarrow \mathbb{R}^+$ is the estimated time to rerun the CE $c$ in the new session. 

\paragraph{Replication Plan Cost}
Using migration and recomputation costs (i.e., \cref{eq:migrate_cost,eq:recompute_cost}), 
    the total cost $w$---with variables to migrate $\mathcal{S}$ and variables to recompute $\mathcal{X} - \mathcal{S}$---is expressed as:
\begin{equation}
\label{eq:cost_model}
    w(\mathcal{S}) = w_M(\mathcal{S}) + w_R(\mathcal{X} - \mathcal{S})
\end{equation}

\subsection{Optimization Problem for State Replication}
\label{sec:opt_problem}
The goal is to find the variables to migrate $\mathcal{S} \subseteq \mathcal{X}$
    that minimizes the cost \cref{eq:cost_model}.
To ensure isomoprhic replication
    in consideration of variable inter-dependencies,
    additional constraints are added.

\paragraph{Constraint for Linked Variables}

Two variables containing references to the same object (which we refer to as \textit{linked variables}, e.g., \texttt{l1} and \texttt{2dlist1} in \Cref{fig:constraint}) must be either both migrated or recomputed, as migrating one and recomputing the other may result in their contained shared reference/alias being broken, as illustrated in \Cref{fig:constraint}.
Let the set of linked variable pairs be denoted as $\mathcal{L}$, then the constraint can be formally expressed as follows:
\begin{equation}
\label{eq:constraint}
    (x_1\in \mathcal{S}\land x_2\in\mathcal{S})\lor (x_1\not\in \mathcal{S}\land x_2\not\in\mathcal{S})\,\,\forall (x_1, x_2)\in \mathcal{L}
\end{equation}

\paragraph{Problem definition}
Using the cost model in \cref{eq:cost_model} and the constraint in \cref{eq:constraint},
    we formally define the state replication problem:

\begin{problem}{\textbf{\textsf{Optimal State Replication}}}
\label{prof:optimization}

\label{optimization_problem_def}
\vspace{-2mm}
\end{problem}
\hspace{-6mm}
\addtolength{\tabcolsep}{-2pt}
\begin{tabularx}{\linewidth}{lp{0.8\linewidth}}
\textsf{Input:} & 1. AHG $\mathcal{G} = \{\mathcal{V}\cup\mathcal{C},\mathcal{E}\}$  \\         & 2. Migration cost function $w_M: 2^\mathcal{X} \rightarrow \mathbb{R}^+$\\
  & 3. Recompute cost function $w_R: 2^\mathcal{X} \rightarrow \mathbb{R}^+$\\    
 & 4. Linked variables $\mathcal{L}\subseteq \mathcal{X}\times \mathcal{X}$\\
 \textsf{Output:} & A replication plan of subset of variables $\mathcal{S} \subseteq \mathcal{X}$ for which we migrate (and another subset $\mathcal{X}-\mathcal{S}$ which we recompute)\\
 \textsf{Objective:} & Minimize replication cost $w_M(\mathcal{S}) + w_R(\mathcal{X}-\mathcal{S})$\\
 \textsf{Constraint:} & Linked variables are either both migrated or recomputed: $(x_1,x_2\in\mathcal{S})\lor (x_1,x_2 \not\in\mathcal{S})\,\,\forall (x_1, x_2)\in \mathcal{L}$\\
\end{tabularx}%
\addtolength{\tabcolsep}{2pt}
\\

\noindent
The next section (\Cref{sec:algorithm_desc})
presents our solution to \Cref{optimization_problem_def}.

\subsection{Solving State Replication Opt.~Problem}
\label{sec:algorithm_desc}

We solve \Cref{optimization_problem_def} by reducing it to a min-cut problem,
    with a $src\text{-}sink$ flow graph constructed from the AHG 
such that each $src\text{-}sink$ cut 
    (a subset of edges, which, when removed from the flow graph, disconnects source $s$ and sink $t$) 
corresponds to a replication plan $\mathcal{S}$, while the cost of the cut is equal to the replication cost $w_M(\mathcal{S}) + w_R(\mathcal{X}-\mathcal{S})$.
Therefore, finding the minimum cost $src\text{-}sink$ cut 
    is equivalent to finding the optimal replication plan.

\paragraph{Flow Graph Construction} A flow graph $H := \{\mathcal{V}_H, \mathcal{E}_H\}$ 
and its edge capacity $\phi: \mathcal{E}_H \rightarrow \mathbb{R}^+$ are defined as follows:
\begin{itemize}
    \item 
    $\mathcal{V}_H = \mathcal{V}_a \cup \mathcal{C} \cup \{src, sink\}$:
    $\mathcal{V}_a$ is active VSes,
    $\mathcal{C}$ is cell executions,
    and $src$ and $sink$ are \emph{dummy} source and sink nodes.
    
    \item $\forall x \in \mathcal{V}_a$,
    $(src, (x, t)) \in \mathcal{E}_H$ and
    $\phi(src, (x, t)) = w_M(x)$: 
    We add an edge from the source to each active VS with a capacity equal to the migration cost of the variable.
    
    \item $\forall c \in \mathcal{C}$,
    $(c, sink) \in \mathcal{E}_H$ and
    $\phi(c, sink) = w_{rerun}(c)$: 
    We add an edge with capacity from each CE to the sink with a capacity equal to the rerun cost of the CE.
    
    \item $\forall c \in \mathcal{C}$, 
    $c \in req(x, t) \rightarrow ((x, t), c) \in \mathcal{E}_H$ and
    $\phi((x, t), c) = \infty$ and 
    $(x, t) \in \mathcal{V}_a$: 
    We add an edge with infinite capacity from an active VS $(x, t)$ to a CE $c$ if $(x, t)$ must be recomputed.
    
    \item 
    $\forall (x_1, x_2) \in \mathcal{L}$, \,
    $((x_1, t_1)\leftrightarrow (x_2, t_2)) \in \mathcal{E}_H$ and 
    $\phi((x_1, t_1) \leftrightarrow (x_2, t_2)) = \infty$:
    We add a \textit{bi-directional} edge with an infinite capacity between each pair of active VSes corresponding to linked variables $x_1$ and $x_2$, e.g., \texttt{l1} and \texttt{2dlist1}.
\end{itemize}
The flow graph $\mathcal{H}$ for the AHG in \Cref{fig:problem_setup_graph} is depicted in \Cref{fig:algorithm}.

\paragraph{Solution}
We can now solve \Cref{optimization_problem_def} by running a $src\text{---}sink$ min-cut solving algorithm (i.e., Ford-Fulkerson~\cite{ford1962flows}) on $H$. 
The set of edges that form the $src\text{---}sink$ min-cut (dashed edges), when removed, disconnects $src$ from $sink$; therefore, it defines a partition (in red) of the nodes into nodes reachable from $src$, $\mathcal{V}_{H_{src}}$ and nodes unreachable from $src$, $\mathcal{V}_{H_{sink}}$. The replication plan can be obtained from the partition:
\begin{itemize}
    \item $\mathcal{S} = \{x\,\vert\,(x,t) \in \mathcal{V}_{H_{sink}} \cap \mathcal{V}_a\}$ are the active variable snapshots (and thus variables) that we want to migrate; in the example, these variables are \texttt{l1}, \texttt{2dlist1}, and \texttt{gen}.
    \item $\mathcal{V}_{H_{src}} \cap \mathcal{C}$ are the CEs which we will rerun post-migration to recompute $\mathcal{X}-\mathcal{S}$. In the example, these CEs are $t_1$, $t_2$, and $t_3$; when rerun, they recompute \texttt{y}, \texttt{z}, and \texttt{x}.\footnote{Rerunning $t_3$ also recomputes \texttt{l1}; however, it will be overwritten with the stored \texttt{l1} in the checkpoint file following the procedure in \Cref{sec:system_workflow}. This is to preserve the link between \texttt{l1} and \texttt{2dlist1}.}
\end{itemize}
By construction of $\mathcal{H}$, the sum of migration and recomputation costs of this configuration $w_M(\{x\,\vert\,  (x,t) \in \mathcal{V}_{H_{sink}}) + w_R(\mathcal{C}_a - (\mathcal{V}_{H_{src}} \cap \mathcal{C}))$ is precisely the cost of the found $src\text{---}sink$ min-cut.

\section{Implementation and Discussion}
\label{sec:implementation}

This section describes \system's implementation details (\Cref{sec:fault_tolerance}) and 
    design considerations (\Cref{sec:design_considerations}).

\subsection{Implementation}
\label{sec:fault_tolerance}


\paragraph{Integrating with Jupyter}

For seamless integration,
\system's data layer is implemented using a magic extension~\cite{jupytermagic}, 
    which is loaded into the kernel upon session initialization.
The cell magic is automatically added to each cell (\Cref{sec:background_abstraction}) 
    to transparently intercept user cell executions,
    perform code analyses, create ID Graphs and object hashes, and so on.

\paragraph{Serialization Protocol}
The Pickle protocol (e.g., \texttt{\_\_reduce\_\_}) is 
employed
for (1) object serialization and (2) definition of reachable objects, i.e., an object \texttt{y} is reachable from a variable \texttt{x} if \texttt{pickle(x)} includes \texttt{y}.
As Pickle is the de-facto standard (in Python) observed by almost all data science libraries (e.g., NumPy, PyTorch~\cite{pytorchcheckpoint}), \system can be used for almost all use cases.



\paragraph{Handling Undeserializable variables} 

Certain variables can be serialized but contain errors in its \textit{deserialization} instructions (which we refer to as \textit{undeserializable} variables), and are typically caused by oversights in incompletely implemented libraries~\cite{bokeh, dataprepeda}. While undetectable via serializability checks prior to checkpointing, \system handles them via fallback recomputation: if \system encounters an error while deserializing a stored variable during session restoration, it will trace the AHG to determine and rerun (only) necessary cell executions to recompute said variable, which is still faster than recomputing the session from scratch.

\subsection{Design Considerations}
\label{sec:design_considerations}

\paragraph{Definition of Session State}

In \system, the session state is formally defined as the contents of the user namespace dictionary (\texttt{user\_ns}),
    which contains key-value pairs of variable names to their values (i.e., reachable objects).
The session state does not include local/module/hidden variables,
    which we do not aim to capture.


\paragraph{Unobservable State / External Functions}

Although the Pickle protocol is
    followed by almost all libraries, 
there could be lesser-known ones 
    with incorrect serialization 
        (e.g., ignoring data defined in a C stack).
To address this, \system can be easily extended to 
allow users to annotate cells/variables 
    to inform our system that they must be recomputed
        for proper reconstruction. 
Mathematically, this has the same effect as setting their recomputation costs to infinity in \cref{eq:recompute_cost}.

\paragraph{Cell Executions with Side Effects}
Certain cell executions may cause external changes
    outside a notebook session
(e.g., filesystem) and may not be desirable to rerun (e.g., uploading items to a repository). 
Our prototype currently does not identify these side effects 
as our focus is read-oriented data science and analytics workloads.
Nevertheless, our system can be extended at least in two ways to prevent them.
\textbf{\textit{(1: Annotation)}}
    We can allow users to add manual annotations to the cells that may
        cause side effects; then,
    our system will never re-run them during replications\footnote{Replication may be unfeasible due to annotations, e.g., an unserializable variable requiring an cell execution annotated 'never-rerun' to recompute. \system can detect these cases as they have infinite min-cut cost (\Cref{sec:algorithm_desc}), upon which the user can be warned to delete the problematic variable to proceed with replicating the remaining (majority of) variables in the state.}
\textbf{\textit{(2: Sandbook)}}
    We can block external changes
by replicating a notebook into a sandbox
    with altered file system access (e.g., chroot~\cite{chroot}) 
and blocked outgoing network (e.g., ufw~\cite{ufw}). 
The sandbox can then be associated with regular file/network accesses upon successful restoration.

\paragraph{Non-deterministic Operations} 

The replication has the same effect as rerunning the cells in the exact same order as they occurred in the past;
    thus, under the existence of nondeterministic operations (e.g., \texttt{randint()}),
the reconstructed variables may have different values than the original ones.
Users can avoid this by using annotations to inform \system to always copy them.


\paragraph{Library Version Compatibility} 

Accurate replication is ensured when external resources (e.g., installed modules, database tables)
    remain the same before and after the replication.
While there are existing tools (i.e., pip freeze~\cite{pipfreeze}) for reproducing computational environments on existing data science platforms (i.e., Jupyter Notebook, Colab)~\cite{wannipurage2022framework, ahmad2022reproducible},
    this work does not incorporate such tools.


\section{Experimental Evaluation}
\label{sec:experiments}

In this section, we empirically study the effectiveness of \system's session replication. 
We make the following claims: 
\begin{enumerate}
\item \textbf{Robust Replication:} Unlike existing mechanisms, \system is capable of replicating almost all notebooks. (\Cref{sec:exp_robust})
\item \textbf{Faster Migration:} \system reduces session migration time to upscaled/downscaled machines by 85\%--98\%/84\%-99\% compared to rerunning all cells and is up to 2.07$\times$/2.00$\times$ faster than the next best alternative, respectively. (\Cref{sec:exp_migrate}) 
\item \textbf{Faster Resumption:} \system reduces session restoration time by 94\%--99\% compared to rerunning all cells and is up to 3.92$\times$ faster than the next best alternative. (\Cref{sec:exp_restore}) 
\item \textbf{Low Runtime Overhead:} \system incurs negligible overhead---amortized runtime and memory overhead of <2.5\% and <10\%, respectively. (\Cref{sec:exp_overhead})
\item \textbf{Low Storage Overhead:} \system's checkpoint sizes are up to 66\% smaller compared to existing tools. (\Cref{sec:exp_storage})
\item \textbf{Adaptability to System Environments:} \system achieves consistent savings across various environments 
    with different network speeds and available compute resources. (\Cref{sec:exp_bandwidth})
\item \textbf{Scalability for Complex Notebooks}: 
    \system's runtime and memory overheads remain negligible (<150ms, <4MB)
    even for complex notebooks with 2000 cells. (\Cref{sec:exp_scalability})
\end{enumerate}

\subsection{Experiment Setup}
\label{sec:exp_setup}

\paragraph{Datasets} 
We select a total of 60 notebooks from 4 datasets:
\begin{itemize}
\item \kaggle~\cite{kaggle}: We select 35 popular notebooks on the topic of EDA (exploratory data analysis) + machine learning from Kaggle created by Grandmaster/Master-level users.
\item \jwst~\cite{jwst}: We select 5 notebooks on the topic of data pipelining from the example notebooks provided on investigating data from the James Webb Space Telescope (JWST).
\item \tutorial~\cite{tutorial}: We select 5 notebooks from the Cornell Virtual Workshop Tutorial. These notebooks are lightweight and introduce tools (i.e., clustering, graph analysis) to the user.
\item \homework~\cite{handsonml, mlnotebooks, mlnotebooksstanford}: 15 in-progress notebooks are chosen from data science exercises. 
They contain out-of-order cell executions, runtime errors, and mistakes (e.g., \texttt{df\_backup=df}\footnote{This creates a shallow copy of \texttt{df}, which does not serve the purpose of backup.}).
\end{itemize}
\cref{tbl:workload} reports our selected notebooks' dataset sizes and runtimes.

\begin{table}[t]

\caption{Summary of datasets for evaluation.}
\vspace{-3mm}
\small
\addtolength{\tabcolsep}{-3pt} 
\midsepremove
\begin{tabular}{lllll}
\toprule
\textbf{Dataset} & \textbf{Notebooks} & \textbf{Runtime (s)} & \textbf{Input data (MB)} & \textbf{Cell count}\\
 \midrule
Kaggle~\cite{kaggle} & 35 & 178-31831 & 107--12,560     &  15--103 \\
JWST~\cite{jwst} & 5 & 25--323 & 2--109 & 21--44 \\
Tutorial~\cite{tutorial} & 5 & 10--96 & 1--139 & 10--48 \\
HW~\cite{handsonml, mlnotebooks, mlnotebooksstanford} & 15 & 9--1203 & 16--439 & 11--160 \\
\bottomrule
\end{tabular}
\midsepdefault
\addtolength{\tabcolsep}{3pt}
\vspace{-2mm}
\label{tbl:workload}
\end{table}

\paragraph{Methods} We evaluate \system against existing tools capable of performing session replication:
\begin{itemize}
    \item \checkpoint~\cite{jupytercheckpoint}: Save (only) cell code and outputs as an \texttt{ipynb} file. 
        All cells are rerun to restore the session state.
    \item \criu~\cite{criu}: Performs a system-level memory dump of the process hosting the notebook session. 
        The session state is restored by loading the memory dump and reviving the process.
    \item \storedill~\cite{jupyterstore}: A checkpointing tool that serializes variables one by one into storage. We use a modified version using Dill~\cite{dill} instead of Pickle~\cite{pickle} for robustness.\footnote{The original implementation of \%store uses Python Pickle~\cite{pickle}, and fails on too many notebooks to give meaningful results.}
    \item \dumpsession~\cite{dumpsession}: Unlike \storedill, 
        \dumpsession packs the entire session state into one single file.
\end{itemize}

\paragraph{Ablation Study} We additionally compare against the following ablated implementations of \system:\begin{itemize}
    \item \helix~\cite{xin2018helix}: 
    We replace our min-cut solution with \textsf{Helix},
    which does not consider linked variables (\Cref{sec:opt_problem}).
    \item \noidgraph: 
    This method omits ID Graphs, relying only on
        AST analysis and object hashes
    for detecting variable accesses and modifications, respectively.
\end{itemize} We consider these methods regarding replication correctness (\Cref{sec:exp_robust}) to gauge the impact of ignoring (1) the linked constraint and (2) implicit accesses and structural modifications, respectively.
\begin{figure}[t]
\usetikzlibrary{patterns}
\begin{subfigure}[b]{\linewidth}
\centering
\begin{tikzpicture}

\pgfplotstableread[col sep=comma,]{
name
\checkpoint
\criu
\storedill
\dumpsession
\system \textbf{(Ours)}
}\datatable

\begin{axis}[
    ybar stacked,
    clip=false,
    xtick={},
    xticklabels={},
    xlabel style={yshift = 4ex},
    width=85mm,
    height=28mm,
    bar width=3mm,
    ymin=0,
    ymax=100,
    ylabel style={yshift = -3ex},
    axis y line*=none,
    axis x line*=none,
    ytick={0, 20, 40, 60, 80, 100},
    yticklabels={0\%, 20\%, 40\%, 60\%, 80\%, 100\%},
    xmin=0.5,
    xmax = 8.5,
    ymajorgrids,
    tick label style={font=\footnotesize},
    legend style={
        font=\scriptsize,
        /tikz/every even column/.append style={column sep=0cm},
        legend columns = 3,
        at={(-0.2, 1.1)}, anchor=south west
    },
    label style={font=\scriptsize},
    ylabel={Success rate (\%)},
    area legend
    ]


                 
    
    \addplot
    [NoOptColor,fill=NoOptColor,x tick label style={xshift=-0.3cm}] table[x=Method,y=rerun] {sections/plots/exp_robust_table.txt};
    \addlegendentry[]{\checkpoint};
    
    \addplot [LRUColor,fill=LRUColor,x tick label style={xshift=-0.3cm}] table[x=Method,y=criu] {sections/plots/exp_robust_table.txt};
    \addlegendentry[]{\criu (same architecture)};

    \addplot [black, dashed, thick, fill=none, 
        x tick label style={xshift=-0.3cm}] 
        table[x=Method,y=criutwo] {sections/plots/exp_robust_table.txt};
    \addlegendentry[]{\criu (cross-architecture)};
    
    \addplot [RandomColor,fill=RandomColor,x tick label style={xshift=-0.3cm}] table[x=Method,y=store] {sections/plots/exp_robust_table.txt};
    \addlegendentry[]{\storedill};
    
    \addplot
    [GreedyColor,fill=GreedyColor,x tick label style={xshift=-0.3cm}] table[x=Method,y=dump] {sections/plots/exp_robust_table.txt};
    \addlegendentry[]{\dumpsession};

    \addplot [BlueColor,fill=BlueColor,x tick label style={xshift=-0.3cm}] table[x=Method,y=helix] {sections/plots/exp_robust_table.txt};
    \addlegendentry[]{\helix};

    \addplot [GreenColor,fill=GreenColor,x tick label style={xshift=-0.3cm}] table[x=Method,y=noidgraph] {sections/plots/exp_robust_table.txt};
    \addlegendentry[]{\noidgraph};
    
    \addplot [HeuristicColor,fill=HeuristicColor,x tick label style={xshift=-0.3cm}] table[x=Method,y=ours] {sections/plots/exp_robust_table.txt};
    \addlegendentry[]{\system \textbf{(Ours)}};

    \node[anchor=south west, red, rotate=90,
        font=\scriptsize\bfseries] 
        at (axis cs: 3.2, -7) 
        {100\% failure};

\end{axis}
    
    
\end{tikzpicture}
\end{subfigure}

\vspace{-4mm}
\caption{Ratio of correct replications.
\systemnosf achieves 100\% correctness, on par with full rerun (\checkpoint).}
\vspace{-2mm}
\label{fig:exp_robust}
\end{figure}
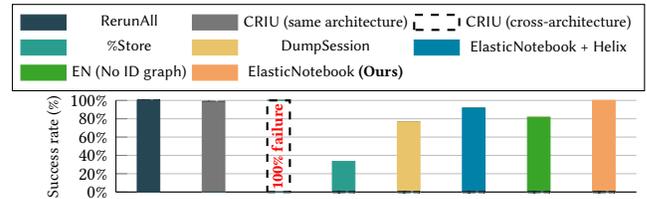

\begin{table}[t]
\caption{Existing work fails for these cases. Ours works.}
\label{tbl:unserializable}
\vspace{-3mm}
\small
\midsepremove
\begin{tabular}{lll}
\toprule
 \textbf{Notebook(s)} & \textbf{Type} & \textbf{Description and purpose}\\
\midrule
NFL~\cite{nfl} & hashlib~\cite{hashlib} & Dropdown list in plot            \\
\midrule
All 5 JWST & mmap~\cite{mmap} & Helps avoid reading large file   \\
notebooks~\cite{jwst}&  & into memory       \\
\midrule
Arxiv~\cite{arxivdata} & generator~\cite{generators} & Speedup iterable comprehension  \\
Plant~\cite{plantdisease} &  & via lazy element generation\\
\bottomrule
\end{tabular}
\midsepdefault

\end{table}


\paragraph{Environment}
We use 
an Azure Standard D32as v5 VM instance with 32 vCPUs and 128 GB RAM.
For the migration experiment (\Cref{sec:exp_migrate}), we migrate sessions from D32as to D64as/D16as with 64/16 vCPUs and 256/64 GB RAM for upscaling/downscaling, respectively.
Input data and checkpoints are read/stored from/to an Azure storage 
    with block blobs configuration (NFS).
Its network bandwidth is 274 MB/s with a read latency of 175 $\mu s$.

\paragraph{Time measurement} We measure (1) \emph{migration time} as the time from starting the checkpointing process to having the state restored (i.e., all variables declared into the namespace) in the destination session and (2) \emph{restoration time} as the time to restore the state from a checkpoint file.
We clear our cache between (1) checkpointing and restoring a notebook and (2) between subsequent runs.
\subfile{plots/exp_faster_migrate}
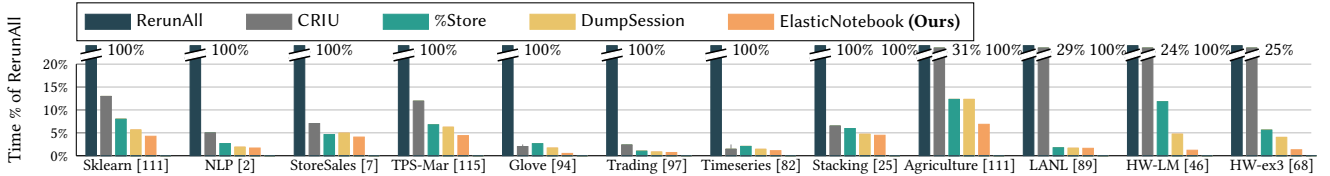
\begin{figure*}[t]
\usetikzlibrary{patterns}
\begin{subfigure}[b]{\linewidth}
\centering
\begin{tikzpicture}

\pgfplotstableread[col sep=comma,]{
name
Sklearn~\cite{sklearntweet}
NLP~\cite{nlpglove}
StoreSales~\cite{storesales}
TPS-Mar~\cite{tpsmar}
Glove~\cite{glove}
Trading~\cite{asset}
Timeseries~\cite{timeseries}
Stacking~\cite{modelstacking}
Agriculture~\cite{sklearntweet}
LANL~\cite{lanl}
HW-LM~\cite{hwlm}
HW-ex3~\cite{hwex3}
}\datatable

\begin{axis}[
    ybar stacked,
    clip=false,
    xtick={2, 9, 16, 23, 30, 37, 44, 51, 58, 65, 72, 79},
    xticklabels from table={\datatable}{name},
                 x tick label style={anchor=center,xshift=2ex, yshift = -1ex},
    xlabel style={yshift = 2.5ex},
    ylabel style={yshift = -3ex, xshift=2ex},
    width=180mm,
    height=28mm,
    bar width=1.4mm,
    ymin=0,
    ymax=0.2,
    axis y line*=none,
    axis x line*=none,
    ytick={0, 0.05, 0.10, 0.15, 0.20},
    yticklabels={0\%, 5\%, 10\%, 15\%, 20\%},
    xmin=0,
    xmax = 83,
    ymajorgrids,
    tick label style={font=\scriptsize},
    legend style={
        font=\footnotesize,
        /tikz/every even column/.append style={column sep=0.5cm},
        legend columns = 5,
        at={(0,1.25)},
        anchor=south west,
    },
    label style={font=\footnotesize},
    ylabel={Time \% of \checkpoint},
    area legend
    ]
                 
    \draw[fill=white,draw=white] (axis cs: 0.3,0.22) -- (axis cs: 1.7,0.23) -- (axis cs: 1.7,0.22) -- (axis cs: 0.3,0.21) -- cycle;
    \draw[draw=black,ultra thick] (axis cs: 1.7,0.23) -- (axis cs: 0.3,0.22);
    \draw[draw=black,ultra thick] (axis cs: 1.7,0.22) -- (axis cs: 0.3,0.21);
    
    \node[anchor=south west] at (axis cs: 1.5,0.20) {\footnotesize 100\%};

    \draw[fill=white,draw=white] (axis cs: 7.3,0.22) -- (axis cs: 8.7,0.23) -- (axis cs: 8.7,0.22) -- (axis cs: 7.3,0.21) -- cycle;
    \draw[draw=black,ultra thick] (axis cs: 8.7,0.23) -- (axis cs: 7.3,0.22);
    \draw[draw=black,ultra thick] (axis cs: 8.7,0.22) -- (axis cs: 7.3,0.21);
    
    \node[anchor=south west] at (axis cs: 8.5,0.20) {\footnotesize 100\%};

    \draw[fill=white,draw=white] (axis cs: 14.3,0.22) -- (axis cs: 15.7,0.23) -- (axis cs: 15.7,0.22) -- (axis cs: 14.3,0.21) -- cycle;
    \draw[draw=black,ultra thick] (axis cs: 15.7,0.23) -- (axis cs: 14.3,0.22);
    \draw[draw=black,ultra thick] (axis cs: 15.7,0.22) -- (axis cs: 14.3,0.21);
    
    \node[anchor=south west] at (axis cs: 15.5,0.20) {\footnotesize 100\%};

    \draw[fill=white,draw=white] (axis cs: 21.3,0.22) -- (axis cs: 22.7,0.23) -- (axis cs: 22.7,0.22) -- (axis cs: 21.3,0.21) -- cycle;
    \draw[draw=black,ultra thick] (axis cs: 22.7,0.23) -- (axis cs: 21.3,0.22);
    \draw[draw=black,ultra thick] (axis cs: 22.7,0.22) -- (axis cs: 21.3,0.21);
    
    \node[anchor=south west] at (axis cs: 22.5,0.20) {\footnotesize 100\%};

    \draw[fill=white,draw=white] (axis cs: 28.3,0.22) -- (axis cs: 29.7,0.23) -- (axis cs: 29.7,0.22) -- (axis cs: 28.3,0.21) -- cycle;
    \draw[draw=black,ultra thick] (axis cs: 29.7,0.23) -- (axis cs: 28.3,0.22);
    \draw[draw=black,ultra thick] (axis cs: 29.7,0.22) -- (axis cs: 28.3,0.21);
    
    \node[anchor=south west] at (axis cs: 29.5,0.20) {\footnotesize 100\%};

    \draw[fill=white,draw=white] (axis cs: 35.3,0.22) -- (axis cs: 36.7,0.23) -- (axis cs: 36.7,0.22) -- (axis cs: 35.3,0.21) -- cycle;
    \draw[draw=black,ultra thick] (axis cs: 36.7,0.23) -- (axis cs: 35.3,0.22);
    \draw[draw=black,ultra thick] (axis cs: 36.7,0.22) -- (axis cs: 35.3,0.21);
    
    \node[anchor=south west] at (axis cs: 36.5,0.20) {\footnotesize 100\%};

    \draw[fill=white,draw=white] (axis cs: 42.3,0.22) -- (axis cs: 43.7,0.23) -- (axis cs: 43.7,0.22) -- (axis cs: 42.3,0.21) -- cycle;
    \draw[draw=black,ultra thick] (axis cs: 43.7,0.23) -- (axis cs: 42.3,0.22);
    \draw[draw=black,ultra thick] (axis cs: 43.7,0.22) -- (axis cs: 42.3,0.21);
    
    \node[anchor=south west] at (axis cs: 43.5,0.20) {\footnotesize 100\%};

    \draw[fill=white,draw=white] (axis cs: 49.3,0.22) -- (axis cs: 50.7,0.23) -- (axis cs: 50.7,0.22) -- (axis cs: 49.3,0.21) -- cycle;
    \draw[draw=black,ultra thick] (axis cs: 50.7,0.23) -- (axis cs: 49.3,0.22);
    \draw[draw=black,ultra thick] (axis cs: 50.7,0.22) -- (axis cs: 49.3,0.21);
    
    \node[anchor=south west] at (axis cs: 50.5,0.20) {\footnotesize 100\%};

    \draw[fill=white,draw=white] (axis cs: 56.5,0.22) -- (axis cs: 57.5,0.23) -- (axis cs: 57.5,0.22) -- (axis cs: 56.5,0.21) -- cycle;
    \draw[draw=black,ultra thick] (axis cs: 57.5,0.23) -- (axis cs: 56.5,0.22);
    \draw[draw=black,ultra thick] (axis cs: 57.5,0.22) -- (axis cs: 56.5,0.21);
    
    \node[anchor=south west] at (axis cs: 53.5,0.20) {\footnotesize 100\%};

    \draw[fill=white,draw=white] (axis cs: 57.5,0.22) -- (axis cs: 58.5,0.23) -- (axis cs: 58.5,0.22) -- (axis cs: 57.5,0.21) -- cycle;
    \draw[draw=black,ultra thick] (axis cs: 58.5,0.23) -- (axis cs: 57.5,0.22);
    \draw[draw=black,ultra thick] (axis cs: 58.5,0.22) -- (axis cs: 57.5,0.21);
    
    \node[anchor=south west] at (axis cs: 58.3,0.20) {\footnotesize 31\%};

    \draw[fill=white,draw=white] (axis cs: 63.5,0.22) -- (axis cs: 64.5,0.23) -- (axis cs: 64.5,0.22) -- (axis cs: 63.5,0.21) -- cycle;
    \draw[draw=black,ultra thick] (axis cs: 64.5,0.23) -- (axis cs: 63.5,0.22);
    \draw[draw=black,ultra thick] (axis cs: 64.5,0.22) -- (axis cs: 63.5,0.21);
    
    \node[anchor=south west] at (axis cs: 60.5,0.20) {\footnotesize 100\%};

    \draw[fill=white,draw=white] (axis cs: 64.5,0.22) -- (axis cs: 65.5,0.23) -- (axis cs: 65.5,0.22) -- (axis cs: 64.5,0.21) -- cycle;
    \draw[draw=black,ultra thick] (axis cs: 65.5,0.23) -- (axis cs: 64.5,0.22);
    \draw[draw=black,ultra thick] (axis cs: 65.5,0.22) -- (axis cs: 64.5,0.21);
    
    \node[anchor=south west] at (axis cs: 65.3,0.20) {\footnotesize 29\%};

    \draw[fill=white,draw=white] (axis cs: 70.5,0.22) -- (axis cs: 71.5,0.23) -- (axis cs: 71.5,0.22) -- (axis cs: 70.5,0.21) -- cycle;
    \draw[draw=black,ultra thick] (axis cs: 71.5,0.23) -- (axis cs: 70.5,0.22);
    \draw[draw=black,ultra thick] (axis cs: 71.5,0.22) -- (axis cs: 70.5,0.21);
    
    \node[anchor=south west] at (axis cs: 67.5,0.20) {\footnotesize 100\%};

    \draw[fill=white,draw=white] (axis cs: 71.5,0.22) -- (axis cs: 72.5,0.23) -- (axis cs: 72.5,0.22) -- (axis cs: 71.5,0.21) -- cycle;
    \draw[draw=black,ultra thick] (axis cs: 72.5,0.23) -- (axis cs: 71.5,0.22);
    \draw[draw=black,ultra thick] (axis cs: 72.5,0.22) -- (axis cs: 71.5,0.21);
    
    \node[anchor=south west] at (axis cs: 72.3,0.20) {\footnotesize 24\%};

     \draw[fill=white,draw=white] (axis cs: 77.5,0.22) -- (axis cs: 78.5,0.23) -- (axis cs: 78.5,0.22) -- (axis cs: 77.5,0.21) -- cycle;
    \draw[draw=black,ultra thick] (axis cs: 78.5,0.23) -- (axis cs: 77.5,0.22);
    \draw[draw=black,ultra thick] (axis cs: 78.5,0.22) -- (axis cs: 77.5,0.21);
    
    \node[anchor=south west] at (axis cs: 74.5,0.20) {\footnotesize 100\%};

    \draw[fill=white,draw=white] (axis cs: 78.5,0.22) -- (axis cs: 79.5,0.23) -- (axis cs: 79.5,0.22) -- (axis cs: 78.5,0.21) -- cycle;
    \draw[draw=black,ultra thick] (axis cs: 79.5,0.23) -- (axis cs: 78.5,0.22);
    \draw[draw=black,ultra thick] (axis cs: 79.5,0.22) -- (axis cs: 78.5,0.21);
    
    \node[anchor=south west] at (axis cs: 79.3,0.20) {\footnotesize 25\%};
    
    \addplot
    [NoOptColor,fill=NoOptColor,x tick label style={xshift=-0.3cm}] table[x=Method,y=rerun] {sections/plots/exp_faster_restore_table.txt};
    \addlegendentry[]{\checkpoint};
    
    \addplot [LRUColor,fill=LRUColor,x tick label style={xshift=-0.3cm}] table[x=Method,y=criu] {sections/plots/exp_faster_restore_table.txt};
    \addlegendentry[]{\criu};
    
    \addplot [RandomColor,fill=RandomColor,x tick label style={xshift=-0.3cm}] table[x=Method,y=store] {sections/plots/exp_faster_restore_table.txt};
    \addlegendentry[]{\storedill};
    
    \addplot
    [GreedyColor,fill=GreedyColor,x tick label style={xshift=-0.3cm}] table[x=Method,y=dump] {sections/plots/exp_faster_restore_table.txt};
    \addlegendentry[]{\dumpsession};
    
    \addplot [HeuristicColor,fill=HeuristicColor,x tick label style={xshift=-0.3cm}] table[x=Method,y=ours] {sections/plots/exp_faster_restore_table.txt};
    \addlegendentry[]{\system \textbf{(Ours)}};

\end{axis}
    
\end{tikzpicture}
\end{subfigure}

\vspace{-4mm}
\caption{\systembf's session restoration time vs.~existing tools. Times normalized w.r.t.~\checkpoint. \systembf speeds up session restore by 94\%-99\%, and is up to 3.92$\times$ faster compared to the next best alternative.
}
\vspace{-2mm}
\label{fig:exp_restore}
\end{figure*}
\begin{table*}[t]

\caption{Runtime and memory overhead of \systembf's workflow monitoring on selected notebooks.}
\vspace{-3mm}
\footnotesize 
\addtolength{\tabcolsep}{-2.5pt} 
\midsepremove
\begin{tabular}{l|p{0.95cm}p{0.95cm}p{0.95cm}p{0.95cm}p{0.95cm}p{0.95cm}p{0.95cm}p{0.95cm}p{0.95cm}p{0.95cm}p{0.95cm}p{0.95cm}}
\toprule
&Sklearn
&NLP
&StoreSales
&TPS-Mar
&Glove
&Trading
&Timeser.
&Stacking
&Agricult.
&LANL
&HW-LM
&HW-ex3\\
\midrule
Notebook runtime (s) &   58.48  & 1016.77 & 283.06       & 178.42 & 696.64 & 687.54 & 204.10 & 788.54 & 269.40 & 1437.87 & 22.54 & 27.29 \\
Total cell monitoring time (s) & 1.26     & 4.30  & 0.81  &  1.34   & 6.43 & 0.46 & 0.60 & 2.13 & 3.08 &   0.19  & 0.50 & 0.09        \\
\textbf{Runtime overhead (\%)} & \textbf{2.14}     & \textbf{0.42} & \textbf{0.28}       & \textbf{0.78} & \textbf{0.92} & \textbf{0.07} & \textbf{0.29} &  \textbf{0.27} &   \textbf{1.14} & \textbf{0.01} & \textbf{2.21}& \textbf{0.32}          \\
\midrule
User Namespace memory usage (MB) & 1021.45    & 325.82 &  6732.17      & 1558.52 & 347.16  & 1363.32  & 130.27 & 20211.51  & 5026.48  &  7641.19  &  31.28 &  19.06     \\
\system memory usage (MB) & 19.16    & 4.73 & 0.14       & 1.69 & 33.25 & 4.09 & 0.28 &  0.33 &  0.06 & 0.14  & 0.99 & 0.47         \\
\textbf{Memory overhead (\%)} & \textbf{1.88}     & \textbf{1.45} &  \textbf{0.002}     &  \textbf{0.11} & \textbf{9.58} & \textbf{0.30} &  \textbf{0.21} &\textbf{0.002} & \textbf{0.001} &  \textbf{0.001} &  \textbf{3.16} &  \textbf{2.45}     \\
\bottomrule
\end{tabular}
\midsepdefault
\addtolength{\tabcolsep}{2.5pt}
\vspace{-2mm}
\label{tbl:overhead}
\end{table*}

\paragraph{Reproducibility} Our implementation of \system, experiment notebooks, and scripts can be found in our Github repository.\footnote{\url{https://github.com/illinoisdata/ElasticNotebook}}

\subsection{Robust Session Replication}
\label{sec:exp_robust}
This section compares the robustness of \system's session replication to existing methods. 
We count the number of isomorphic (thus, \textit{correct}) replications (\Cref{sec:problem_statement}) achieved with each method on the 60 notebooks and report the results in \cref{fig:exp_robust}.

\textbf{\systembf} correctly replicates all sessions, on par with full rerun from checkpoint file (which almost always works). Notably, it replicates 19, 25, and 2 notebooks containing unserializable variables, variable aliases, and undeserializable variables (\Cref{sec:fault_tolerance}), respectively. \dumpsessionbf and \storedillbf fail on 19/60 notebooks containing unserializable variables, many of which are used to enhance data science workflow efficiency (examples in \Cref{tbl:unserializable}); \system successfully replicates them as it can bypass the serialization of these variables through recomputation. \storedillbf additionally fails on 21/60 notebooks (total 40/60) without unserializable variables but contain variable aliases (i.e., Timeseries~\cite{agriculture} notebook, Cell 15, linked components of a Matplotlib~\cite{matplotlib} plot---\texttt{f,fig,ax}) \textcolor{red}; it serializes variables into individual files, which breaks object references and isomorphism. \system's linked variables constraint (\Cref{sec:opt_problem}) ensures that it does not do so. \helixbf fails to correctly replicate 5/60 notebooks containing variable aliases due to its lacking of the linked variable constraint. \noidgraphbf fails to correctly replicate 11/60 sessions due to it missing indirect accesses and structural modifications causing incorrect construction of the AHG, which in turn leads it to recompute some variables value-incorrectly.
\criubf fails on one notebook~\cite{amex} which contains an invisible file; however, unlike \system's failures, this failure is currently a fundamental limitation in CRIU~\cite{invisiblefile}.

\paragraph{Robust Migration across System Architectures}
We additionally performed session replication from our D32as VM (x64 architecture) to a D32pds V5 VM instance (arm64 architecture). The CRIU images cannot be replicated across machines with different architectures.
In contrast, \system does not have such a limitation.

\subsection{Faster Session Migration}
\label{sec:exp_migrate}

This section compares the efficiency of \system's session migration to existing methods. We choose 10 notebooks with no unserializable variables 
    (otherwise, existing methods fail) 
to compare the end-to-end session migration time 
achieved by different methods. 
We report upscaling and downscaling results in \cref{fig:exp_migrate} and \cref{fig:exp_migrate_2}, respectively.

The design goal of \system is to reduce session replication time through balancing variable storage and recomputation, which is successfully reflected as follows.
\system is able to reduce session migration time to the upscaled/downscaled VMs by 85\%--98\%/84\%-99\% compared \checkpoint.
Compared to \dumpsession, \storedill, and \criu, which 
store all variables in the checkpoint file, 
\system upscales/downscales up to 2.07$\times$/2.00$\times$ faster than the best of the three.
\dumpsession, while being the next best alternative for upscaling/downscaling on 8/9 notebooks, falls short in robustness as demonstrated in \Cref{sec:exp_robust}.
\storedill's individual reading and writing of each variable results in high overhead from multiple calls to the NFS for each migration.
\criu is the slowest non-rerun method for upscaling/downscaling on 6/7 notebooks, due to the size of its memory dump (higher I/O during migration) being up to 10$\times$ larger compared to checkpoint files from native tools (\Cref{sec:exp_storage}).
\subsection{Faster Session Restoration}
\label{sec:exp_restore}

In this section, we compare the efficiency of \system's session restoration to existing methods. We generate checkpoint files using each method, then compare the time taken to restore the session from the checkpoint files on the 10 notebooks from \Cref{sec:exp_migrate}.
For \system, we set the coefficient $\alpha$ to 0.05 (\Cref{sec:cost_model}) to emphasize session restoration time heavily.


We report the results in \cref{fig:exp_migrate}.
\system's restoration time is 94\%--99\% faster compared to full rerun.
Compared to the baselines, 
\system is 3.92$\times$ faster than 
the next best alternative. 
These fast restoration can be attributed to \system capable of adapting to the new optimization objective, unlike the baselines: for example, on the Sklearn~\cite{sklearntweet} notebook, instead of rerunning cell 3 (\texttt{df = pd.read\_csv(...)}) to re-read the dataframe \texttt{df} into the session as in the migration-centric plan, the restoration-centric plan opts to store \texttt{df} instead.
The reasoning is that despite the sum of serialization and deserialization times of \texttt{df} being greater than the re-reading time with \texttt{pd.read\_csv} (6.19s + 1.17s > 5.5s), the deserialization time by itself is less than the re-reading time (1.17s < 5.5s); hence, storing \texttt{df} is the optimal choice.

\subsection{Low Runtime Overhead}
\label{sec:exp_overhead}
This section investigates the overhead of \system's notebook workflow monitoring.
We measure \system's total time spent in pre/post-processing steps before/after each cell execution for updating the AHG and cell runtimes (\textit{Total cell monitoring time}), and total storage space taken to store the AHG, ID Graphs, and hashes at checkpoint time (\textit{\system memory usage}).

We report the results in \cref{tbl:overhead}. \system's cell monitoring incurs a maximum and median runtime overhead of (only) 2.21\% and 0.6\%; thus, \system can be seamlessly integrated into existing workflow. 
\system is similarly memory-efficient as its stored items (AHG, ID Graphs, and hashes) are all metadata largely independent of the size of items in the session: the median memory overhead is 0.25\%, with the worst case being 9.58\%. 

\paragraph{Fine-grained Analysis} To study the per-cell time and memory overheads during experimental notebook usage, we examined three notebooks from \homework category to confirm the maximum time and memory overheads were 92ms and 4.9MB, respectively. 
We report details in \Cref{sec:appendix_percell}. 

\subsection{Lower Storage Overhead}
\label{sec:exp_storage}
This section measures the storage cost of \system's checkpoint files: we compare the migration-centric checkpoint file sizes from \system and those from other baseline methods.
\begin{figure}[t]
\usetikzlibrary{patterns}
\centering
\begin{subfigure}[b]{\linewidth}
\begin{tikzpicture}

\pgfplotstableread[col sep=comma,]{
name
NLP\cite{nlpglove}
TPS\cite{tpsmar}
Trading\cite{asset}
Timeseries\cite{timeseries}
Agriculture\cite{sklearntweet}
HW-LM~\cite{hwlm}
}\datatable

\begin{axis}[
    ybar stacked,
    clip=false,
    xtick={2, 9, 15, 22, 29.5, 37},
    xticklabels from table={\datatable}{name},
                 x tick label style={anchor=center,xshift=2ex, yshift = -1ex},
    xlabel style={yshift = 2.5ex},
    ylabel style={yshift = -3ex},
    width=90mm,
    height=28mm,
    bar width=1.4mm,
    ymin=0,
    ymax=2,
    axis y line*=none,
    axis x line*=none,
    ytick={0.5, 1, 1.5, 2},
    yticklabels={50\%, 100\%, 150\%, 200\%},
    xmin=-1,
    xmax = 41,
    ymajorgrids,
    tick label style={font=\scriptsize},
    legend style={
        font=\footnotesize,
        /tikz/every even column/.append style={column sep=0.08cm},
        legend columns = 3,
        at={(-0.14,1.25)},
        anchor=south west,
    },
    label style={font=\footnotesize},
    ylabel={Size \% of DS},
    area legend
    ]
                 

    \draw[fill=white,draw=white] (axis cs: 1.3,2.2) -- (axis cs: 2.7,2.3) -- (axis cs: 2.7,2.2) -- (axis cs: 1.3,2.1) -- cycle;
    \draw[draw=black,ultra thick] (axis cs: 2.7,2.3) -- (axis cs: 1.3,2.2);
    \draw[draw=black,ultra thick] (axis cs: 2.7,2.2) -- (axis cs: 1.3,2.1);
    
    \node[anchor=south west] at (axis cs: 2.5,2.0) {\scriptsize 1048\%};
    \node[anchor=south west] at (axis cs: -1.5,0.0) {\scriptsize 0.5\%};

    \draw[fill=white,draw=white] (axis cs: 8.3,2.2) -- (axis cs: 9.7,2.3) -- (axis cs: 9.7,2.2) -- (axis cs: 8.3,2.1) -- cycle;
    \draw[draw=black,ultra thick] (axis cs: 9.7,2.3) -- (axis cs: 8.3,2.2);
    \draw[draw=black,ultra thick] (axis cs: 9.7,2.2) -- (axis cs: 8.3,2.1);
    
    \node[anchor=south west] at (axis cs: 9.5,2.0) {\scriptsize 283\%};
        \node[anchor=south west] at (axis cs: 5.5,0.0) {\scriptsize 0.4\%};

    \draw[fill=white,draw=white] (axis cs: 15.3,2.2) -- (axis cs: 16.7,2.3) -- (axis cs: 16.7,2.2) -- (axis cs: 15.3,2.1) -- cycle;
    \draw[draw=black,ultra thick] (axis cs: 16.7,2.3) -- (axis cs: 15.3,2.2);
    \draw[draw=black,ultra thick] (axis cs: 16.7,2.2) -- (axis cs: 15.3,2.1);
    
    \node[anchor=south west] at (axis cs: 16.5,2.0) {\scriptsize 388\%};
        \node[anchor=south west] at (axis cs: 12.5,0.0) {\scriptsize 5.6\%};

    \draw[fill=white,draw=white] (axis cs: 22.3,2.2) -- (axis cs: 23.7,2.3) -- (axis cs: 23.7,2.2) -- (axis cs: 22.3,2.1) -- cycle;
    \draw[draw=black,ultra thick] (axis cs: 23.7,2.3) -- (axis cs: 22.3,2.2);
    \draw[draw=black,ultra thick] (axis cs: 23.7,2.2) -- (axis cs: 22.3,2.1);
    
    \node[anchor=south west] at (axis cs: 23.5,2.0) {\scriptsize 467\%};
        \node[anchor=south west] at (axis cs: 19.5,0.0) {\scriptsize 2.6\%};

        \node[anchor=south west] at (axis cs: 26.5,0.0) {\scriptsize 0.1\%};

    \draw[fill=white,draw=white] (axis cs: 36.3,2.2) -- (axis cs: 37.7,2.3) -- (axis cs: 37.7,2.2) -- (axis cs: 36.3,2.1) -- cycle;
    \draw[draw=black,ultra thick] (axis cs: 37.7,2.3) -- (axis cs: 36.3,2.2);
    \draw[draw=black,ultra thick] (axis cs: 37.7,2.2) -- (axis cs: 36.3,2.1);
    
    \node[anchor=south west] at (axis cs: 37.5,2.0) {\scriptsize 1137\%};
    \node[anchor=south west] at (axis cs: 33.5,0.0) {\scriptsize 0.5\%};

    

    

    

    

        \addplot [NoOptColor,fill=NoOptColor,x tick label style={xshift=-0.3cm}] table[x=Method,y=checkpoint] {sections/plots/exp_smaller_space_table.txt};
    \addlegendentry[]{\checkpoint};
    
    \addplot [LRUColor,fill=LRUColor,x tick label style={xshift=-0.3cm}] table[x=Method,y=criu] {sections/plots/exp_smaller_space_table.txt};
    \addlegendentry[]{\criu};
    
    \addplot [RandomColor,fill=RandomColor,x tick label style={xshift=-0.3cm}] table[x=Method,y=store] {sections/plots/exp_smaller_space_table.txt};
    \addlegendentry[]{\storedill};
    
    \addplot
    [GreedyColor,fill=GreedyColor,x tick label style={xshift=-0.3cm}] table[x=Method,y=dump] {sections/plots/exp_smaller_space_table.txt};
    \addlegendentry[]{\dumpsession};

    \addplot [HeuristicColor,fill=HeuristicColor,x tick label style={xshift=-0.3cm}] table[x=Method,y=ours] {sections/plots/exp_smaller_space_table.txt};
    \addlegendentry[]{\system \textbf{(Ours)}};

\end{axis}
    
\end{tikzpicture}
\end{subfigure}

\vspace{-4mm}
\caption{\systembf's checkpoint file size vs. existing tools. Times normalized w.r.t. output from \dumpsession. \systembf's checkpoint file size is up to 67\% smaller compared to those from existing tools (excluding \checkpoint).
}
\vspace{-2mm}
\label{fig:exp_smaller_space}
\end{figure}
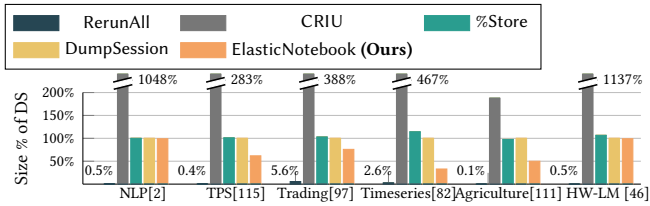

We report select results in \cref{fig:exp_smaller_space}.
\system's AHG allows it to choose between storing and recomputing each variable, reflected in \system's checkpoint files being up to 67\% smaller compared to \dumpsession's.
For example, on the Agriculture~\cite{agriculture} notebook, \system recomputes the train-test splits of the input dataframes \texttt{X} and \texttt{Y} (Cell 5, \texttt{x\_train, x\_test,... = train\_test\_split(X, Y)}) instead of storing them in the checkpoint file: this saves considerable storage space (2.5GB) in addition to speeding up migration.
Conversely, \criu's checkpoint file sizes can be 10$\times$ larger than \system's as it additionally dumps memory occupied by the Python process itself and imported modules, no matter necessary or not, into the checkpoint file.
Output sizes from \checkpoint (i.e., notebook metadata size consisting of cell code and outputs) are provided for comparison. While metadata are significantly smaller than checkpoint files, the storage benefit is offset by significantly slower session recovery times (\Cref{sec:exp_restore}).
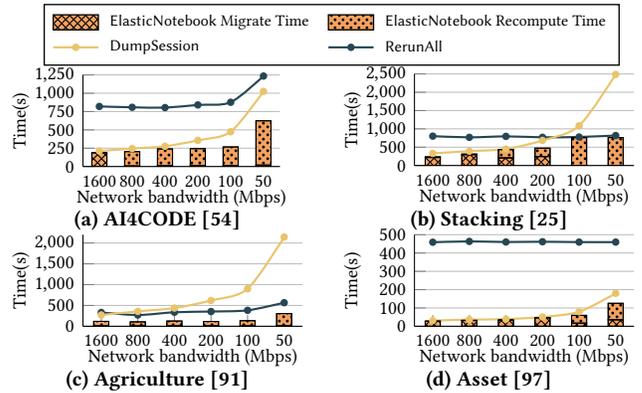
\begin{figure}[t]

\centering
\begin{subfigure}[b]{0.48\linewidth}
\begin{tikzpicture}

\begin{axis}[
    ybar stacked,
    xtick=data,
    width=42mm,
    height=28mm,
    bar width=2mm,
    ymin=0,
    ymax=1250,
    axis y line*=none,
    axis x line*=none,
    xtick={1,2,3, 4, 5, 6},
    xticklabel style   = {align=center},
    xticklabels = {1600, 800, 400, 200, 100, 50},
    ytick={0, 250, 500, 750, 1000, 1250},
    xlabel=Network bandwidth (Mbps),
    xlabel style={yshift = 2.5ex},
    ylabel style={yshift=-3ex},
    xmin = 0.5,
    xmax = 6.5,
    tick label style={font=\footnotesize},
    legend style={
        at={(-0.2,1.1)},anchor=south west,column sep=2pt,
        draw=black,fill=none,line width=.5pt,
        /tikz/every even column/.append style={column sep=5pt},
        font=\scriptsize,
    },
    legend cell align={left},
    legend columns=2,
    label style={font=\footnotesize},
    ylabel={Time(s)},
    ymajorgrids,
    area legend,
    legend image code/.code={%
    \draw[#1, draw=none] (0cm,-0.1cm) rectangle (0.6cm,0.1cm);}
]

\addplot[black,fill=HeuristicColor, postaction={
        pattern=crosshatch
    }]
table[x=x,y=y] {
x y
1 185.37
2 5.65
3 11.88
4 12.96
5 13.3
6 17.81
};

\addplot[black,fill=HeuristicColor, postaction={
        pattern=crosshatch dots
    }]
table[x=x,y=y] {
x y
1 0.01
2 197.75
3 227.87
4 231.03
5 253.70
6 606.30
};

\addplot[line legend,GreedyColor, mark = *, mark size=1pt,thick,smooth] 
    coordinates {(1, 25.11) (2, 41.26) (3, 36.79) (4, 112.9) (5, 207.52) (6, 401.31)};

\addplot[line legend,NoOptColor, mark = *, mark size=1pt,thick,smooth] 
    coordinates {(1, 609.39) (2, 563.61) (3, 527.46) (4, 483.71) (5, 402.51) (6, 209.06)};
\addlegendentry{\system Migrate Time}
\addlegendentry{\system Recompute Time}
\addlegendentry{\dumpsession}
\addlegendentry{\checkpoint}
\end{axis}
\end{tikzpicture}
\vspace{-7mm}
\caption{AI4CODE~\cite{ai4codeeda}}
\vspace{-2mm}
\label{fig:exp_bandwidth_ai4code}
\end{subfigure}
\hfill
\begin{subfigure}[b]{0.48\linewidth}
\begin{tikzpicture}

\begin{axis}[
    ybar stacked,
    xtick=data,
    width=45mm,
    height=28mm,
    bar width=2mm,
    ymin=0,
    ymax=2500,
    axis y line*=none,
    axis x line*=none,
    xticklabel style   = {align=center},
    xticklabels = {1600, 800, 400, 200, 100, 50},
    ytick={0, 500, 1000, 1500, 2000, 2500},
    xlabel=Network bandwidth (Mbps),
    xlabel style={yshift = 2.5ex},
    ylabel style={yshift=-3ex},
    xmin = 0.5,
    xmax = 6.5,
    tick label style={font=\footnotesize},
    legend style={
        at={(-0.2,1.1)},anchor=south west,column sep=2pt,
        draw=black,fill=none,line width=.5pt,
        /tikz/every even column/.append style={column sep=5pt},
        font=\scriptsize,
    },
    legend cell align={left},
    legend columns=2,
    label style={font=\footnotesize},
    ylabel={Time(s)},
    ymajorgrids,
    area legend,
    legend image code/.code={%
    \draw[#1, draw=none] (0cm,-0.1cm) rectangle (0.6cm,0.1cm);}
]

\addplot[black,fill=HeuristicColor, postaction={
        pattern=crosshatch
    }]
table[x=x,y=y] {
x y
1 218.27
2 294.27
3 202.94
4 245.21
5 0.8
6 0.93
};

\addplot[black,fill=HeuristicColor, postaction={
        pattern=crosshatch dots
    }]
table[x=x,y=y] {
x y
1 17.41
2 17.72
3 231.26
4 227.44
5 760.57
6 763.20
};

\addplot[line legend,GreedyColor, mark = *, mark size=1pt,thick,smooth] 
    coordinates {(1, 95.21) (2, 79.55) (3, 15.32) (4, 211.54) (5, 322.1) (6, 1721.44)};

\addplot[line legend,NoOptColor, mark = *, mark size=1pt,thick,smooth] 
    coordinates {(1, 470.66) (2, 377.17) (3, 345.7) (4, 85.41) (5, -304.95) (6, -1667.02)};

\end{axis}
\end{tikzpicture}
\vspace{-3mm}
\caption{Stacking~\cite{modelstacking}}
\vspace{-2mm}
\label{fig:exp_bandwidth_stacking}
\end{subfigure}
\begin{subfigure}[b]{0.48\linewidth}
\begin{tikzpicture}

\begin{axis}[
    ybar stacked,
    xtick=data,
    width=45mm,
    height=28mm,
    bar width=2mm,
    ymin=0,
    ymax=2200,
    axis y line*=none,
    axis x line*=none,
    xtick={1,2,3, 4, 5, 6},
    xticklabel style   = {align=center},
    xticklabels = {1600, 800, 400, 200, 100, 50},
    ytick={0,500, 1000, 1500, 2000},
    xlabel=Network bandwidth (Mbps),
    xlabel style={yshift = 2.5ex},
    ylabel style={yshift=-3ex},
    xmin = 0.5,
    xmax = 6.5,
    tick label style={font=\footnotesize},
    legend style={
        at={(-0.2,1.1)},anchor=south west,column sep=2pt,
        draw=black,fill=none,line width=.5pt,
        /tikz/every even column/.append style={column sep=5pt},
        font=\scriptsize,
    },
    legend cell align={left},
    legend columns=2,
    label style={font=\footnotesize},
    ylabel={Time(s)},
    ymajorgrids,
    area legend,
    legend image code/.code={%
    \draw[#1, draw=none] (0cm,-0.1cm) rectangle (0.6cm,0.1cm);}
]

\addplot[black,fill=HeuristicColor, postaction={
        pattern=crosshatch
    }]
table[x=x,y=y] {
x y
1 2.82
2 2.01
3 2.99
4 3.01
5 3.13
6 4.59
};

\addplot[black,fill=HeuristicColor, postaction={
        pattern=crosshatch dots
    }]
table[x=x,y=y] {
x y
1 114.47
2 103.82
3 122.66
4 112.53
5 131.98
6 301.07
};

\addplot[line legend,GreedyColor, mark = *, mark size=1pt,thick,smooth] 
    coordinates {(1, 153.79) (2, 252.64) (3, 311.13) (4, 503.36) (5, 765.3) (6, 1836.99)};

\addplot[line legend,NoOptColor, mark = *, mark size=1pt,thick,smooth] 
    coordinates {(1, 60.67) (2, -88.25) (3, -100.38) (4, -265.05) (5, -517.62) (6, -1576.28)};

\end{axis}
\end{tikzpicture}
\vspace{-3mm}
\caption{Agriculture~\cite{agriculture}}
\end{subfigure}
\hfill
\begin{subfigure}[b]{0.48\linewidth}
\begin{tikzpicture}

\begin{axis}[
    ybar stacked,
    xtick=data,
    width=45mm,
    height=28mm,
    bar width=2mm,
    ymin=0,
    ymax=500,
    axis y line*=none,
    axis x line*=none,
    xtick={1,2,3, 4, 5, 6},
    xticklabel style   = {align=center},
    xticklabels = {1600, 800, 400, 200, 100, 50},
    ytick={0, 100, 200, 300, 400, 500},
    xlabel=Network bandwidth (Mbps),
    xlabel style={yshift = 2.5ex},
    ylabel style={yshift=-3ex},
    xmin = 0.5,
    xmax = 6.5,
    tick label style={font=\footnotesize},
    legend style={
        at={(-0.2,1.1)},anchor=south west,column sep=2pt,
        draw=black,fill=none,line width=.5pt,
        /tikz/every even column/.append style={column sep=5pt},
        font=\scriptsize,
    },
    legend cell align={left},
    legend columns=4,
    label style={font=\footnotesize},
    ylabel={Time(s)},
    ymajorgrids,
    area legend,
    legend image code/.code={%
    \draw[#1, draw=none] (0cm,-0.1cm) rectangle (0.6cm,0.1cm);}
]

\addplot[black,fill=HeuristicColor, postaction={
        pattern=crosshatch
    }]
table[x=x,y=y] {
x y
1 26.78
2 31.48
3 32.75
4 45.94
5 16.36
6 34.92
};

\addplot[black,fill=HeuristicColor, postaction={
        pattern=crosshatch dots
    }]
table[x=x,y=y] {
x y
1 0.97
2 0.97
3 1.00
4 0.99
5 43.25
6 90.36
};

\addplot[line legend,GreedyColor, mark = *, mark size=1pt,thick,smooth] 
    coordinates {(1, 3.31) (2, 3.79) (3, 5.17) (4, 3.77) (5, 19.46) (6, 54.32)};

\addplot[line legend,NoOptColor, mark = *, mark size=1pt,thick,smooth] 
    coordinates {(1, 427.93) (2, 427.17) (3, 421.21) (4, 410.94) (5, 380.5) (6, 280)};

\end{axis}
\end{tikzpicture}
\vspace{-3mm}
\caption{Asset~\cite{asset}}
\end{subfigure}

\vspace{-4mm}
\caption{\systembf adapts to different environments for its replication plan. 
The lower the network bandwidth, 
    the more variables are recomputed.}
\label{fig:exp_bandwidth}
\vspace{-2mm}
\end{figure}
\begin{figure}[t]

\centering
\begin{subfigure}[b]{0.48\linewidth}
\begin{tikzpicture}

\begin{axis}[
    xtick=data,
    width=45mm,
    height=28mm,
    ymin=0,
    ymax=4000,
    axis y line*=none,
    axis x line*=none,
    xtick={1,2,3,4,5, 6},
    xticklabel style   = {align=center},
    xticklabels = {1600, 800, 400, 200, 100, 50},
    ytick={0, 1000, 2000, 3000, 4000},
    yticklabels={0, 1, 2, 3, 4},
    xlabel=No. cell executions,
    xlabel style={yshift = 2.5ex},
    ylabel style={yshift=-4ex},
    xmin = 0,
    xmax = 20,
    xtick = {0, 5,10,15.0,20.0},
    xticklabels = {0, 500,1000,1500,2000},
    tick label style={font=\footnotesize},
    legend style={
        at={(-0.2,1.1)},anchor=south west,column sep=2pt,
        draw=black,fill=white,
        /tikz/every even column/.append style={column sep=5pt},
        font=\scriptsize,
    },
    legend cell align={left},
    legend columns=4,
    label style={font=\footnotesize},
    ylabel={Overhead (MB)},
    ymajorgrids,
]

\addplot[GreenColor, thick, mark = *, mark size=1pt]
table[x=x,y=y] {
x y
1 173
2 324
3 476
4 627
5 779
6 931
7 1081
8 1233
9 1385
10 1536
11 1687
12 1839
13 1991
14 2141
15 2294
16 2444
17 2595
18 2748
19 2900
20 3049
};
\addplot[BlueColor, thick, mark = *, mark size=1pt]
table[x=x,y=y] {
x y
1 107
2 212
3 316
4 422
5 526
6 632
7 737
8 842
9 930
10 1035
11 1139
12 1245
13 1351
14 1455
15 1561
16 1665
17 1771
18 1858
19 1965
20 2069
};
\addplot[YellowColor, thick, mark = *, mark size=1pt]
table[x=x,y=y] {
x y
1 174
2 287
3 404
4 574
5 691
6 806
7 920
8 1095
9 1210
10 1326
11 1498
12 1611
13 1724
14 1830
15 2012
16 2129
17 2245
18 2356
19 2529
20 2644
};
\addlegendentry{Interactive~\cite{interactive}}
\addlegendentry{Twitter~\cite{twitternetworks}}
\addlegendentry{Sklearn~\cite{sklearntweet}}



\end{axis}
\end{tikzpicture}
\vspace{-6.5mm}
\caption{AHG size}
\end{subfigure}
\begin{subfigure}[b]{0.48\linewidth}
\begin{tikzpicture}

\begin{axis}[
    xtick=data,
    width=45mm,
    height=28mm,
    ymin=0,
    ymax=150,
    axis y line*=none,
    axis x line*=none,
    xtick={1,2,3,4,5, 6},
    xticklabel style   = {align=center},
    xticklabels = {1600, 800, 400, 200, 100, 50},
    ytick={0, 50, 100, 150},
    yticklabels={0, 50, 100, 150},
    xlabel=No. cell executions,
    xlabel style={yshift = 2.5ex},
    ylabel style={yshift=-4ex},
    xmin = 0,
    xmax = 20,
    xtick = {0, 5,10,15.0,20.0},
    xticklabels = {0, 500,1000,1500,2000},
    tick label style={font=\footnotesize},
    legend style={
        at={(-0.2,1.1)},anchor=south west,column sep=2pt,inner ysep = 0.5pt,
        draw=black,fill=white,
        /tikz/every even column/.append style={column sep=5pt},
        font=\scriptsize,
    },
    legend cell align={left},
    legend columns=4,
    label style={font=\footnotesize},
    ylabel={Time (ms)},
    ymajorgrids,
]

\addplot[GreenColor, thick, mark = *, mark size=1pt]
table[x=x,y=y] {
x y
1 3.7
2 5.7
3 8.1
4 9.7
5 10.8
6 14.6
7 23.6
8 16.8
9 19.0
10 24.2
11 24.4
12 26.6
13 26.5
14 30.7
15 33.4
16 32.3
17 34.2
18 38.4
19 38.5
20 43.2
};
\addplot[BlueColor, thick, mark = *, mark size=1pt]
table[x=x,y=y] {
x y
1 6.2
2 10.6
3 15.7
4 19.4
5 24.2
6 28.9
7 35.8
8 39.1
9 41.7
10 45.4
11 50.5
12 68.0
13 66.4
14 67.7
15 71.6
16 76.7
17 80.5
18 88.0
19 91.0
20 96.9
};
\addplot[YellowColor, thick, mark = *, mark size=1pt]
table[x=x,y=y] {
x y
1 9.4
2 17.9
3 25.4
4 27.9
5 38.3
6 24.4
7 50.5
8 57.2
9 66.8
10 80.0
11 71.9
12 81.9
13 91.8
14 92.5
15 105.0
16 123.5
17 125.0
18 128.2
19 122.8
20 130.7
};



\end{axis}
\end{tikzpicture}
\vspace{-6.5mm}
\caption{Optimization Time}
\end{subfigure}
\vspace{-3.5mm}

\caption{Scalability of \systembf with cell execution count. 
The size of AHG increases linearly.
    Replication plan optimization time increases sub-linearly.
}
\label{fig:exp_scalability}
\vspace{-2mm}
\end{figure}
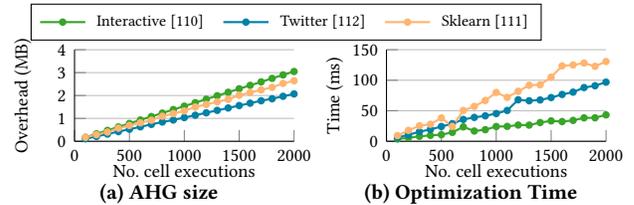
\subsection{Performance Gains Across Environments}
\label{sec:exp_bandwidth}

This section demonstrates \system's operation in environments with varying specifications. We perform a parameter sweep on the NFS network bandwidth via rate limiting~\cite{wondershaper} and compare the migration time of \system, \dumpsession (migrating all variables), and \checkpoint.

We report the results in \Cref{fig:exp_bandwidth}.
\system's balancing of variables storage and recomputation ensures that it is always at least as fast as the faster of \dumpsession and \checkpoint.
Notably, \system can adapt to the relative availability between network bandwidth and compute power: as the bandwidth decreases, the replication plan is changed accordingly to migrate more variables through recomputation rather than storage.
For example, on the Stacking~\cite{modelstacking} notebook, at regular bandwidth (>400Mbps), \system's replication plan includes migrating most of the session state, opting only to recompute certain train/test splits (i.e., Cell 37, \texttt{Y\_train}, \texttt{Y\_validation}).
At <400 Mbps, \system modifies its plan to recompute instead of store a computationally expensive processed dataframe (Cell 39, \texttt{latest\_record}).
At <100 Mbps, \system modifies its plan again to only store the imported class and function definitions (i.e., \texttt{XGBRegressor}, \texttt{mean\_squared\_error} in Cell 1) while recomputing the rest of the notebook.

\subsection{Scaling to Complex Workloads}
\label{sec:exp_scalability}

In this section, we test the scalability of \system's session replication on complex notebook sessions with a large number of cell executions and re-executions.
Specifically, we choose 3 tutorial notebooks, on which we randomly re-execute cells and measure the (1) size of \system's AHG and (2) optimization time for computing the replication plan at up to 2000 cell re-executions\footnote{This is twice the length of the longest observed notebook on Kaggle~\cite{ai4code}.}.

We report the results in \cref{fig:exp_scalability}. The memory consumption of \system's AHG exhibits linear scaling vs. the number of cell executions reaching only <4MB at 2000 cell re-executions, which is negligible compared to the memory consumption of the notebook session (>1GB) itself.
\system's optimization time for computing the replication plan similarly exhibits linear scaling, reaching a negligible <150ms at 2000 cell re-executions: \system's chosen algorithm for solving min-cut, Ford-Fulkerson~\cite{ford1962flows}, has time complexity $O(Ef)$, where $E$ is the number of edges in the AHG and $f$ is the cost of the optimal replication plan: The former scales linearly while the latter is largely constant.


\section{Related Work}
\label{sec:related}

\paragraph{Intermediate Result Reuse in Data Science}

The storage of intermediate results has been explored in various contexts in Data Science due to the incremental and feed-forward nature of tasks, which allows outputs from prior operations to be useful for speeding up future operations~\cite{vartak2018mistique, garcia2020hindsight, xin2018helix, zhang2016materialization, koop2017dataflow, xin2021enhancing, hex, li2023s}. Examples include caching to speed up model training replay for ML model diagnosis~\cite{vartak2018mistique, garcia2020hindsight}, caching to speedup materialized view refresh workloaods~\cite{li2023s}, caching to speed up anticipated future dataframe operations in notebook workflows~\cite{xin2021enhancing}, and storage of cell outputs to facilitate graphical exploration of the notebook's execution history for convenient cell re-runs~\cite{koop2017dataflow, hex}. There are related works~\cite{xin2018helix, zhang2016materialization} which algorithmically explore the most efficient way to (re)compute a state given currently stored items; compared to our work, while Helix~\cite{xin2018helix} similarly features balancing loading and recomputation, its model lacks the linked variable constraint which may result in silently incorrect replication if directly applied to the computational notebook problem setting.

\paragraph{Data-level Session Replication}

Session replication on Jupyter-based platforms can be performed with serialization libraries~\cite{pickle, dill, marshal, pythonjson, bson}.
There exists a variety of checkpoint tools built on these serialization libraries: IPython's \%Store ~\cite{jupyterstore} is a Pickle-based~\cite{pickle} interface for saving variables to a key-value store; 
however, it breaks object references as linked variables are serialized into separate files.
The Dill-based~\cite{dill} DumpSession~\cite{dumpsession} correctly resolves object references, yet it still fails if the session contains unserializable objects.
Tensorflow~\cite{tensorflowcheckpoint} and Pytorch~\cite{pytorchcheckpoint} offer periodical checkpointing during ML model training limited to objects within the same library.
Jupyter's native checkpointing mechanism~\cite{jupytercheckpoint} only saves cell metadata and often fails to exactly restore a session due to the common presence of hidden states.
Compared to existing data-level tools, session replication with \system is both more efficient and robust: the Application History Graph enables balancing state storage and recomputation, which achieves considerable speedup while avoiding failure on unserializable objects.


\paragraph{System-Level Session Replication}

Session replication can similarly be performed using system-level checkpoint/restart (C/R) tools, on which there is much existing work~\cite{chen2010shelp, bala2012fault, sidiroglou2009assure, haproxy, di2013optimization, bala2012fault, belgaum2018cloud, meshram2013fault, li2010frem}.
Applicable tools include DMTCP~\cite{ansel2009dmtcp} and CRIU~\cite{criu};
recently, CRUM~\cite{garg2018crum} and CRAC~\cite{jain2020crac} have explored extending C/R to CUDA applications.
Elsa~\cite{juric2021checkpoint} integrates CRIU with JupyterHub to enable C/R of JupyterHub servers.
Compared to \system, system-level tools are less efficient and robust due to their large memory dump sizes and limited cross-platform portability, respectively.


\paragraph{Lineage Tracing}
Lineage tracing has seen extensive use in state management to enable recomputation of data for more efficient storage of state or fault tolerance~\cite{li2014tachyon, zaharia2010spark,cores2013improving, to2018survey, vartak2018mistique, phani2021lima, guo2012burrito}
Recently, the usage of data lineage in computational notebooks has enabled multi-version notebook replay~\cite{manne2022chex}, recommending notebook interactions~\cite{macke2020fine}, and creating reproducible notebook containers~\cite{ahmad2022reproducible}, and program slicing, i.e., finding the minimal set of code to run to compute certain variable(s)~\cite{guo2012burrito, pimentel2017noworkflow, koop2017dataflow, head2019managing, shankar2022bolt}.
This work adopts lineage tracing techniques to capturing inter-variable dependencies (the Application History Graph) for optimization; to the best of our knowledge, existing works on Python programs focus on capturing value modifications (via equality comparisons); however, our techniques additionally identifies and captures \textit{strucal changes} via the ID graph, which is crucial for preserving variable aliases and avoiding silent errors during state replication.


\paragraph{Replicating Execution Environment}
An identical execution environment may be necessary for session replication on a different machine.
There is some recent work exploring environment replication for Jupyter Notebook via containerizing input files and modules~\cite{ahmad2022reproducible, wannipurage2022framework}.
While useful in conjunction with \system, we consider these works to be largely orthogonal.

\paragraph{Notebook Parameterization and Scripts}
There exists works on executing notebooks in parameterized form for systematic experimentation (e.g., in the form of a script ~\cite{nbconvert, chockchowwat2023transactional} or papermill~\cite{papermill}).
While \system is designed for use within interactive notebook interfaces, it is similarly applicable for the migration of parameterized notebook execution results.
\section{Conclusion}

In this work,
    we have proposed \system,
a new computational notebook system
    that newly offers elastic scaling and checkpointing/restoration.
To achieve this,
    \system introduces a transparent data management layer
between the user interface and the underlying kernel,
    enabling robust, efficient, and platform-independent
    state replication for notebook sessions.
Its core contributions
    include (1) low-overhead, on-the-fly application history construction
    and (2) a new optimization for
        combining copying and re-computation of
        variables that comprise session states.
We have demonstrated that \system can reduce upscaling, downscaling, and restoration times by 85\%-98\%, 84\%-99\%, and 94\%-99\%, respectively, 
on real-world data science notebooks with negligible runtime and memory overheads of <2.5\% and <10\%, respectively.

In the future,
    we plan to achieve higher efficiency and usability
        by tracing state changes at a finer level.
Specifically,
    we will introduce \emph{micro-cells}
        to capture code blocks inside a cell
    that repeatedly runs (e.g., \texttt{for}-loop for machine learning training).
Then, the system will automatically store intermediate models
        (along with other metadata)
    that will enable live migration and checkpointing/restoration for long-running cell executions.

\begin{acks}
The authors are grateful to Chandra Chekuri and Kent Quanrud for assistance with the derivation of the reduction to min-cut employed in \system. This work is supported in part by the National Center for Supercomputing Applications and Microsoft Azure.
\end{acks}

\clearpage

\bibliographystyle{ACM-Reference-Format}
\bibliography{main}


\begin{thebibliography}{124}


\ifx \showCODEN    \undefined \def \showCODEN     #1{\unskip}     \fi
\ifx \showDOI      \undefined \def \showDOI       #1{#1}\fi
\ifx \showISBNx    \undefined \def \showISBNx     #1{\unskip}     \fi
\ifx \showISBNxiii \undefined \def \showISBNxiii  #1{\unskip}     \fi
\ifx \showISSN     \undefined \def \showISSN      #1{\unskip}     \fi
\ifx \showLCCN     \undefined \def \showLCCN      #1{\unskip}     \fi
\ifx \shownote     \undefined \def \shownote      #1{#1}          \fi
\ifx \showarticletitle \undefined \def \showarticletitle #1{#1}   \fi
\ifx \showURL      \undefined \def \showURL       {\relax}        \fi
\providecommand\bibfield[2]{#2}
\providecommand\bibinfo[2]{#2}
\providecommand\natexlab[1]{#1}
\providecommand\showeprint[2][]{arXiv:#2}

\bibitem[\protect\citeauthoryear{Ahmad, Manne, and Malik}{Ahmad
  et~al\mbox{.}}{2022}]%
        {ahmad2022reproducible}
\bibfield{author}{\bibinfo{person}{Raza Ahmad}, \bibinfo{person}{Naga~Nithin
  Manne}, {and} \bibinfo{person}{Tanu Malik}.} \bibinfo{year}{2022}\natexlab{}.
\newblock \showarticletitle{Reproducible Notebook Containers using Application
  Virtualization}. In \bibinfo{booktitle}{\emph{2022 IEEE 18th International
  Conference on e-Science (e-Science)}}. IEEE, \bibinfo{pages}{1--10}.
\newblock


\bibitem[\protect\citeauthoryear{AndresHG}{AndresHG}{2021}]%
        {nlpglove}
\bibfield{author}{\bibinfo{person}{AndresHG}.} \bibinfo{year}{2021}\natexlab{}.
\newblock \bibinfo{title}{NLP, GloVe, BERT, TF-IDF, LSTM... Explained}.
\newblock
  \bibinfo{howpublished}{\url{https://www.kaggle.com/code/andreshg/nlp-glove-bert-tf-idf-lstm-explained/notebook}}.
\newblock


\bibitem[\protect\citeauthoryear{Ansel, Arya, and Cooperman}{Ansel
  et~al\mbox{.}}{2009}]%
        {ansel2009dmtcp}
\bibfield{author}{\bibinfo{person}{Jason Ansel}, \bibinfo{person}{Kapil Arya},
  {and} \bibinfo{person}{Gene Cooperman}.} \bibinfo{year}{2009}\natexlab{}.
\newblock \showarticletitle{DMTCP: Transparent checkpointing for cluster
  computations and the desktop}. In \bibinfo{booktitle}{\emph{2009 IEEE
  International Symposium on Parallel \& Distributed Processing}}. IEEE,
  \bibinfo{pages}{1--12}.
\newblock


\bibitem[\protect\citeauthoryear{Azure}{Azure}{2023a}]%
        {azuremlstudio}
\bibfield{author}{\bibinfo{person}{Microsoft Azure}.}
  \bibinfo{year}{2023}\natexlab{a}.
\newblock \bibinfo{title}{Azure ML Studio}.
\newblock
  \bibinfo{howpublished}{\url{https://learn.microsoft.com/en-us/azure/machine-learning/how-to-run-jupyter-notebooks}}.
\newblock


\bibitem[\protect\citeauthoryear{Azure}{Azure}{2023b}]%
        {azurepayasyougo}
\bibfield{author}{\bibinfo{person}{Microsoft Azure}.}
  \bibinfo{year}{2023}\natexlab{b}.
\newblock \bibinfo{title}{Microsoft Azure pay-as-you-go}.
\newblock
  \bibinfo{howpublished}{\url{https://azure.microsoft.com/en-us/pricing/purchase-options/pay-as-you-go/}}.
\newblock


\bibitem[\protect\citeauthoryear{Bala and Chana}{Bala and Chana}{2012}]%
        {bala2012fault}
\bibfield{author}{\bibinfo{person}{Anju Bala} {and} \bibinfo{person}{Inderveer
  Chana}.} \bibinfo{year}{2012}\natexlab{}.
\newblock \showarticletitle{Fault tolerance-challenges, techniques and
  implementation in cloud computing}.
\newblock \bibinfo{journal}{\emph{International Journal of Computer Science
  Issues (IJCSI)}} \bibinfo{volume}{9}, \bibinfo{number}{1}
  (\bibinfo{year}{2012}), \bibinfo{pages}{288}.
\newblock


\bibitem[\protect\citeauthoryear{Bayar}{Bayar}{2022}]%
        {storesales}
\bibfield{author}{\bibinfo{person}{Ekrem Bayar}.}
  \bibinfo{year}{2022}\natexlab{}.
\newblock \bibinfo{title}{Store Sales TS Forecasting - A Comprehensive Guide}.
\newblock
  \bibinfo{howpublished}{\url{https://www.kaggle.com/code/ekrembayar/store-sales-ts-forecasting-a-comprehensive-guide/notebook}}.
\newblock


\bibitem[\protect\citeauthoryear{Belgaum, Soomro, Alansari, and Alam}{Belgaum
  et~al\mbox{.}}{2018}]%
        {belgaum2018cloud}
\bibfield{author}{\bibinfo{person}{Mohammad~Riyaz Belgaum},
  \bibinfo{person}{Safeeullah Soomro}, \bibinfo{person}{Zainab Alansari}, {and}
  \bibinfo{person}{Muhammad Alam}.} \bibinfo{year}{2018}\natexlab{}.
\newblock \showarticletitle{Cloud service ranking using checkpoint-based load
  balancing in real-time scheduling of cloud computing}.
\newblock In \bibinfo{booktitle}{\emph{Progress in advanced computing and
  intelligent engineering}}. \bibinfo{publisher}{Springer},
  \bibinfo{pages}{667--676}.
\newblock


\bibitem[\protect\citeauthoryear{Bergstra and Bengio}{Bergstra and
  Bengio}{2012}]%
        {bergstra2012random}
\bibfield{author}{\bibinfo{person}{James Bergstra} {and}
  \bibinfo{person}{Yoshua Bengio}.} \bibinfo{year}{2012}\natexlab{}.
\newblock \showarticletitle{Random search for hyper-parameter optimization.}
\newblock \bibinfo{journal}{\emph{Journal of machine learning research}}
  \bibinfo{volume}{13}, \bibinfo{number}{2} (\bibinfo{year}{2012}).
\newblock


\bibitem[\protect\citeauthoryear{Bert~Hubert}{Bert~Hubert}{2020}]%
        {wondershaper}
\bibfield{author}{\bibinfo{person}{Simon~Séhier Bert~Hubert, Jacco~Geul}.}
  \bibinfo{year}{2020}\natexlab{}.
\newblock \bibinfo{title}{WonderShaper}.
\newblock
  \bibinfo{howpublished}{\url{https://github.com/magnific0/wondershaper}}.
\newblock


\bibitem[\protect\citeauthoryear{Brachmann and Spoth}{Brachmann and
  Spoth}{2020}]%
        {brachmann2020your}
\bibfield{author}{\bibinfo{person}{Michael Brachmann} {and}
  \bibinfo{person}{William Spoth}.} \bibinfo{year}{2020}\natexlab{}.
\newblock \showarticletitle{Your notebook is not crumby enough, REPLace it}. In
  \bibinfo{booktitle}{\emph{Conference on Innovative Data Systems Research
  (CIDR)}}.
\newblock


\bibitem[\protect\citeauthoryear{Chen, Jin, Zou, Zhou, Qiang, and Hu}{Chen
  et~al\mbox{.}}{2010}]%
        {chen2010shelp}
\bibfield{author}{\bibinfo{person}{Gang Chen}, \bibinfo{person}{Hai Jin},
  \bibinfo{person}{Deqing Zou}, \bibinfo{person}{Bing~Bing Zhou},
  \bibinfo{person}{Weizhong Qiang}, {and} \bibinfo{person}{Gang Hu}.}
  \bibinfo{year}{2010}\natexlab{}.
\newblock \showarticletitle{Shelp: Automatic self-healing for multiple
  application instances in a virtual machine environment}. In
  \bibinfo{booktitle}{\emph{2010 IEEE International Conference on Cluster
  Computing}}. IEEE, \bibinfo{pages}{97--106}.
\newblock


\bibitem[\protect\citeauthoryear{Chockchowwat, Li, and Park}{Chockchowwat
  et~al\mbox{.}}{2023}]%
        {chockchowwat2023transactional}
\bibfield{author}{\bibinfo{person}{Supawit Chockchowwat},
  \bibinfo{person}{Zhaoheng Li}, {and} \bibinfo{person}{Yongjoo Park}.}
  \bibinfo{year}{2023}\natexlab{}.
\newblock \showarticletitle{Transactional Python for Durable Machine Learning:
  Vision, Challenges, and Feasibility}. In
  \bibinfo{booktitle}{\emph{Proceedings of the Seventh Workshop on Data
  Management for End-to-End Machine Learning}}. \bibinfo{pages}{1--5}.
\newblock


\bibitem[\protect\citeauthoryear{Choudhary}{Choudhary}{2023a}]%
        {mlnotebooks}
\bibfield{author}{\bibinfo{person}{Chhaya Choudhary}.}
  \bibinfo{year}{2023}\natexlab{a}.
\newblock \bibinfo{title}{Machine Learning and Deep learning Notebooks}.
\newblock
  \bibinfo{howpublished}{\url{https://github.com/chhayac/Machine-Learning-Notebooks}}.
\newblock


\bibitem[\protect\citeauthoryear{Choudhary}{Choudhary}{2023b}]%
        {customerchurn}
\bibfield{author}{\bibinfo{person}{Chhaya Choudhary}.}
  \bibinfo{year}{2023}\natexlab{b}.
\newblock \bibinfo{title}{This project is about customer churn prediction.}
\newblock
  \bibinfo{howpublished}{\url{https://github.com/chhayac/Machine-Learning-Notebooks/blob/master/customer_churn_prediction.ipynb}}.
\newblock


\bibitem[\protect\citeauthoryear{Contributors}{Contributors}{2023}]%
        {bokeh}
\bibfield{author}{\bibinfo{person}{Bokeh Contributors}.}
  \bibinfo{year}{2023}\natexlab{}.
\newblock \bibinfo{title}{Bokeh - Interaction}.
\newblock
  \bibinfo{howpublished}{\url{https://docs.bokeh.org/en/latest/docs/user_guide/interaction.html}}.
\newblock


\bibitem[\protect\citeauthoryear{Cores, Rodr{\'\i}guez, Mart{\'\i}n,
  Gonz{\'a}lez, and Osorio}{Cores et~al\mbox{.}}{2013}]%
        {cores2013improving}
\bibfield{author}{\bibinfo{person}{Iv{\'a}n Cores}, \bibinfo{person}{Gabriel
  Rodr{\'\i}guez}, \bibinfo{person}{Mar{\'a}~J Mart{\'\i}n},
  \bibinfo{person}{Patricia Gonz{\'a}lez}, {and} \bibinfo{person}{Roberto~R
  Osorio}.} \bibinfo{year}{2013}\natexlab{}.
\newblock \showarticletitle{Improving scalability of application-level
  checkpoint-recovery by reducing checkpoint sizes}.
\newblock \bibinfo{journal}{\emph{New Generation Computing}}
  \bibinfo{volume}{31} (\bibinfo{year}{2013}), \bibinfo{pages}{163--185}.
\newblock


\bibitem[\protect\citeauthoryear{CRIU}{CRIU}{2023a}]%
        {invisiblefile}
\bibfield{author}{\bibinfo{person}{CRIU}.} \bibinfo{year}{2023}\natexlab{a}.
\newblock \bibinfo{title}{CRIU - Invisible file}.
\newblock \bibinfo{howpublished}{\url{https://criu.org/Invisible_files}}.
\newblock


\bibitem[\protect\citeauthoryear{CRIU}{CRIU}{2023b}]%
        {criu}
\bibfield{author}{\bibinfo{person}{CRIU}.} \bibinfo{year}{2023}\natexlab{b}.
\newblock \bibinfo{title}{Linux CRIU}.
\newblock \bibinfo{howpublished}{\url{https://criu.org/Main_Page}}.
\newblock


\bibitem[\protect\citeauthoryear{Crotty, Galakatos, Zgraggen, Binnig, and
  Kraska}{Crotty et~al\mbox{.}}{2015}]%
        {crotty2015vizdom}
\bibfield{author}{\bibinfo{person}{Andrew Crotty}, \bibinfo{person}{Alex
  Galakatos}, \bibinfo{person}{Emanuel Zgraggen}, \bibinfo{person}{Carsten
  Binnig}, {and} \bibinfo{person}{Tim Kraska}.}
  \bibinfo{year}{2015}\natexlab{}.
\newblock \showarticletitle{Vizdom: interactive analytics through pen and
  touch}.
\newblock \bibinfo{journal}{\emph{Proceedings of the VLDB Endowment}}
  \bibinfo{volume}{8}, \bibinfo{number}{12} (\bibinfo{year}{2015}),
  \bibinfo{pages}{2024--2027}.
\newblock


\bibitem[\protect\citeauthoryear{Culler}{Culler}{2023}]%
        {idleculler}
\bibfield{author}{\bibinfo{person}{JupyterHub~Idle Culler}.}
  \bibinfo{year}{2023}\natexlab{}.
\newblock \bibinfo{title}{JupyterHub Idle Culler}.
\newblock
  \bibinfo{howpublished}{\url{https://github.com/jupyterhub/jupyterhub-idle-culler}}.
\newblock


\bibitem[\protect\citeauthoryear{Cunha, Real, Souza, Silva, and Netto}{Cunha
  et~al\mbox{.}}{2021}]%
        {cunha2021context}
\bibfield{author}{\bibinfo{person}{Renato~LF Cunha}, \bibinfo{person}{Lucas
  C~Villa Real}, \bibinfo{person}{Renan Souza}, \bibinfo{person}{Bruno Silva},
  {and} \bibinfo{person}{Marco~AS Netto}.} \bibinfo{year}{2021}\natexlab{}.
\newblock \showarticletitle{Context-aware Execution Migration Tool for Data
  Science Jupyter Notebooks on Hybrid Clouds}. In
  \bibinfo{booktitle}{\emph{2021 IEEE 17th International Conference on eScience
  (eScience)}}. IEEE, \bibinfo{pages}{30--39}.
\newblock


\bibitem[\protect\citeauthoryear{Developer}{Developer}{2023}]%
        {cuda}
\bibfield{author}{\bibinfo{person}{Nvidia Developer}.}
  \bibinfo{year}{2023}\natexlab{}.
\newblock \bibinfo{title}{Nvidia - CUDA}.
\newblock
  \bibinfo{howpublished}{\url{https://developer.nvidia.com/cuda-toolkit}}.
\newblock


\bibitem[\protect\citeauthoryear{Di, Robert, Vivien, Kondo, Wang, and
  Cappello}{Di et~al\mbox{.}}{2013}]%
        {di2013optimization}
\bibfield{author}{\bibinfo{person}{Sheng Di}, \bibinfo{person}{Yves Robert},
  \bibinfo{person}{Fr{\'e}d{\'e}ric Vivien}, \bibinfo{person}{Derrick Kondo},
  \bibinfo{person}{Cho-Li Wang}, {and} \bibinfo{person}{Franck Cappello}.}
  \bibinfo{year}{2013}\natexlab{}.
\newblock \showarticletitle{Optimization of cloud task processing with
  checkpoint-restart mechanism}. In \bibinfo{booktitle}{\emph{Proceedings of
  the International Conference on High Performance Computing, Networking,
  Storage and Analysis}}. \bibinfo{pages}{1--12}.
\newblock


\bibitem[\protect\citeauthoryear{DimitreOliveira}{DimitreOliveira}{2019}]%
        {modelstacking}
\bibfield{author}{\bibinfo{person}{DimitreOliveira}.}
  \bibinfo{year}{2019}\natexlab{}.
\newblock \bibinfo{title}{Model stacking, feature engineering and EDA}.
\newblock
  \bibinfo{howpublished}{https://www.kaggle.com/code/dimitreoliveira/model-stacking-feature-engineering-and-eda/notebook}.
\newblock


\bibitem[\protect\citeauthoryear{Docker}{Docker}{[n.d.]}]%
        {docker-swarm}
\bibfield{author}{\bibinfo{person}{Docker}.} \bibinfo{year}{[n.d.]}\natexlab{}.
\newblock \bibinfo{title}{Docker documentation - Swarm mode overview}.
\newblock \bibinfo{howpublished}{\url{https://docs.docker.com/engine/swarm/}}.
\newblock


\bibitem[\protect\citeauthoryear{Dunne, Henry~Riche, Lee, Metoyer, and
  Robertson}{Dunne et~al\mbox{.}}{2012}]%
        {dunne2012graphtrail}
\bibfield{author}{\bibinfo{person}{Cody Dunne}, \bibinfo{person}{Nathalie
  Henry~Riche}, \bibinfo{person}{Bongshin Lee}, \bibinfo{person}{Ronald
  Metoyer}, {and} \bibinfo{person}{George Robertson}.}
  \bibinfo{year}{2012}\natexlab{}.
\newblock \showarticletitle{GraphTrail: Analyzing large multivariate,
  heterogeneous networks while supporting exploration history}. In
  \bibinfo{booktitle}{\emph{Proceedings of the SIGCHI conference on human
  factors in computing systems}}. \bibinfo{pages}{1663--1672}.
\newblock


\bibitem[\protect\citeauthoryear{dwd daniel}{dwd daniel}{2022}]%
        {ufw}
\bibfield{author}{\bibinfo{person}{dwd daniel}.}
  \bibinfo{year}{2022}\natexlab{}.
\newblock \bibinfo{title}{UncomplicatedFirewall}.
\newblock
  \bibinfo{howpublished}{\url{https://wiki.ubuntu.com/UncomplicatedFirewall}}.
\newblock


\bibitem[\protect\citeauthoryear{Eichmann, Zgraggen, Binnig, and
  Kraska}{Eichmann et~al\mbox{.}}{2020}]%
        {eichmann2020idebench}
\bibfield{author}{\bibinfo{person}{Philipp Eichmann}, \bibinfo{person}{Emanuel
  Zgraggen}, \bibinfo{person}{Carsten Binnig}, {and} \bibinfo{person}{Tim
  Kraska}.} \bibinfo{year}{2020}\natexlab{}.
\newblock \showarticletitle{Idebench: A benchmark for interactive data
  exploration}. In \bibinfo{booktitle}{\emph{Proceedings of the 2020 ACM SIGMOD
  International Conference on Management of Data}}.
  \bibinfo{pages}{1555--1569}.
\newblock


\bibitem[\protect\citeauthoryear{et~al.}{et~al.}{2018}]%
        {pytorchcheckpoint}
\bibfield{author}{\bibinfo{person}{Lightning~AI et al.}}
  \bibinfo{year}{2018}\natexlab{}.
\newblock \bibinfo{title}{PyTorch ModelCheckpoint}.
\newblock
  \bibinfo{howpublished}{\url{https://pytorch-lightning.readthedocs.io/en/stable/api/pytorch_lightning.callbacks.ModelCheckpoint.html}}.
\newblock


\bibitem[\protect\citeauthoryear{FORD-FULKERSON}{FORD-FULKERSON}{1962}]%
        {ford1962flows}
\bibfield{author}{\bibinfo{person}{LRDR FORD-FULKERSON}.}
  \bibinfo{year}{1962}\natexlab{}.
\newblock \bibinfo{title}{Flows in Networks}.
\newblock
\newblock


\bibitem[\protect\citeauthoryear{Foundation}{Foundation}{2023a}]%
        {ast}
\bibfield{author}{\bibinfo{person}{Python~Software Foundation}.}
  \bibinfo{year}{2023}\natexlab{a}.
\newblock \bibinfo{title}{Python - AST}.
\newblock
  \bibinfo{howpublished}{\url{https://docs.python.org/3/library/ast.html}}.
\newblock


\bibitem[\protect\citeauthoryear{Foundation}{Foundation}{2023b}]%
        {generators}
\bibfield{author}{\bibinfo{person}{Python~Software Foundation}.}
  \bibinfo{year}{2023}\natexlab{b}.
\newblock \bibinfo{title}{Python - Generators}.
\newblock
  \bibinfo{howpublished}{\url{https://wiki.python.org/moin/Generators}}.
\newblock


\bibitem[\protect\citeauthoryear{Foundation}{Foundation}{2023c}]%
        {hashlib}
\bibfield{author}{\bibinfo{person}{Python~Software Foundation}.}
  \bibinfo{year}{2023}\natexlab{c}.
\newblock \bibinfo{title}{Python Hashlib}.
\newblock
  \bibinfo{howpublished}{\url{https://docs.python.org/3/library/hashlib.html}}.
\newblock


\bibitem[\protect\citeauthoryear{Foundation}{Foundation}{2023d}]%
        {pythonjson}
\bibfield{author}{\bibinfo{person}{Python~Software Foundation}.}
  \bibinfo{year}{2023}\natexlab{d}.
\newblock \bibinfo{title}{Python JSON}.
\newblock
  \bibinfo{howpublished}{\url{https://docs.python.org/3/library/json.html}}.
\newblock


\bibitem[\protect\citeauthoryear{Foundation}{Foundation}{2023e}]%
        {marshal}
\bibfield{author}{\bibinfo{person}{Python~Software Foundation}.}
  \bibinfo{year}{2023}\natexlab{e}.
\newblock \bibinfo{title}{Python Marshal}.
\newblock
  \bibinfo{howpublished}{\url{https://docs.python.org/3/library/marshal.html}}.
\newblock


\bibitem[\protect\citeauthoryear{Foundation}{Foundation}{2023f}]%
        {mmap}
\bibfield{author}{\bibinfo{person}{Python~Software Foundation}.}
  \bibinfo{year}{2023}\natexlab{f}.
\newblock \bibinfo{title}{Python Mmap}.
\newblock
  \bibinfo{howpublished}{\url{https://docs.python.org/3/library/mmap.html}}.
\newblock


\bibitem[\protect\citeauthoryear{Foundation}{Foundation}{2023g}]%
        {pythonreduce}
\bibfield{author}{\bibinfo{person}{Python~Software Foundation}.}
  \bibinfo{year}{2023}\natexlab{g}.
\newblock \bibinfo{title}{Python Object Reduction}.
\newblock
  \bibinfo{howpublished}{\url{https://docs.python.org/3/library/pickle.html\#object.__reduce__}}.
\newblock


\bibitem[\protect\citeauthoryear{Foundation}{Foundation}{2023h}]%
        {pickle}
\bibfield{author}{\bibinfo{person}{Python~Software Foundation}.}
  \bibinfo{year}{2023}\natexlab{h}.
\newblock \bibinfo{title}{Python Pickle Documentation}.
\newblock
  \bibinfo{howpublished}{\url{https://docs.python.org/3/library/pickle.html}}.
\newblock


\bibitem[\protect\citeauthoryear{Foundation}{Foundation}{2023i}]%
        {dill}
\bibfield{author}{\bibinfo{person}{The Uncertainty~Quantification Foundation}.}
  \bibinfo{year}{2023}\natexlab{i}.
\newblock \bibinfo{title}{Dill - PyPi}.
\newblock \bibinfo{howpublished}{\url{https://pypi.org/project/dill/}}.
\newblock


\bibitem[\protect\citeauthoryear{Foundation}{Foundation}{2023j}]%
        {dumpsession}
\bibfield{author}{\bibinfo{person}{The Uncertainty~Quantification Foundation}.}
  \bibinfo{year}{2023}\natexlab{j}.
\newblock \bibinfo{title}{Dill dump session}.
\newblock
  \bibinfo{howpublished}{\url{https://dill.readthedocs.io/en/latest/dill.html}}.
\newblock


\bibitem[\protect\citeauthoryear{Gao}{Gao}{2020}]%
        {pythonwatchpoints}
\bibfield{author}{\bibinfo{person}{Tian Gao}.} \bibinfo{year}{2020}\natexlab{}.
\newblock \bibinfo{title}{Python Watchpoints}.
\newblock \bibinfo{howpublished}{\url{https://pypi.org/project/watchpoints/ }}.
\newblock


\bibitem[\protect\citeauthoryear{Garcia, Liu, Sreekanti, Yan, Dandamudi,
  Gonzalez, Hellerstein, and Sen}{Garcia et~al\mbox{.}}{2020}]%
        {garcia2020hindsight}
\bibfield{author}{\bibinfo{person}{Rolando Garcia}, \bibinfo{person}{Eric Liu},
  \bibinfo{person}{Vikram Sreekanti}, \bibinfo{person}{Bobby Yan},
  \bibinfo{person}{Anusha Dandamudi}, \bibinfo{person}{Joseph~E Gonzalez},
  \bibinfo{person}{Joseph~M Hellerstein}, {and} \bibinfo{person}{Koushik Sen}.}
  \bibinfo{year}{2020}\natexlab{}.
\newblock \showarticletitle{Hindsight logging for model training}.
\newblock \bibinfo{journal}{\emph{arXiv preprint arXiv:2006.07357}}
  (\bibinfo{year}{2020}).
\newblock


\bibitem[\protect\citeauthoryear{Garg, Mohan, Sullivan, and Cooperman}{Garg
  et~al\mbox{.}}{2018}]%
        {garg2018crum}
\bibfield{author}{\bibinfo{person}{Rohan Garg}, \bibinfo{person}{Apoorve
  Mohan}, \bibinfo{person}{Michael Sullivan}, {and} \bibinfo{person}{Gene
  Cooperman}.} \bibinfo{year}{2018}\natexlab{}.
\newblock \showarticletitle{CRUM: Checkpoint-restart support for CUDA's unified
  memory}. In \bibinfo{booktitle}{\emph{2018 IEEE International Conference on
  Cluster Computing (CLUSTER)}}. IEEE, \bibinfo{pages}{302--313}.
\newblock


\bibitem[\protect\citeauthoryear{GDB}{GDB}{2022}]%
        {gdbcheckpoints}
\bibfield{author}{\bibinfo{person}{GDB}.} \bibinfo{year}{2022}\natexlab{}.
\newblock \bibinfo{title}{GDB Watchpoints}.
\newblock
  \bibinfo{howpublished}{\url{https://sourceware.org/gdb/download/onlinedocs/gdb/Set-Watchpoints.html}}.
\newblock


\bibitem[\protect\citeauthoryear{Geron}{Geron}{2023a}]%
        {hwlm}
\bibfield{author}{\bibinfo{person}{Aurélien Geron}.}
  \bibinfo{year}{2023}\natexlab{a}.
\newblock \bibinfo{title}{Chapter 4 – Training Models}.
\newblock
  \bibinfo{howpublished}{\url{https://github.com/ageron/handson-ml3/blob/main/04_training_linear_models.ipynb}}.
\newblock


\bibitem[\protect\citeauthoryear{Geron}{Geron}{2023b}]%
        {handsonml}
\bibfield{author}{\bibinfo{person}{Aurélien Geron}.}
  \bibinfo{year}{2023}\natexlab{b}.
\newblock \bibinfo{title}{Machine Learning Notebooks, 3rd edition}.
\newblock \bibinfo{howpublished}{\url{https://github.com/ageron/handson-ml3}}.
\newblock


\bibitem[\protect\citeauthoryear{Google}{Google}{2023a}]%
        {colab}
\bibfield{author}{\bibinfo{person}{Google}.} \bibinfo{year}{2023}\natexlab{a}.
\newblock \bibinfo{title}{Google Colab}.
\newblock \bibinfo{howpublished}{\url{https://colab.research.google.com/}}.
\newblock


\bibitem[\protect\citeauthoryear{Google}{Google}{2023b}]%
        {colabpayasyougo}
\bibfield{author}{\bibinfo{person}{Google}.} \bibinfo{year}{2023}\natexlab{b}.
\newblock \bibinfo{title}{Google Colab pay-as-you-go}.
\newblock
  \bibinfo{howpublished}{\url{https://colab.research.google.com/signup}}.
\newblock


\bibitem[\protect\citeauthoryear{Google}{Google}{2023c}]%
        {tensorflowcheckpoint}
\bibfield{author}{\bibinfo{person}{Google}.} \bibinfo{year}{2023}\natexlab{c}.
\newblock \bibinfo{title}{Tensorflow Checkpoint}.
\newblock
  \bibinfo{howpublished}{\url{https://www.tensorflow.org/guide/checkpoint}}.
\newblock


\bibitem[\protect\citeauthoryear{Google and X}{Google and X}{2022}]%
        {ai4code}
\bibfield{author}{\bibinfo{person}{Google} {and} \bibinfo{person}{X}.}
  \bibinfo{year}{2022}\natexlab{}.
\newblock \bibinfo{title}{Google AI4Code – Understand Code in Python
  Notebooks}.
\newblock \bibinfo{howpublished}{https://www.kaggle.com/competitions/AI4Code}.
\newblock


\bibitem[\protect\citeauthoryear{Guo and Seltzer}{Guo and Seltzer}{2012}]%
        {guo2012burrito}
\bibfield{author}{\bibinfo{person}{Philip~J Guo} {and} \bibinfo{person}{Margo~I
  Seltzer}.} \bibinfo{year}{2012}\natexlab{}.
\newblock \showarticletitle{Burrito: Wrapping your lab notebook in
  computational infrastructure}.
\newblock  (\bibinfo{year}{2012}).
\newblock


\bibitem[\protect\citeauthoryear{HAProxy}{HAProxy}{2023}]%
        {haproxy}
\bibfield{author}{\bibinfo{person}{HAProxy}.} \bibinfo{year}{2023}\natexlab{}.
\newblock \bibinfo{title}{HAProxy}.
\newblock \bibinfo{howpublished}{\url{http://www.haproxy.org/}}.
\newblock


\bibitem[\protect\citeauthoryear{Hasija}{Hasija}{2022}]%
        {ai4codeeda}
\bibfield{author}{\bibinfo{person}{Sanskar Hasija}.}
  \bibinfo{year}{2022}\natexlab{}.
\newblock \bibinfo{title}{AI4Code Detailed EDA}.
\newblock
  \bibinfo{howpublished}{\url{https://www.kaggle.com/code/odins0n/ai4code-detailed-eda}}.
\newblock


\bibitem[\protect\citeauthoryear{Head, Hohman, Barik, Drucker, and DeLine}{Head
  et~al\mbox{.}}{2019}]%
        {head2019managing}
\bibfield{author}{\bibinfo{person}{Andrew Head}, \bibinfo{person}{Fred Hohman},
  \bibinfo{person}{Titus Barik}, \bibinfo{person}{Steven~M Drucker}, {and}
  \bibinfo{person}{Robert DeLine}.} \bibinfo{year}{2019}\natexlab{}.
\newblock \showarticletitle{Managing messes in computational notebooks}. In
  \bibinfo{booktitle}{\emph{Proceedings of the 2019 CHI Conference on Human
  Factors in Computing Systems}}. \bibinfo{pages}{1--12}.
\newblock


\bibitem[\protect\citeauthoryear{Hex~Technologies}{Hex~Technologies}{2023}]%
        {hex}
\bibfield{author}{\bibinfo{person}{Inc. Hex~Technologies}.}
  \bibinfo{year}{2023}\natexlab{}.
\newblock \bibinfo{title}{Hex 2.0: Reactivity, Graphs, and a little bit of
  Magic}.
\newblock
  \bibinfo{howpublished}{\url{https://hex.tech/blog/hex-two-point-oh/}}.
\newblock


\bibitem[\protect\citeauthoryear{IBM}{IBM}{2022}]%
        {ibmwatson}
\bibfield{author}{\bibinfo{person}{IBM}.} \bibinfo{year}{2022}\natexlab{}.
\newblock \bibinfo{title}{IBM Watson Studio Service}.
\newblock
  \bibinfo{howpublished}{\url{https://www.ibm.com/docs/en/knowledge-accelerators/1.0.0?topic=catalog-jupyter-notebook}}.
\newblock


\bibitem[\protect\citeauthoryear{Inc.}{Inc.}{2023a}]%
        {kaggle}
\bibfield{author}{\bibinfo{person}{Kaggle Inc.}}
  \bibinfo{year}{2023}\natexlab{a}.
\newblock \bibinfo{title}{Kaggle}.
\newblock \bibinfo{howpublished}{\url{https://www.kaggle.com/}}.
\newblock


\bibitem[\protect\citeauthoryear{Inc.}{Inc.}{2023b}]%
        {kaggleforums}
\bibfield{author}{\bibinfo{person}{Kaggle Inc.}}
  \bibinfo{year}{2023}\natexlab{b}.
\newblock \bibinfo{title}{Kaggle Forums - Product Feedback}.
\newblock
  \bibinfo{howpublished}{\url{https://www.kaggle.com/discussions/product-feedback}}.
\newblock


\bibitem[\protect\citeauthoryear{Inc.}{Inc.}{2023c}]%
        {kaggletimeout}
\bibfield{author}{\bibinfo{person}{Kaggle Inc.}}
  \bibinfo{year}{2023}\natexlab{c}.
\newblock \bibinfo{title}{Kaggle Notebook Specifications}.
\newblock
  \bibinfo{howpublished}{\url{https://www.kaggle.com/docs/notebooks\#technical-specifications}}.
\newblock


\bibitem[\protect\citeauthoryear{Institute}{Institute}{2023}]%
        {jwst}
\bibfield{author}{\bibinfo{person}{Space Telescope~Science Institute}.}
  \bibinfo{year}{2023}\natexlab{}.
\newblock \bibinfo{title}{JWST Data Analysis Example}.
\newblock
  \bibinfo{howpublished}{\url{https://jwst-docs.stsci.edu/jwst-post-pipeline-data-analysis/data-analysis-example-jupyter-notebooks}}.
\newblock


\bibitem[\protect\citeauthoryear{Jain and Cooperman}{Jain and
  Cooperman}{2020}]%
        {jain2020crac}
\bibfield{author}{\bibinfo{person}{Twinkle Jain} {and} \bibinfo{person}{Gene
  Cooperman}.} \bibinfo{year}{2020}\natexlab{}.
\newblock \showarticletitle{Crac: Checkpoint-restart architecture for cuda with
  streams and uvm}. In \bibinfo{booktitle}{\emph{SC20: International Conference
  for High Performance Computing, Networking, Storage and Analysis}}. IEEE,
  \bibinfo{pages}{1--15}.
\newblock


\bibitem[\protect\citeauthoryear{Johnson}{Johnson}{2020}]%
        {johnson2020benefits}
\bibfield{author}{\bibinfo{person}{Jeremiah~W Johnson}.}
  \bibinfo{year}{2020}\natexlab{}.
\newblock \showarticletitle{Benefits and pitfalls of jupyter notebooks in the
  classroom}. In \bibinfo{booktitle}{\emph{Proceedings of the 21st Annual
  Conference on Information Technology Education}}. \bibinfo{pages}{32--37}.
\newblock


\bibitem[\protect\citeauthoryear{Jupyter}{Jupyter}{2023}]%
        {jupyter}
\bibfield{author}{\bibinfo{person}{Project Jupyter}.}
  \bibinfo{year}{2023}\natexlab{}.
\newblock \bibinfo{title}{Jupyter Notebook}.
\newblock \bibinfo{howpublished}{\url{https://jupyter.org/}}.
\newblock


\bibitem[\protect\citeauthoryear{Juric, Stetzler, and Slater}{Juric
  et~al\mbox{.}}{2021}]%
        {juric2021checkpoint}
\bibfield{author}{\bibinfo{person}{Mario Juric}, \bibinfo{person}{Steven
  Stetzler}, {and} \bibinfo{person}{Colin~T Slater}.}
  \bibinfo{year}{2021}\natexlab{}.
\newblock \showarticletitle{Checkpoint, Restore, and Live Migration for Science
  Platforms}.
\newblock \bibinfo{journal}{\emph{arXiv preprint arXiv:2101.05782}}
  (\bibinfo{year}{2021}).
\newblock


\bibitem[\protect\citeauthoryear{Koop and Patel}{Koop and Patel}{2017}]%
        {koop2017dataflow}
\bibfield{author}{\bibinfo{person}{David Koop} {and} \bibinfo{person}{Jay
  Patel}.} \bibinfo{year}{2017}\natexlab{}.
\newblock \showarticletitle{Dataflow notebooks: encoding and tracking
  dependencies of cells}. In \bibinfo{booktitle}{\emph{9th USENIX Workshop on
  the Theory and Practice of Provenance (TaPP 2017)}}.
\newblock


\bibitem[\protect\citeauthoryear{Krasser}{Krasser}{2023a}]%
        {mlnotebooksstanford}
\bibfield{author}{\bibinfo{person}{Martin Krasser}.}
  \bibinfo{year}{2023}\natexlab{a}.
\newblock \bibinfo{title}{Machine learning notebooks}.
\newblock
  \bibinfo{howpublished}{\url{https://github.com/krasserm/machine-learning-notebook}}.
\newblock


\bibitem[\protect\citeauthoryear{Krasser}{Krasser}{2023b}]%
        {hwex3}
\bibfield{author}{\bibinfo{person}{Martin Krasser}.}
  \bibinfo{year}{2023}\natexlab{b}.
\newblock \bibinfo{title}{Multi-class Classification}.
\newblock
  \bibinfo{howpublished}{\url{https://github.com/krasserm/machine-learning-notebooks/blob/master/ml-ex3.ipynb}}.
\newblock


\bibitem[\protect\citeauthoryear{Kubernetes}{Kubernetes}{[n.d.]}]%
        {kubernetes}
\bibfield{author}{\bibinfo{person}{Kubernetes}.}
  \bibinfo{year}{[n.d.]}\natexlab{}.
\newblock \bibinfo{title}{Kubernetes}.
\newblock \bibinfo{howpublished}{\url{https://kubernetes.io/}}.
\newblock


\bibitem[\protect\citeauthoryear{Lab}{Lab}{2022}]%
        {dataprepeda}
\bibfield{author}{\bibinfo{person}{SFU Database~System Lab}.}
  \bibinfo{year}{2022}\natexlab{}.
\newblock \bibinfo{title}{Dataprep - Low-Code Data Preparation}.
\newblock \bibinfo{howpublished}{\url{https://dataprep.ai/}}.
\newblock


\bibitem[\protect\citeauthoryear{Lagator}{Lagator}{2020}]%
        {arxivdata}
\bibfield{author}{\bibinfo{person}{Colin Lagator}.}
  \bibinfo{year}{2020}\natexlab{}.
\newblock \bibinfo{title}{Arxiv Data Processing}.
\newblock
  \bibinfo{howpublished}{\url{https://www.kaggle.com/code/colinlagator/arxiv-data-processing}}.
\newblock


\bibitem[\protect\citeauthoryear{Li, Ghodsi, Zaharia, Shenker, and Stoica}{Li
  et~al\mbox{.}}{2014}]%
        {li2014tachyon}
\bibfield{author}{\bibinfo{person}{Haoyuan Li}, \bibinfo{person}{Ali Ghodsi},
  \bibinfo{person}{Matei Zaharia}, \bibinfo{person}{Scott Shenker}, {and}
  \bibinfo{person}{Ion Stoica}.} \bibinfo{year}{2014}\natexlab{}.
\newblock \showarticletitle{Tachyon: Reliable, memory speed storage for cluster
  computing frameworks}. In \bibinfo{booktitle}{\emph{Proceedings of the ACM
  Symposium on Cloud Computing}}. \bibinfo{pages}{1--15}.
\newblock


\bibitem[\protect\citeauthoryear{Li and Lan}{Li and Lan}{2010}]%
        {li2010frem}
\bibfield{author}{\bibinfo{person}{Yawei Li} {and} \bibinfo{person}{Zhiling
  Lan}.} \bibinfo{year}{2010}\natexlab{}.
\newblock \showarticletitle{FREM: A fast restart mechanism for general
  checkpoint/restart}.
\newblock \bibinfo{journal}{\emph{IEEE Trans. Comput.}} \bibinfo{volume}{60},
  \bibinfo{number}{5} (\bibinfo{year}{2010}), \bibinfo{pages}{639--652}.
\newblock


\bibitem[\protect\citeauthoryear{Li, Pi, and Park}{Li et~al\mbox{.}}{2023}]%
        {li2023s}
\bibfield{author}{\bibinfo{person}{Zhaoheng Li}, \bibinfo{person}{Xinyu Pi},
  {and} \bibinfo{person}{Yongjoo Park}.} \bibinfo{year}{2023}\natexlab{}.
\newblock \showarticletitle{S/C: Speeding up Data Materialization with Bounded
  Memory}.
\newblock \bibinfo{journal}{\emph{arXiv preprint arXiv:2303.09774}}
  (\bibinfo{year}{2023}).
\newblock


\bibitem[\protect\citeauthoryear{Linux}{Linux}{2023}]%
        {chroot}
\bibfield{author}{\bibinfo{person}{Arch Linux}.}
  \bibinfo{year}{2023}\natexlab{}.
\newblock \bibinfo{title}{chroot}.
\newblock
  \bibinfo{howpublished}{\url{https://wiki.archlinux.org/title/chroot}}.
\newblock


\bibitem[\protect\citeauthoryear{Liu and Heer}{Liu and Heer}{2014}]%
        {liu2014effects}
\bibfield{author}{\bibinfo{person}{Zhicheng Liu} {and} \bibinfo{person}{Jeffrey
  Heer}.} \bibinfo{year}{2014}\natexlab{}.
\newblock \showarticletitle{The effects of interactive latency on exploratory
  visual analysis}.
\newblock \bibinfo{journal}{\emph{IEEE transactions on visualization and
  computer graphics}} \bibinfo{volume}{20}, \bibinfo{number}{12}
  (\bibinfo{year}{2014}), \bibinfo{pages}{2122--2131}.
\newblock


\bibitem[\protect\citeauthoryear{Macke, Gong, Lee, Head, Xin, and
  Parameswaran}{Macke et~al\mbox{.}}{2020}]%
        {macke2020fine}
\bibfield{author}{\bibinfo{person}{Stephen Macke}, \bibinfo{person}{Hongpu
  Gong}, \bibinfo{person}{Doris Jung-Lin Lee}, \bibinfo{person}{Andrew Head},
  \bibinfo{person}{Doris Xin}, {and} \bibinfo{person}{Aditya Parameswaran}.}
  \bibinfo{year}{2020}\natexlab{}.
\newblock \showarticletitle{Fine-grained lineage for safer notebook
  interactions}.
\newblock \bibinfo{journal}{\emph{arXiv preprint arXiv:2012.06981}}
  (\bibinfo{year}{2020}).
\newblock


\bibitem[\protect\citeauthoryear{Manne, Satpati, Malik, Bagchi, Gehani, and
  Chaudhary}{Manne et~al\mbox{.}}{2022}]%
        {manne2022chex}
\bibfield{author}{\bibinfo{person}{Naga~Nithin Manne}, \bibinfo{person}{Shilvi
  Satpati}, \bibinfo{person}{Tanu Malik}, \bibinfo{person}{Amitabha Bagchi},
  \bibinfo{person}{Ashish Gehani}, {and} \bibinfo{person}{Amitabh Chaudhary}.}
  \bibinfo{year}{2022}\natexlab{}.
\newblock \showarticletitle{CHEX: Multiversion Replay with Ordered
  Checkpoints}.
\newblock \bibinfo{journal}{\emph{arXiv preprint arXiv:2202.08429}}
  (\bibinfo{year}{2022}).
\newblock


\bibitem[\protect\citeauthoryear{Meshram, Sambare, and Zade}{Meshram
  et~al\mbox{.}}{2013}]%
        {meshram2013fault}
\bibfield{author}{\bibinfo{person}{Anjali~D Meshram}, \bibinfo{person}{AS
  Sambare}, {and} \bibinfo{person}{SD Zade}.} \bibinfo{year}{2013}\natexlab{}.
\newblock \showarticletitle{Fault tolerance model for reliable cloud
  computing}.
\newblock \bibinfo{journal}{\emph{International Journal on Recent and
  Innovation Trends in Computing and Communication}} \bibinfo{volume}{1},
  \bibinfo{number}{7} (\bibinfo{year}{2013}), \bibinfo{pages}{600--603}.
\newblock


\bibitem[\protect\citeauthoryear{MongoDB}{MongoDB}{2023}]%
        {bson}
\bibfield{author}{\bibinfo{person}{Inc. MongoDB}.}
  \bibinfo{year}{2023}\natexlab{}.
\newblock \bibinfo{title}{BSON}.
\newblock
  \bibinfo{howpublished}{\url{https://pymongo.readthedocs.io/en/stable/api/bson/index.html}}.
\newblock


\bibitem[\protect\citeauthoryear{Moritz, Nishihara, Wang, Tumanov, Liaw, Liang,
  Elibol, Yang, Paul, Jordan, et~al\mbox{.}}{Moritz et~al\mbox{.}}{2018}]%
        {moritz2018ray}
\bibfield{author}{\bibinfo{person}{Philipp Moritz}, \bibinfo{person}{Robert
  Nishihara}, \bibinfo{person}{Stephanie Wang}, \bibinfo{person}{Alexey
  Tumanov}, \bibinfo{person}{Richard Liaw}, \bibinfo{person}{Eric Liang},
  \bibinfo{person}{Melih Elibol}, \bibinfo{person}{Zongheng Yang},
  \bibinfo{person}{William Paul}, \bibinfo{person}{Michael~I Jordan},
  {et~al\mbox{.}}} \bibinfo{year}{2018}\natexlab{}.
\newblock \showarticletitle{Ray: A distributed framework for emerging
  $\{$AI$\}$ applications}. In \bibinfo{booktitle}{\emph{13th $\{$USENIX$\}$
  Symposium on Operating Systems Design and Implementation ($\{$OSDI$\}$ 18)}}.
  \bibinfo{pages}{561--577}.
\newblock


\bibitem[\protect\citeauthoryear{Mulla}{Mulla}{2020}]%
        {timeseries}
\bibfield{author}{\bibinfo{person}{Rob Mulla}.}
  \bibinfo{year}{2020}\natexlab{}.
\newblock \bibinfo{title}{Time Series forecasting with Prophet}.
\newblock
  \bibinfo{howpublished}{\url{https://www.kaggle.com/code/robikscube/time-series-forecasting-with-prophet}}.
\newblock


\bibitem[\protect\citeauthoryear{Petersohn, Macke, Xin, Ma, Lee, Mo, Gonzalez,
  Hellerstein, Joseph, and Parameswaran}{Petersohn et~al\mbox{.}}{2020}]%
        {petersohn2020towards}
\bibfield{author}{\bibinfo{person}{Devin Petersohn}, \bibinfo{person}{Stephen
  Macke}, \bibinfo{person}{Doris Xin}, \bibinfo{person}{William Ma},
  \bibinfo{person}{Doris Lee}, \bibinfo{person}{Xiangxi Mo},
  \bibinfo{person}{Joseph~E Gonzalez}, \bibinfo{person}{Joseph~M Hellerstein},
  \bibinfo{person}{Anthony~D Joseph}, {and} \bibinfo{person}{Aditya
  Parameswaran}.} \bibinfo{year}{2020}\natexlab{}.
\newblock \showarticletitle{Towards scalable dataframe systems}.
\newblock \bibinfo{journal}{\emph{arXiv preprint arXiv:2001.00888}}
  (\bibinfo{year}{2020}).
\newblock


\bibitem[\protect\citeauthoryear{Phani, Rath, and Boehm}{Phani
  et~al\mbox{.}}{2021}]%
        {phani2021lima}
\bibfield{author}{\bibinfo{person}{Arnab Phani}, \bibinfo{person}{Benjamin
  Rath}, {and} \bibinfo{person}{Matthias Boehm}.}
  \bibinfo{year}{2021}\natexlab{}.
\newblock \showarticletitle{LIMA: Fine-grained Lineage Tracing and Reuse in
  Machine Learning Systems}. In \bibinfo{booktitle}{\emph{Proceedings of the
  2021 International Conference on Management of Data}}.
  \bibinfo{pages}{1426--1439}.
\newblock


\bibitem[\protect\citeauthoryear{Pimentel, Murta, Braganholo, and
  Freire}{Pimentel et~al\mbox{.}}{2017}]%
        {pimentel2017noworkflow}
\bibfield{author}{\bibinfo{person}{Joao~Felipe Pimentel},
  \bibinfo{person}{Leonardo Murta}, \bibinfo{person}{Vanessa Braganholo}, {and}
  \bibinfo{person}{Juliana Freire}.} \bibinfo{year}{2017}\natexlab{}.
\newblock \showarticletitle{noWorkflow: a tool for collecting, analyzing, and
  managing provenance from python scripts}.
\newblock \bibinfo{journal}{\emph{Proceedings of the VLDB Endowment}}
  \bibinfo{volume}{10}, \bibinfo{number}{12} (\bibinfo{year}{2017}).
\newblock


\bibitem[\protect\citeauthoryear{pip developers}{pip developers}{2023}]%
        {pipfreeze}
\bibfield{author}{\bibinfo{person}{The pip developers}.}
  \bibinfo{year}{2023}\natexlab{}.
\newblock \bibinfo{title}{Pip Freeze}.
\newblock
  \bibinfo{howpublished}{\url{https://pip.pypa.io/en/stable/cli/pip_freeze/}}.
\newblock


\bibitem[\protect\citeauthoryear{Poppe, Guo, Lang, Arora, Oslake, Xu, and
  Kalhan}{Poppe et~al\mbox{.}}{2022}]%
        {poppe2022moneyball}
\bibfield{author}{\bibinfo{person}{Olga Poppe}, \bibinfo{person}{Qun Guo},
  \bibinfo{person}{Willis Lang}, \bibinfo{person}{Pankaj Arora},
  \bibinfo{person}{Morgan Oslake}, \bibinfo{person}{Shize Xu}, {and}
  \bibinfo{person}{Ajay Kalhan}.} \bibinfo{year}{2022}\natexlab{}.
\newblock \showarticletitle{Moneyball: proactive auto-scaling in Microsoft
  Azure SQL database serverless}.
\newblock \bibinfo{journal}{\emph{Proceedings of the VLDB Endowment}}
  \bibinfo{volume}{15}, \bibinfo{number}{6} (\bibinfo{year}{2022}),
  \bibinfo{pages}{1279--1287}.
\newblock


\bibitem[\protect\citeauthoryear{Posit~Software}{Posit~Software}{2023}]%
        {rstudio}
\bibfield{author}{\bibinfo{person}{PBC Posit~Software, PBC formerly~RStudio}.}
  \bibinfo{year}{2023}\natexlab{}.
\newblock \bibinfo{title}{Posit RStudio}.
\newblock \bibinfo{howpublished}{\url{https://posit.co/}}.
\newblock


\bibitem[\protect\citeauthoryear{Preda}{Preda}{2019}]%
        {lanl}
\bibfield{author}{\bibinfo{person}{Gabriel Preda}.}
  \bibinfo{year}{2019}\natexlab{}.
\newblock \bibinfo{title}{LANL Earthquake EDA and Prediction}.
\newblock
  \bibinfo{howpublished}{\url{https://www.kaggle.com/code/gpreda/lanl-earthquake-eda-and-prediction}}.
\newblock


\bibitem[\protect\citeauthoryear{Rahman}{Rahman}{2022}]%
        {nfl}
\bibfield{author}{\bibinfo{person}{Kalilur Rahman}.}
  \bibinfo{year}{2022}\natexlab{}.
\newblock \bibinfo{title}{NFL Data Bowl 2023 - Offensive Plays EDA}.
\newblock
  \bibinfo{howpublished}{\url{https://www.kaggle.com/code/kalilurrahman/nfl-data-bowl-2023-offensive-plays-eda/notebook}}.
\newblock


\bibitem[\protect\citeauthoryear{Rahul}{Rahul}{2020}]%
        {agriculture}
\bibfield{author}{\bibinfo{person}{DS Rahul}.} \bibinfo{year}{2020}\natexlab{}.
\newblock \bibinfo{title}{Agricultural Drought Prediction}.
\newblock
  \bibinfo{howpublished}{\url{https://www.kaggle.com/code/dsrhul/agricultural-drought-prediction}}.
\newblock


\bibitem[\protect\citeauthoryear{Raj}{Raj}{2022}]%
        {amex}
\bibfield{author}{\bibinfo{person}{Mani Raj}.} \bibinfo{year}{2022}\natexlab{}.
\newblock \bibinfo{title}{Amex Dataset}.
\newblock
  \bibinfo{howpublished}{\url{https://www.kaggle.com/code/manirajheerakar/amex-dataset}}.
\newblock


\bibitem[\protect\citeauthoryear{Services}{Services}{2023}]%
        {awsjupyterhub}
\bibfield{author}{\bibinfo{person}{Amazon~Web Services}.}
  \bibinfo{year}{2023}\natexlab{}.
\newblock \bibinfo{title}{AWS JupyterHub}.
\newblock
  \bibinfo{howpublished}{\url{https://docs.aws.amazon.com/emr/latest/ReleaseGuide/emr-jupyterhub.html}}.
\newblock


\bibitem[\protect\citeauthoryear{Shahules}{Shahules}{2022}]%
        {glove}
\bibfield{author}{\bibinfo{person}{Shahules}.} \bibinfo{year}{2022}\natexlab{}.
\newblock \bibinfo{title}{Basic EDA,Cleaning and GloVe}.
\newblock
  \bibinfo{howpublished}{\url{https://www.kaggle.com/code/shahules/basic-eda-cleaning-and-glove/notebook}}.
\newblock


\bibitem[\protect\citeauthoryear{Shankar, Macke, Chasins, Head, and
  Parameswaran}{Shankar et~al\mbox{.}}{2022}]%
        {shankar2022bolt}
\bibfield{author}{\bibinfo{person}{Shreya Shankar}, \bibinfo{person}{Stephen
  Macke}, \bibinfo{person}{Sarah Chasins}, \bibinfo{person}{Andrew Head}, {and}
  \bibinfo{person}{Aditya Parameswaran}.} \bibinfo{year}{2022}\natexlab{}.
\newblock \showarticletitle{Bolt-on, compact, and rapid program slicing for
  notebooks}.
\newblock \bibinfo{journal}{\emph{Proceedings of the VLDB Endowment}}
  \bibinfo{volume}{15}, \bibinfo{number}{13} (\bibinfo{year}{2022}),
  \bibinfo{pages}{4038--4047}.
\newblock


\bibitem[\protect\citeauthoryear{shreyas thorat30}{shreyas thorat30}{2023}]%
        {plantdisease}
\bibfield{author}{\bibinfo{person}{shreyas thorat30}.}
  \bibinfo{year}{2023}\natexlab{}.
\newblock \bibinfo{title}{Plant disease classification SDP}.
\newblock
  \bibinfo{howpublished}{\url{https://www.kaggle.com/code/shreyasthorat30/plant-disease-classification-sdp}}.
\newblock


\bibitem[\protect\citeauthoryear{Shtrauss}{Shtrauss}{2022}]%
        {asset}
\bibfield{author}{\bibinfo{person}{Andrey Shtrauss}.}
  \bibinfo{year}{2022}\natexlab{}.
\newblock \bibinfo{title}{Building an Asset Trading Strategy}.
\newblock
  \bibinfo{howpublished}{\url{https://www.kaggle.com/code/shtrausslearning/building-an-asset-trading-strategy/notebook}}.
\newblock


\bibitem[\protect\citeauthoryear{Sidiroglou, Laadan, Perez, Viennot, Nieh, and
  Keromytis}{Sidiroglou et~al\mbox{.}}{2009}]%
        {sidiroglou2009assure}
\bibfield{author}{\bibinfo{person}{Stelios Sidiroglou}, \bibinfo{person}{Oren
  Laadan}, \bibinfo{person}{Carlos Perez}, \bibinfo{person}{Nicolas Viennot},
  \bibinfo{person}{Jason Nieh}, {and} \bibinfo{person}{Angelos~D Keromytis}.}
  \bibinfo{year}{2009}\natexlab{}.
\newblock \showarticletitle{Assure: automatic software self-healing using
  rescue points}.
\newblock \bibinfo{journal}{\emph{ACM SIGARCH Computer Architecture News}}
  \bibinfo{volume}{37}, \bibinfo{number}{1} (\bibinfo{year}{2009}),
  \bibinfo{pages}{37--48}.
\newblock


\bibitem[\protect\citeauthoryear{StackOverflow}{StackOverflow}{2019}]%
        {colabstackoverflow}
\bibfield{author}{\bibinfo{person}{StackOverflow}.}
  \bibinfo{year}{2019}\natexlab{}.
\newblock \bibinfo{title}{Colab Session Timeout}.
\newblock
  \bibinfo{howpublished}{\url{https://stackoverflow.com/questions/57113226/how-can-i-prevent-google-colab-from-disconnecting}}.
\newblock


\bibitem[\protect\citeauthoryear{Stitchfix}{Stitchfix}{2017}]%
        {nodebooks}
\bibfield{author}{\bibinfo{person}{Stitchfix}.}
  \bibinfo{year}{2017}\natexlab{}.
\newblock \bibinfo{title}{Nodebooks}.
\newblock \bibinfo{howpublished}{\url{https://github.com/stitchfix/nodebook}}.
\newblock


\bibitem[\protect\citeauthoryear{Team}{Team}{2023a}]%
        {nbconvert}
\bibfield{author}{\bibinfo{person}{Jupyter~Development Team}.}
  \bibinfo{year}{2023}\natexlab{a}.
\newblock \bibinfo{title}{nbconvert - Jupyter Notebook Conversion}.
\newblock \bibinfo{howpublished}{\url{https://github.com/jupyter/nbconvert}}.
\newblock


\bibitem[\protect\citeauthoryear{Team}{Team}{2023b}]%
        {papermill}
\bibfield{author}{\bibinfo{person}{Nteract Team}.}
  \bibinfo{year}{2023}\natexlab{b}.
\newblock \bibinfo{title}{Welcome to papermill}.
\newblock
  \bibinfo{howpublished}{\url{https://papermill.readthedocs.io/en/latest/}}.
\newblock


\bibitem[\protect\citeauthoryear{Team}{Team}{2023c}]%
        {ipython}
\bibfield{author}{\bibinfo{person}{The IPython~Development Team}.}
  \bibinfo{year}{2023}\natexlab{c}.
\newblock \bibinfo{title}{IPython Interactive Computing}.
\newblock \bibinfo{howpublished}{\url{https://ipython.org/}}.
\newblock


\bibitem[\protect\citeauthoryear{Team}{Team}{2023d}]%
        {jupytercheckpoint}
\bibfield{author}{\bibinfo{person}{The IPython~Development Team}.}
  \bibinfo{year}{2023}\natexlab{d}.
\newblock \bibinfo{title}{Jupyter checkpoint}.
\newblock
  \bibinfo{howpublished}{\url{https://jupyter-server.readthedocs.io/en/latest/developers/contents.html}}.
\newblock


\bibitem[\protect\citeauthoryear{Team}{Team}{2023e}]%
        {jupytermagic}
\bibfield{author}{\bibinfo{person}{The IPython~Development Team}.}
  \bibinfo{year}{2023}\natexlab{e}.
\newblock \bibinfo{title}{Jupyter Magics Class}.
\newblock
  \bibinfo{howpublished}{\url{https://ipython.readthedocs.io/en/stable/config/custommagics.html}}.
\newblock


\bibitem[\protect\citeauthoryear{Team}{Team}{2023f}]%
        {jupyterstore}
\bibfield{author}{\bibinfo{person}{The IPython~Development Team}.}
  \bibinfo{year}{2023}\natexlab{f}.
\newblock \bibinfo{title}{Jupyter store magic}.
\newblock
  \bibinfo{howpublished}{\url{https://ipython.readthedocs.io/en/stable/config/extensions/storemagic.html}}.
\newblock


\bibitem[\protect\citeauthoryear{Team}{Team}{2023g}]%
        {matplotlib}
\bibfield{author}{\bibinfo{person}{The Matplotlib~Development Team}.}
  \bibinfo{year}{2023}\natexlab{g}.
\newblock \bibinfo{title}{Matplotlib}.
\newblock \bibinfo{howpublished}{\url{https://matplotlib.org/}}.
\newblock


\bibitem[\protect\citeauthoryear{To, Soto, and Markl}{To et~al\mbox{.}}{2018}]%
        {to2018survey}
\bibfield{author}{\bibinfo{person}{Quoc-Cuong To}, \bibinfo{person}{Juan Soto},
  {and} \bibinfo{person}{Volker Markl}.} \bibinfo{year}{2018}\natexlab{}.
\newblock \showarticletitle{A survey of state management in big data processing
  systems}.
\newblock \bibinfo{journal}{\emph{The VLDB Journal}} \bibinfo{volume}{27},
  \bibinfo{number}{6} (\bibinfo{year}{2018}), \bibinfo{pages}{847--872}.
\newblock


\bibitem[\protect\citeauthoryear{University}{University}{2021a}]%
        {tutorial}
\bibfield{author}{\bibinfo{person}{Cornell University}.}
  \bibinfo{year}{2021}\natexlab{a}.
\newblock \bibinfo{title}{Cornell Virtual Workshop Tutorial Notebooks}.
\newblock
  \bibinfo{howpublished}{\url{https://github.com/CornellCAC/CVW_PyDataSci2}}.
\newblock


\bibitem[\protect\citeauthoryear{University}{University}{2021b}]%
        {interactive}
\bibfield{author}{\bibinfo{person}{Cornell University}.}
  \bibinfo{year}{2021}\natexlab{b}.
\newblock \bibinfo{title}{Investigating Tweet Timelines Using Interactive Bokeh
  Scatterplots}.
\newblock
  \bibinfo{howpublished}{\url{https://github.com/CornellCAC/CVW_PyDataSci2/blob/master/code/interactive_visualization_with_bokeh.ipynb}}.
\newblock


\bibitem[\protect\citeauthoryear{University}{University}{2021c}]%
        {sklearntweet}
\bibfield{author}{\bibinfo{person}{Cornell University}.}
  \bibinfo{year}{2021}\natexlab{c}.
\newblock \bibinfo{title}{SKLearn Tweet Classification}.
\newblock
  \bibinfo{howpublished}{\url{https://github.com/CornellCAC/CVW_PyDataSci2/blob/master/code/sklearn_tweet_classification.ipynb}}.
\newblock


\bibitem[\protect\citeauthoryear{University}{University}{2021d}]%
        {twitternetworks}
\bibfield{author}{\bibinfo{person}{Cornell University}.}
  \bibinfo{year}{2021}\natexlab{d}.
\newblock \bibinfo{title}{Twitter Networks}.
\newblock
  \bibinfo{howpublished}{\url{https://github.com/CornellCAC/CVW_PyDataSci2/blob/master/code/twitter_networks.ipynb}}.
\newblock


\bibitem[\protect\citeauthoryear{Vartak, F.~da Trindade, Madden, and
  Zaharia}{Vartak et~al\mbox{.}}{2018}]%
        {vartak2018mistique}
\bibfield{author}{\bibinfo{person}{Manasi Vartak}, \bibinfo{person}{Joana~M
  F.~da Trindade}, \bibinfo{person}{Samuel Madden}, {and}
  \bibinfo{person}{Matei Zaharia}.} \bibinfo{year}{2018}\natexlab{}.
\newblock \showarticletitle{Mistique: A system to store and query model
  intermediates for model diagnosis}. In \bibinfo{booktitle}{\emph{Proceedings
  of the 2018 International Conference on Management of Data}}.
  \bibinfo{pages}{1285--1300}.
\newblock


\bibitem[\protect\citeauthoryear{Verbitski, Gupta, Saha, Brahmadesam, Gupta,
  Mittal, Krishnamurthy, Maurice, Kharatishvili, and Bao}{Verbitski
  et~al\mbox{.}}{2017}]%
        {verbitski2017amazon}
\bibfield{author}{\bibinfo{person}{Alexandre Verbitski},
  \bibinfo{person}{Anurag Gupta}, \bibinfo{person}{Debanjan Saha},
  \bibinfo{person}{Murali Brahmadesam}, \bibinfo{person}{Kamal Gupta},
  \bibinfo{person}{Raman Mittal}, \bibinfo{person}{Sailesh Krishnamurthy},
  \bibinfo{person}{Sandor Maurice}, \bibinfo{person}{Tengiz Kharatishvili},
  {and} \bibinfo{person}{Xiaofeng Bao}.} \bibinfo{year}{2017}\natexlab{}.
\newblock \showarticletitle{Amazon aurora: Design considerations for high
  throughput cloud-native relational databases}. In
  \bibinfo{booktitle}{\emph{Proceedings of the 2017 ACM International
  Conference on Management of Data}}. \bibinfo{pages}{1041--1052}.
\newblock


\bibitem[\protect\citeauthoryear{Vlad}{Vlad}{2022}]%
        {tpsmar}
\bibfield{author}{\bibinfo{person}{Devlikamov Vlad}.}
  \bibinfo{year}{2022}\natexlab{}.
\newblock \bibinfo{title}{[TPS-Mar] Fast workflow using scikit-learn-intelex}.
\newblock
  \bibinfo{howpublished}{\url{https://www.kaggle.com/code/lordozvlad/tps-mar-fast-workflow-using-scikit-learn-intelex/notebook}}.
\newblock


\bibitem[\protect\citeauthoryear{Wagenmakers and Farrell}{Wagenmakers and
  Farrell}{2004}]%
        {wagenmakers2004aic}
\bibfield{author}{\bibinfo{person}{Eric-Jan Wagenmakers} {and}
  \bibinfo{person}{Simon Farrell}.} \bibinfo{year}{2004}\natexlab{}.
\newblock \showarticletitle{AIC model selection using Akaike weights}.
\newblock \bibinfo{journal}{\emph{Psychonomic bulletin \& review}}
  \bibinfo{volume}{11}, \bibinfo{number}{1} (\bibinfo{year}{2004}),
  \bibinfo{pages}{192--196}.
\newblock


\bibitem[\protect\citeauthoryear{Wannipurage, Marru, and Pierce}{Wannipurage
  et~al\mbox{.}}{2022}]%
        {wannipurage2022framework}
\bibfield{author}{\bibinfo{person}{Dimuthu Wannipurage},
  \bibinfo{person}{Suresh Marru}, {and} \bibinfo{person}{Marlon Pierce}.}
  \bibinfo{year}{2022}\natexlab{}.
\newblock \showarticletitle{A Framework to capture and reproduce the Absolute
  State of Jupyter Notebooks}.
\newblock \bibinfo{journal}{\emph{arXiv preprint arXiv:2204.07452}}
  (\bibinfo{year}{2022}).
\newblock


\bibitem[\protect\citeauthoryear{Xin, Macke, Ma, Liu, Song, and
  Parameswaran}{Xin et~al\mbox{.}}{2018}]%
        {xin2018helix}
\bibfield{author}{\bibinfo{person}{Doris Xin}, \bibinfo{person}{Stephen Macke},
  \bibinfo{person}{Litian Ma}, \bibinfo{person}{Jialin Liu},
  \bibinfo{person}{Shuchen Song}, {and} \bibinfo{person}{Aditya Parameswaran}.}
  \bibinfo{year}{2018}\natexlab{}.
\newblock \showarticletitle{Helix: Holistic optimization for accelerating
  iterative machine learning}.
\newblock \bibinfo{journal}{\emph{arXiv preprint arXiv:1812.05762}}
  (\bibinfo{year}{2018}).
\newblock


\bibitem[\protect\citeauthoryear{Xin, Petersohn, Tang, Wu, Gonzalez,
  Hellerstein, Joseph, and Parameswaran}{Xin et~al\mbox{.}}{2021}]%
        {xin2021enhancing}
\bibfield{author}{\bibinfo{person}{Doris Xin}, \bibinfo{person}{Devin
  Petersohn}, \bibinfo{person}{Dixin Tang}, \bibinfo{person}{Yifan Wu},
  \bibinfo{person}{Joseph~E Gonzalez}, \bibinfo{person}{Joseph~M Hellerstein},
  \bibinfo{person}{Anthony~D Joseph}, {and} \bibinfo{person}{Aditya~G
  Parameswaran}.} \bibinfo{year}{2021}\natexlab{}.
\newblock \showarticletitle{Enhancing the interactivity of dataframe queries by
  leveraging think time}.
\newblock \bibinfo{journal}{\emph{arXiv preprint arXiv:2103.02145}}
  (\bibinfo{year}{2021}).
\newblock


\bibitem[\protect\citeauthoryear{xxHash}{xxHash}{2023}]%
        {xxhash}
\bibfield{author}{\bibinfo{person}{xxHash}.} \bibinfo{year}{2023}\natexlab{}.
\newblock \bibinfo{title}{xxHash - Extremely fast non-cryptographic hash
  algorithm}.
\newblock \bibinfo{howpublished}{\url{https://github.com/Cyan4973/xxHash}}.
\newblock


\bibitem[\protect\citeauthoryear{Yandex}{Yandex}{2023}]%
        {catboost}
\bibfield{author}{\bibinfo{person}{Yandex}.} \bibinfo{year}{2023}\natexlab{}.
\newblock \bibinfo{title}{CatBoost - open-source gradient boosting library}.
\newblock \bibinfo{howpublished}{\url{https://catboost.ai/}}.
\newblock


\bibitem[\protect\citeauthoryear{Zaharia, Chowdhury, Franklin, Shenker, and
  Stoica}{Zaharia et~al\mbox{.}}{2010}]%
        {zaharia2010spark}
\bibfield{author}{\bibinfo{person}{Matei Zaharia}, \bibinfo{person}{Mosharaf
  Chowdhury}, \bibinfo{person}{Michael~J Franklin}, \bibinfo{person}{Scott
  Shenker}, {and} \bibinfo{person}{Ion Stoica}.}
  \bibinfo{year}{2010}\natexlab{}.
\newblock \showarticletitle{Spark: Cluster computing with working sets}. In
  \bibinfo{booktitle}{\emph{2nd USENIX Workshop on Hot Topics in Cloud
  Computing (HotCloud 10)}}.
\newblock


\bibitem[\protect\citeauthoryear{Zgraggen, Zeleznik, and Drucker}{Zgraggen
  et~al\mbox{.}}{2014}]%
        {zgraggen2014panoramicdata}
\bibfield{author}{\bibinfo{person}{Emanuel Zgraggen}, \bibinfo{person}{Robert
  Zeleznik}, {and} \bibinfo{person}{Steven~M Drucker}.}
  \bibinfo{year}{2014}\natexlab{}.
\newblock \showarticletitle{PanoramicData: Data analysis through pen \& touch}.
\newblock \bibinfo{journal}{\emph{IEEE transactions on visualization and
  computer graphics}} \bibinfo{volume}{20}, \bibinfo{number}{12}
  (\bibinfo{year}{2014}), \bibinfo{pages}{2112--2121}.
\newblock


\bibitem[\protect\citeauthoryear{Zhang, Kumar, and R{\'e}}{Zhang
  et~al\mbox{.}}{2016}]%
        {zhang2016materialization}
\bibfield{author}{\bibinfo{person}{Ce Zhang}, \bibinfo{person}{Arun Kumar},
  {and} \bibinfo{person}{Christopher R{\'e}}.} \bibinfo{year}{2016}\natexlab{}.
\newblock \showarticletitle{Materialization optimizations for feature selection
  workloads}.
\newblock \bibinfo{journal}{\emph{ACM Transactions on Database Systems (TODS)}}
  \bibinfo{volume}{41}, \bibinfo{number}{1} (\bibinfo{year}{2016}),
  \bibinfo{pages}{1--32}.
\newblock


\end{thebibliography}
\appendix
\section{Appendix}
\begin{figure*}[t]

\centering
\begin{subfigure}[b]{0.24\linewidth}
\begin{tikzpicture}

\begin{axis}[
    xtick=data,
    width=45mm,
    height=32mm,
    ymin=0,
    ymax=30,
    axis y line*=none,
    axis x line*=none,
    xtick={1,2,3,4,5, 6},
    xticklabel style   = {align=center},
    xticklabels = {1600, 800, 400, 200, 100, 50},
    ytick={0, 10, 20, 30},
    yticklabels={0, 10, 20, 30},
    xlabel=No. cell executions,
    xlabel style={yshift = 2ex},
    ylabel style={yshift=-4ex},
    xmin = 0,
    xmax = 20,
    xtick = {0, 5,10,15.0,20.0},
    xticklabels = {0, 20, 40, 60, 80},
    tick label style={font=\footnotesize},
    legend style={
        at={(-0.2,1.1)},anchor=south west,column sep=2pt,
        draw=black,fill=white,
        /tikz/every even column/.append style={column sep=5pt},
        font=\scriptsize,
    },
    legend cell align={left},
    legend columns=4,
    label style={font=\footnotesize},
    ylabel={Overhead (MB)},
    ymajorgrids,
]

\addplot[GreenColor, densely dashed, thick, mark = *, mark size=1pt]
table[x=x,y=y] {
x y
1 0.585
2 0.607
3 0.628
4 0.644
5 4.932
6 4.372
7 0.820
8 0.833
9 0.872
10 0.886
11 0.902
12 1.487
13 17.585
14 18.417
15 24.496
16 26.343
17 26.380
18 26.396
19 26.422
20 26.442
};
\addplot[BlueColor, thick, mark = *, mark size=1pt]
table[x=x,y=y] {
x y
1 0.027
2 0.034
3 0.049
4 0.060
5 3.425
6 2.818
7 0.119
8 0.127
9 0.155
10 0.162
11 0.171
12 0.650
13 0.716
14 0.716
15 0.885
16 0.894
17 0.914
18 0.923
19 0.929
20 0.950
};
\addlegendentry{User namespace memory usage}
\addlegendentry{\system memory usage}



\end{axis}
\end{tikzpicture}
\vspace{-6mm}
\caption{Mem. overhead, \cite{hwlm}}
\end{subfigure}
\begin{subfigure}[b]{0.24\linewidth}
\begin{tikzpicture}

\begin{axis}[
    xtick=data,
    width=45mm,
    height=32mm,
    ymin=0,
    ymax=40,
    axis y line*=none,
    axis x line*=none,
    xtick={1,2,3,4,5, 6},
    xticklabel style   = {align=center},
    xticklabels = {1600, 800, 400, 200, 100, 50},
    ytick={0, 10, 20, 30, 40},
    yticklabels={0, 10, 20, 30, 40},
    xlabel=No. cell executions,
    xlabel style={yshift = 2ex},
    ylabel style={yshift=-4ex},
    xmin = 0,
    xmax = 20,
    xtick = {0, 6.66, 13.33, 20},
    xticklabels = {0, 5, 10, 15},
    tick label style={font=\footnotesize},
    legend style={
        at={(-0.2,1.1)},anchor=south west,column sep=2pt,
        draw=black,fill=white,
        /tikz/every even column/.append style={column sep=5pt},
        font=\scriptsize,
    },
    legend cell align={left},
    legend columns=4,
    label style={font=\footnotesize},
    ylabel={Overhead (MB)},
    ymajorgrids,
]

\addplot[GreenColor, densely dashed, thick, mark = *, mark size=1pt]
table[x=x,y=y] {
x y
1.33 0.530
2.66 32.545
4 16.545
5.33 16.592
6.66 16.666
8 16.639
9.33 17.703
10.66 17.186
12 17.360
13.33 17.183
14.66 33.327
16 35.372
17.33 35.177
18.66 35.222
20 35.225
};
\addplot[BlueColor, thick, mark = *, mark size=1pt]
table[x=x,y=y] {
x y
1.33 0.011
2.66 0.017
4 0.026
5.33 0.065
6.66 0.032
8 0.576
9.33 0.055
10.66 0.143
12 0.065
13.33 0.068
14.66 0.073
16 0.077
17.33 0.077
18.66 0.080
20 0.082
};



\end{axis}
\end{tikzpicture}
\vspace{-2mm}
\caption{Mem. overhead, \cite{hwex3}}
\end{subfigure}
\begin{subfigure}[b]{0.24\linewidth}
\begin{tikzpicture}

\begin{axis}[
    xtick=data,
    width=45mm,
    height=32mm,
    ymin=0,
    ymax=20,
    axis y line*=none,
    axis x line*=none,
    xtick={1,2,3,4,5, 6},
    xticklabel style   = {align=center},
    xticklabels = {1600, 800, 400, 200, 100, 50},
    ytick={0, 5, 10, 15, 20},
    yticklabels={0, 5, 10, 15, 20},
    xlabel=No. cell executions,
    xlabel style={yshift = 2ex},
    ylabel style={yshift=-4ex},
    xmin = 0,
    xmax = 25,
    xtick = {0, 5,10,15.0,20.0,25.0},
    xticklabels = {0, 5, 10, 15, 20, 25},
    tick label style={font=\footnotesize},
    legend style={
        at={(-0.2,1.1)},anchor=south west,column sep=2pt,
        draw=black,fill=white,
        /tikz/every even column/.append style={column sep=5pt},
        font=\scriptsize,
    },
    legend cell align={left},
    legend columns=4,
    label style={font=\footnotesize},
    ylabel={Overhead (MB)},
    ymajorgrids,
]

\addplot[GreenColor, thick, mark = *, mark size=1pt]
table[x=x,y=y] {
x y
1 0.591
2 17.049
3 17.051
4 17.051
5 17.051
6 17.063
7 17.067
8 17.075
9 16.319
10 17.644
11 17.656
12 17.671
13 17.677
14 18.689
15 17.692
16 17.694
17 5.391
18 5.481
19 10.805
20 14.161
21 18.797
22 19.084
23 19.038
24 19.060
24 19.060
};
\addplot[BlueColor, thick, mark = *, mark size=1pt]
table[x=x,y=y] {
x y
1 0.022
2 0.024
3 0.025
4 0.026
5 0.027
6 0.028
7 0.029
8 0.030
9 0.031
10 0.032
11 0.038
12 0.042
13 0.044
14 0.044
15 0.122
16 0.123
17 0.124
18 0.125
19 0.126
20 0.281
21 0.284
22 0.298
23 0.528
24 0.465
25 0.473
};



\end{axis}
\end{tikzpicture}
\vspace{-2mm}
\caption{Mem. overhead, \cite{customerchurn}}
\end{subfigure}
\begin{subfigure}[b]{0.24\linewidth}
\begin{tikzpicture}

\begin{axis}[
    xtick=data,
    width=45mm,
    height=32mm,
    ymin=0,
    ymax=100,
    axis y line*=none,
    axis x line*=none,
    xtick={1,2,3,4,5, 6},
    xticklabel style   = {align=center},
    xticklabels = {1600, 800, 400, 200, 100, 50},
    ytick={0, 25, 50, 75, 100},
    yticklabels={0, 25, 50, 75, 100},
    xlabel=No. cell executions,
    xlabel style={yshift = 2ex},
    ylabel style={yshift=-4ex},
    xmin = 0,
    xmax = 20,
    xtick = {0, 5,10,15.0,20.0},
    xticklabels = {0, 25\%, 50\%, 75\%, 100\%},
    tick label style={font=\footnotesize},
    legend style={
        at={(-0.2,1.1)},anchor=south west,column sep=2pt,
        draw=black,fill=white,
        /tikz/every even column/.append style={column sep=5pt},
        font=\scriptsize,
    },
    legend cell align={left},
    legend columns=4,
    label style={font=\footnotesize},
    ylabel={Time (ms)},
    ymajorgrids,
]

\addplot[GreenColor, thick, mark = *, mark size=1pt]
table[x=x,y=y] {
x y
1 0.35
2 0.39
3 0.47
4 0.47
5 13.4
6 92.2
7 0.94
8 0.98
9 0.16
10 1.14
11 1.18
12 2.93
13 5.28
14 5.12
15 7.45
16 6.53
17 7.89
18 7.06
19 7.13
20 7.51
};
\addplot[BlueColor, thick, mark = *, mark size=1pt]
table[x=x,y=y] {
x y
1 0.17
2 0.22
3 0.35
4 0.69
5 0.41
6 0.73
7 0.40
8 0.39
9 0.31
10 0.63
11 0.66
12 0.67
13 0.67
14 0.67
15 0.30
};
\addplot[YellowColor, thick, mark = *, mark size=1pt]
table[x=x,y=y] {
x y
1 0.22
2 0.22
3 0.24
4 0.52
5 0.25
6 0.51
7 0.25
8 0.47
9 0.33
10 0.48
11 1.6
12 2.7
13 1.7
14 2.0
15 1.7
16 4.6
17 4.7
18 5.6
19 24.6
20 0.70
};

\addlegendentry{\cite{hwlm}}
\addlegendentry{\cite{hwex3}}
\addlegendentry{\cite{customerchurn}}



\end{axis}
\end{tikzpicture}
\vspace{-6mm}
\caption{Per-cell time overhead}
\end{subfigure}
\vspace{-3mm}

\caption{Runtime and memory overhead of \system during notebook use on selected homework notebooks. Memory overhead is consistently low, and per-cell runtime overhead is negligible for most cell executions.}

\label{fig:exp_overhead_homework}
\end{figure*}
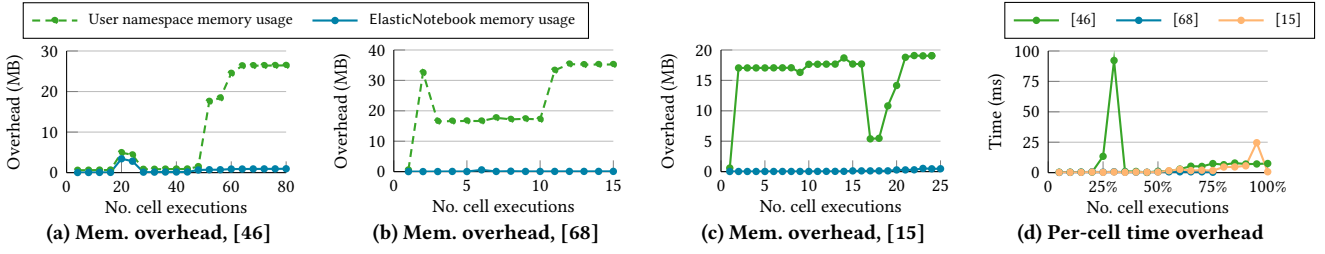
\subfile{plots/exp_faster_migrate_downscale}
\begin{figure}[t]
\begin{subfigure}[b]{\linewidth}
\centering
\begin{tikzpicture}[>={LaTeX[width=1mm,length=1mm]},->]

  \node(ce1) [draw=purple, thick,anchor=north west, minimum width = 4mm, minimum height=4mm, circle, inner sep = 0.4mm]
 at ($(notebook.north east) + (2.2, 0.3)$) {\footnotesize\textcolor{purple}{\bm{$c_{t_1}$}}};

\node(x1) [draw=purple,densely dashed,thick, fill=Lightgrey, opacity = 0.7, anchor=center, minimum height=3mm, inner sep = 0.6mm]
at ($(ce1) + (-0.8, -0.55)$) {\footnotesize \textcolor{purple}{\textbf{(}\texttt{\textbf{x}}\textbf{,} \bm{$t_1$}\textbf{)}}};
\node(y1) [draw=black, anchor=center, minimum height=3mm, inner sep = 0.6mm]
at ($(ce1) + (0.8, -0.55)$) {\footnotesize (\texttt{y}, $t_1$)};

  \node(ce2) [draw=purple, thick,anchor=center, minimum width = 4mm, minimum height=4mm, circle, inner sep = 0.4mm]
 at ($(ce1) + (0, -1.1)$) {\footnotesize \textcolor{purple}{\bm{$c_{t_2}$}}};

\node(z1) [draw=purple,fill opacity = 0.2, thick, text opacity = 1, anchor=center,  minimum height=3mm, inner sep = 0.6mm]
at ($(ce2) + (0, -0.55)$) {\footnotesize\textcolor{purple}{\textbf{(}\texttt{\textbf{z}}\textbf{,} \bm{$t_2$}\textbf{)}}};

  \node(ce3) [draw=black, anchor=center, minimum width = 4mm, minimum height=4mm, circle, inner sep = 0.4mm]
 at ($(z1) + (0, -0.55)$) {\footnotesize $c_{t_3}$};

\node(x2) [draw=black, anchor=center, minimum height=3mm, inner sep = 0.6mm]
at ($(ce3) + (-0.8, -0.55)$) {\footnotesize (\texttt{x}, $t_3$)};
\node(l1) [draw=black, anchor=center,  minimum height=3mm, inner sep = 0.6mm]
at ($(ce3) + (0.8, -0.55)$) {\footnotesize (\texttt{l1}, $t_3$)};

  \node(ce4) [draw=RandomColor, anchor=center, thick, minimum width = 4mm, minimum height=4mm, circle, inner sep = 0.4mm]
 at ($(ce3) + (0, -1.1)$) {\footnotesize \textcolor{RandomColor}{\bm{$c_{t_4}$}}};

 \node(gen1) [draw=RandomColor,densely dashed, fill=Lightgrey, thick,opacity = 0.7, anchor=center, minimum height=3mm, inner sep = 0.6mm]
at ($(ce4) + (-0.8, -0.55)$) {\footnotesize \textcolor{RandomColor}{\textbf{(}\texttt{\textbf{gen}}\textbf{,} \bm{$t_4$}\textbf{)}}};
\node(2dlist) [draw=black, anchor=center,  minimum height=3mm, inner sep = 0.6mm]
at ($(ce4) + (0.8, -0.55)$) {\footnotesize (\texttt{2dlist}, $t_4$)};

   \node(ce5) [draw=RandomColor, thick, anchor=center, minimum width = 4mm, minimum height=4mm, circle, inner sep = 0.4mm]
 at ($(ce4) + (0, -1.1)$) {\footnotesize \textcolor{RandomColor}{\bm{$c_{t_5}$}}};

 \node(gen2) [draw=RandomColor,thick,fill opacity = 0.2,text opacity = 1, anchor=center, minimum height=3mm, inner sep = 0.6mm]
at ($(ce5) + (0, -0.55)$) {\footnotesize \textcolor{RandomColor}{\textbf{(}\texttt{\textbf{gen}}\textbf{,} \bm{$t_5$}\textbf{)}}};

 \draw[->, purple,thick] 
 ($(ce1.south)$) --
($(x1.north)$);
 \draw[->] 
 ($(ce1.south)$) --
($(y1.north)$);

 \draw[->, purple,thick] 
 ($(x1.south)$) --
($(ce2.north)$);
 \draw[->] 
 ($(y1.south)$) --
($(ce2.north)$);

 \draw[->, purple,thick] 
 ($(ce2.south)$) --
($(z1.north)$);

 \draw[-]
 ($(x1.south)$) --
($(x1) + (0, -1.65)$);
 \draw[->] 
 ($(x1) + (0, -1.65)$) --
($(ce3.west)$);

 \draw[->] 
 ($(z1.south)$) --
($(ce3.north)$);

 \draw[->] 
 ($(ce3.south)$) --
($(x2.north)$);
 \draw[->] 
 ($(ce3.south)$) --
($(l1.north)$);

 \draw[->] 
 ($(x2.south)$) --
($(ce4.north)$);

 \draw[->,RandomColor,thick] 
 ($(ce4.south)$) --
($(gen1.north)$);
 \draw[->] 
 ($(ce4.south)$) --
($(2dlist.north)$);

 \draw[->,RandomColor,thick] 
 ($(gen1.south)$) --
($(ce5.north)$);
 \draw[->,RandomColor,thick] 
 ($(ce5.south)$) --
($(gen2.north)$);

  \node(ce11) [draw=black, anchor=center, minimum width = 4mm, minimum height=4mm, circle, inner sep = 0.4mm]
 at ($(ce1) + (4, 0)$) {\footnotesize $c_{t_1}$};

\node(x11) [draw=black,densely dashed, fill=Lightgrey, opacity = 0.7, anchor=center, minimum height=3mm, inner sep = 0.6mm]
at ($(ce11) + (-0.8, -0.55)$) {\footnotesize(\texttt{x}, $t_1$)};
\node(y11) [draw=black, anchor=center, minimum height=3mm, inner sep = 0.6mm]
at ($(ce11) + (0.8, -0.55)$) {\footnotesize(\texttt{y}, $t_1$)};

  \node(ce21) [draw=purple, thick,anchor=center, minimum width = 4mm, minimum height=4mm, circle, inner sep = 0.4mm]
 at ($(ce11) + (0, -1.1)$) {\footnotesize \textcolor{purple}{\bm{$c_{t_2}$}}};

\node(z11) [draw=purple,fill opacity = 0.2, thick, text opacity = 1, anchor=center,  minimum height=3mm, inner sep = 0.6mm]
at ($(ce21) + (0, -0.55)$) {\footnotesize\textcolor{purple}{\textbf{(}\texttt{\textbf{z}}\textbf{,} \bm{$t_2$}\textbf{)}}};

  \node(ce31) [draw=black, anchor=center, minimum width = 4mm, minimum height=4mm, circle, inner sep = 0.4mm]
 at ($(z11) + (0, -0.55)$) {\footnotesize $c_{t_3}$};

\node(x21) [draw=black, anchor=center, minimum height=3mm, inner sep = 0.6mm]
at ($(ce31) + (-0.8, -0.55)$) {\footnotesize(\texttt{x}, $t_3$)};
\node(l11) [draw=black, anchor=center,  minimum height=3mm, inner sep = 0.6mm]
at ($(ce31) + (0.8, -0.55)$) {\footnotesize(\texttt{l1}, $t_3$)};

  \node(ce41) [draw=RandomColor, anchor=center, thick, minimum width = 4mm, minimum height=4mm, circle, inner sep = 0.4mm]
 at ($(ce31) + (0, -1.1)$) {\footnotesize \textcolor{RandomColor}{\bm{$c_{t_4}$}}};

 \node(gen11) [draw=RandomColor,densely dashed, fill=Lightgrey, thick,opacity = 0.7, anchor=center, minimum height=3mm, inner sep = 0.6mm]
at ($(ce41) + (-0.8, -0.55)$) {\footnotesize \textcolor{RandomColor}{\textbf{(}\texttt{\textbf{gen}}\textbf{,} \bm{$t_4$}\textbf{)}}};
\node(2dlist11) [draw=black, anchor=center,  minimum height=3mm, inner sep = 0.6mm]
at ($(ce41) + (0.8, -0.55)$) {\footnotesize(\texttt{2dlist}, $t_4$)};

   \node(ce51) [draw=black, anchor=center, minimum width = 4mm, minimum height=4mm, circle, inner sep = 0.4mm]
 at ($(ce41) + (0, -1.1)$) {\footnotesize $c_{t_5}$};

 \draw[->]
 ($(ce11.south)$) --
($(x11.north)$);
 \draw[->] 
 ($(ce11.south)$) --
($(y11.north)$);

 \draw[->] 
 ($(y11.south)$) --
($(ce21.north)$);

 \draw[->, purple,thick] 
 ($(ce21.south)$) --
($(z11.north)$);

 \draw[-]
 ($(x11.south)$) --
($(x11) + (0, -1.65)$);
 \draw[->] 
 ($(x11) + (0, -1.65)$) --
($(ce31.west)$);

 \draw[->] 
 ($(z11.south)$) --
($(ce31.north)$);

 \draw[->] 
 ($(ce31.south)$) --
($(x21.north)$);
 \draw[->] 
 ($(ce31.south)$) --
($(l11.north)$);

 \draw[->] 
 ($(x21.south)$) --
($(ce41.north)$);

 \draw[->, RandomColor,thick] 
 ($(ce41.south)$) --
($(gen11.north)$);
 \draw[->] 
 ($(ce41.south)$) --
($(2dlist11.north)$);

 \draw[->] 
 ($(gen11.south)$) --
($(ce51.north)$);

 \draw[-, gray, densely dashed] 
 ($(ce1.north) + (1.8, 0)$) --
($(ce1.north) + (1.8, -4.7)$);

\node[anchor=north west, align=left] at ($(ce1.north) + (-2.0, 0)$) {\Huge$\mathcal{G}$};
\node[anchor=north west, align=left] at ($(ce11.north) + (-2.0, 0)$) {\Huge$\mathcal{G}^*$};

\node[anchor=north west, align=center] at ($(ce11.north) + (-3.0, -4.9)$) {\small \textcolor{purple}{$req^*(x) = \{c_{t_2}\} \subseteq req(x) = \{c_{t_1}, c_{t_2}\}$}\\[-0.2em]\small \textcolor{RandomColor}{$req^*(gen) = \{c_{t_4}\} \subseteq req(gen) = \{c_{t_4}, c_{t_5}\}$}};



\node(examplex1) [draw=black,densely dashed, fill=Lightgrey, opacity = 0.7, anchor=north west, minimum height=3mm, inner sep = 0.6mm]
at ($(gen2.south west) + (-1.5, -0.5)$) {\footnotesize (\texttt{x}, $t_1$)};
\node[anchor=west, align=left] at ($(examplex1.east) + (0,0)$) {\small (Overwritten/deleted)\\[-0.3em]\small Variable Snapshot};

  \node(examplex2) [draw=black, anchor=west, minimum width = 4mm, minimum height=4mm, circle, inner sep = 0.4mm]
 at ($(examplex1.east) + (2.8, 0)$) {\footnotesize $c_{t_1}$};
\node[anchor=west, align=left] at ($(examplex2.east) + (0,0)$) {\small Cell \\[-0.2em]\small Execution};

\node(examplex3) [draw=black, anchor=west, minimum height=3mm, inner sep = 0.6mm]
at ($(examplex2.east) + (1.5, 0)$) {\footnotesize(\texttt{x}, $t_1$)};
\node[anchor=west, align=left] at ($(examplex3.east) + (0,0)$) {\small Active\\[-0.3em]\small Variable Snapshot};

\end{tikzpicture}

\end{subfigure}
\vspace{-7mm}

\caption{AHG $\mathcal{G}$ may contain false positives compared to the true AHG $\mathcal{G}^*$.
The correctness is still ensured,
    while the efficiency may be affected due to extra cells re-running,
        for example, when recomputing \texttt{z} (green) and \texttt{gen} (red).
}
\label{fig:reconstruct_correctness}
\end{figure}
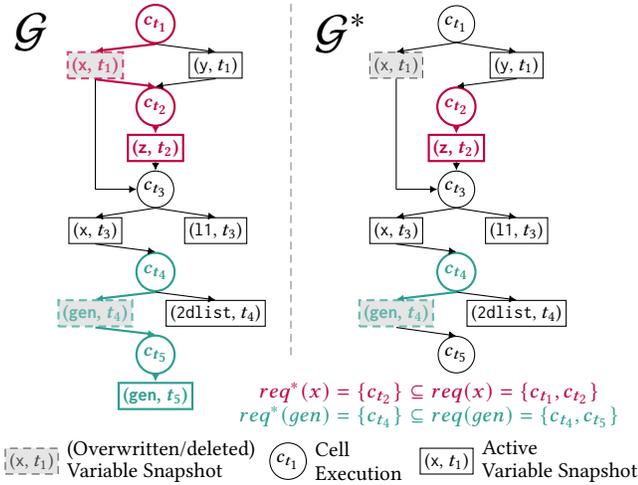

\subsection{Low Per-cell overhead}
\label{sec:appendix_percell}

We report the results for per-cell time and memory overheads on 3 \homework notebooks in \cref{fig:exp_overhead_homework}. \system's memory and per-cell monitoring overhead are consistently under 10\% and 1ms, respectively.
There are occasionally 'spikes' when certain cells declaring/modifying complex variables are executed; for example, the 60\% and 91ms memory and time overheads of cell 28 in \cite{handsonml} is attributed to constructing the ID Graph for a complex nested list. However, even in this worst case, the time overhead is still well under the 500ms threshold suggested for interactive data engines~\cite{liu2014effects}, while the memory overhead is of a low absolute value (4MB) compared to the size of the (not yet loaded) datasets, thus having negligible user impact.
\subsection{Proof of Theorem \ref{theorem:correctness}}
\label{sec:appendix_proof}
An illustration of our proof is provided in \Cref{fig:reconstruct_correctness}.
\begin{proof}
As there are no false negatives, the true AHG $\mathcal{G}^*$ is contained within the approximate AHG $\mathcal{G}$, i.e., $\mathcal{G}^* \subseteq \mathcal{G}$ (\Cref{fig:reconstruct_correctness}).
Let $x$ be a an arbitrary variable, and $(x, t_{\mathcal{G}})$, $(x, t_{\mathcal{G}^*})$ be its active VSs in $\mathcal{G}$ and $\mathcal{G}^*$ respectively. There is $t_{\mathcal{G}} \geq t_{\mathcal{G}^*}$: if $t_{\mathcal{G}} > t_{\mathcal{G}^*}$ (due to falsely implied non-overwrite modifications, i.e., \texttt{gen} in \Cref{fig:reconstruct_correctness}) then there must be a path from $(x, t_{\mathcal{G}})$ to $(x, t_{\mathcal{G}^*})$: $(x, t_{\mathcal{G}}), c_{t_{\mathcal{G}}}, (x, t_{k_1}), c_{t_{k_1}}$ ,..., $(x, t_{k_l}), c_{t_{k_l}}, (x, t_{\mathcal{G}^*})$, where $t_{\mathcal{G}} < t_{k_1} < ... < t_{k_l} < t_{\mathcal{G}^*}$ and $c_{t_{k_1}},...,c_{t_{k_l}}$ all contain false non-overwrite modifications to $x$.
Therefore, the subtree rooted at $(x, t_{\mathcal{G}})$ in $\mathcal{G}$ must be contained the subtree rooted at $(x, t_{\mathcal{G}^*})$ in $\mathcal{G}^*$, hence $req^*(x, t_{\mathcal{G}^*})\subseteq req(x, t_{\mathcal{G}})$.
\end{proof}

\subsection{Handling Large Pandas Dataframes}

To avoid hashing large Pandas dataframes after each cell execution, \system uses the dataframes' underlying \texttt{writeable} flag as a dirty bit to detect in-place changes: before each cell execution, the \texttt{writeable} flag is set to \texttt{False}, and the dataframe is identified as modified if the flag has been flipped to \texttt{True} after the cell execution.

\balance

\end{document}